\documentclass[msc]{ppgccufmg}    

\usepackage[english]{babel}       
\usepackage[utf8]{inputenc}
\usepackage[T1]{fontenc}
\usepackage{type1ec}
\usepackage{graphicx}
\usepackage{tabulary}
\usepackage{rotating}
\usepackage[table,xcdraw]{xcolor}
\usepackage[show]{chato-notes}

\usepackage[a4paper,
  portuguese,
  bookmarks=true,
  bookmarksnumbered=true,
  linktocpage,
  colorlinks,
  citecolor=black,
  urlcolor=black,
  linkcolor=black,
  filecolor=black,
  ]{hyperref}
\usepackage[square]{natbib}
\usepackage{listings}
\usepackage{multirow}
\usepackage{framed}
\usepackage{pdflscape}
\usepackage{subfig}
\usepackage{epstopdf}
\usepackage{epsfig}
\usepackage{float}
\usepackage{amssymb}
\usepackage{pifont}
\usepackage{esint}
\usepackage{url}
\usepackage{lscape}

\begin{document}



\ppgccufmg{
  title={Characterizing Interconnections and Linguistic Patterns in Twitter},
  authorrev={Afonso, Johnnatan Messias Peixoto},
  cutter={A257c},
  cdu={519.6*04 (043)},
  university={Universidade Federal de Minas Gerais},
  course={Computer Science},
  address={Belo Horizonte},
  date={2017-06},
  advisor={Fabrício Benevenuto de Souza},
  keywords={Computação -- Teses, Redes sociais on-line, Twitter, Dados demográficos, Igualdade, Tipologia (Linguística), Gênero e raça},
 abstract={Abstract}{abstract},
 dedication={dedication},
   ack={acknowledgments}
}

 \chapter{Introduction} \label{chap:intro}

Nowadays, millions of people constantly use online social networking sites, such as Facebook and Twitter. In the third quarter of 2016, Facebook and Twitter had $1.79$ billion\footnote{\url{https://goo.gl/8GC6Ii}} and $317$ million\footnote{\url{http://www.statista.com/statistics/282087/number-of-monthly-active-twitter-users/}} monthly active users, respectively, sharing content about their daily lives and things that happen around them. These systems have revolutionized the way we communicate, by organizing our offline social relationships in a digital form.  

Increasingly, our society has been treating actions in the online and offline space in an indistinguishable way. It is now common to see cases in which content posted on Facebook or Twitter is used to vet job candidates, support divorce litigation, or terminate employees \cite{dutta2010s,ronson2016so}. Although these platforms provide a democratic space for conversations and debates, they also open space for key unsolved issues of our society, such as gender and race inequalities. 

The massive popularity of online social media also provides the opportunity to detect useful characteristics and patterns about users and their interconnections. For instance, patterns are valuable for marketing and advertisement companies which capture users behavior and needs in order to promote products, specifically on a target group. In terms of groups, demographics constitutes a significant factor to cluster people and understand their behavior. 

Twitter is a micro-blogging platform in which the main form of communication and interaction is based on posting text messages. It is also a very large system with lots of users. Consequently, it gains the attention of companies and researchers to explore who are these users and what they post on this social network. Although it provides a log of a number of human interaction and social connections, its data is unstructured and lacks information about demographics of users, such as gender, race, and age. So, the existing efforts that attempt to explore demographics of users on this space rely on different inference strategies. These efforts include \cite{blevins2015jane,karimi2016,liu2013s,mislove2011understanding}, and \cite{nilizadeh2016twitter}.

Despite the magnitude of Twitter and the amount of the provided information, inferring such demographic aspects of users is challenging and the existing works have several limitations due to the difficulty to identify and gather such demographic aspects. However, the potential to measure demographic aspects is immense and valuable for different research purposes and applications, such as design systems, to understand the social aspects of our society, and how different demographic groups interact with each other. This dissertation focuses on this context. In the next section, we present the main motivations of our work.

\section{Motivation}

Exploring demographic aspects is important both from the perspective of systems and from the sociological point of view. In the first, demographic aspects allow us to understand groups of users that perform certain actions and interactions within a system. As from the sociological point of view, demographic aspects also help us to understand the gender and race differences in online interactions and, consequently, to better understand our society.

Demographic aspects also allow us to study the linguistic differences as well as diversity in topics of interest between gender and race. We believe that studying the differences in how demographic groups write content about specific topics would help systems developers to provide more transparency to users in their systems. The ability of understanding transparency is for algorithms that operate on large-scale crowdsourced data, the goal is to make the outputs of the algorithms (and the algorithm itself) transparent, as it is also important to understand the non-uniformities in the inputs to the algorithms.

Recent efforts show gender disparities on online systems which highlight a concern regarding the inequalities and the main barriers that keep a given demographic from rising beyond a certain level in a hierarchy. For example, women are under-represented in the content production in Wikipedia. Further, black tenants have much fewer chances of renting a place on Airbnb\footnote{\url{http://www.airbnb.com}}. Some drivers for both UberX\footnote{\url{http://www.uber.com}} and Lyft\footnote{\url{http://www.lyft.com}} discriminate on the basis of the perceived race of the traveler. There is also evidence where disadvantages against female users are found among the users with high visibility \cite{nilizadeh2016twitter}, similar to the barrier females face in attaining higher positions in companies (``glass ceiling'' effect) \cite{cotter2001glass}. Therefore, identifying inequalities and asymmetries in demographics in the online world is crucial for the development of features that can promote equality and diversity in these systems. 

Despite the recent surge of interest in demographic aspects of users from social networks, inferring demographic aspects is challenging and limited due to the difficult to identify and gather such aspects from the users in these systems. The main challenge is that demographic information is not directly available through Twitter API and it needs to be inferred by other means. Hence, in our work, we crawl a large scale sample of active Twitter users and then we identify the gender and race of about $1.6$ million users located in the United States by using Face++\footnote{\url{http://www.faceplusplus.com}}, a face recognition software able to recognize gender and race of identifiable faces in the user's profile pictures. Actually, the state of the art algorithms, for pattern recognition and image processing, can provide, with high accuracy, gender, race, and even the age of an individual via his/her image. In the next section, we discuss our goals.

\section{Goals}

The main goal of this dissertation is the use of demographic aspects along four axes:

\begin{itemize}
    \item Investigate inequalities in terms of visibility: We are interested in analyzing the association of the demographic aspects with visibility and discovering possible inequalities.
    \item Investigate linguistic aspects and topic interests: We want to investigate how demographic groups differ from each other in terms of linguistic aspects and topic interests. In other words, we want to find out the extent to which different demographic groups differs from each other in terms the way they post content in Twitter and also the extent to which they have interest in some specific topics.
    \item Characterize interconnections: We want to examine the proportion of connections and interactions among each demographic group.
    \item Use demographic aspects to design transparent systems: Demographics is an important aspect of providing transparency. Therefore, we want to design a system to provide demographic transparency of trending topic in Twitter.
\end{itemize}

In the next section, we present the contributions of this dissertation.

\section{Contributions}

This dissertation presents the following contributions regarding each of our previous goals.

Our first contribution is to provide a clear insight into how groups of users from different demographics connect on Twitter. As an example of finding, we show that white and male users tend to be more followed. Further, our findings reinforce previous observations about disadvantages against female users in Twitter and identify advantages for white users, in comparison with those identified as black and asian. We also show that users perceived as men experience a ``glass ceiling'' when they are identified as black users.

When we compare the difference between the groups, we find interesting trends, such as the higher interest in zodiac signs by females than by males. Also, white users are more interested in politics, writers, and organizations when compared to asians, and more interested in technology, movie, and politics when compared to black users. On the other hand, black users are more into artists, life, and music topics. We find that asians are more interested in artists, actors, and music than whites and tend to have higher interest for movie, companies, and technology when compared to blacks.

Regarding the interconnection of demographic groups, we found that males and females connections take their part of the responsibility on gender inequality, and that race plays a more important role for connections with white males. Our analysis among these groups explain part of causes of such inequalities and offer hints for the promotion of equality in the online space.

Finally, we believe our findings not only contribute to the social science literature, but we also hope that they can contribute to the recommendation systems design that can act to diminish such inequalities. Thus, we designed the \textit{Who Makes Trends?}\footnote{\url{http://twitter-app.mpi-sws.org/who-makes-trends}} Web-based service in order to provide demographic transparency of trending topics in Twitter.

The results presented in this dissertation are part of the following papers (in chronological order of publication):
\begin{itemize}
    \item \textbf{Messias, J.}, Vikatos, P., and Benevenuto, F. (2017). White, Man, and Highly Followed: Gender and Race Inequalities in Twitter. In Proceedings of the IEEE/WIC/ACM International Conference on Web Intelligence (WI). \cite{messiasWI2017}
    \item Vikatos, P., \textbf{Messias, J.}, Miranda, M., and Benevenuto, F. (2017). Linguistic Diversities of Demographic Groups in Twitter. In Proceedings of the 28th ACM Conference on Hypertext and Social Media (Hypertext). \cite{vikatos2017}
    \item Reis, J. C. S., Kwak, H., An, J., \textbf{Messias, J.}, and Benevenuto, F. (2017). Demographics of News Sharing in the U.S. Twittersphere. In Proceedings of the 28th ACM Conference on Hypertext and Social Media (Hypertext). \cite{reis2017}
    \item Chakraborty, A., \textbf{Messias, J.}, Benevenuto, F., Ghosh, S., Ganguly, N., and Gummadi, K. P. (2017). Who Makes Trends? Understanding Demographic Biases in Crowdsourced Recommendations. In Proceedings of the 11th International AAAI Conference on Web and Social Media (ICWSM). \cite{chakraborty2017@icwsm}
    \item Kulshrestha, J., Eslami, M., \textbf{Messias, J.}, Zafar, M. B., Ghosh, S., Gummadi, K. P., and Karahalios, K. (2017). Quantifying Search Bias: Investigating Sources of Bias for Political Searches in Social Media. In Proceedings of the 20th ACM Conference on Computer-Supported Cooperative Work and Social Computing (CSCW). \cite{juhi2017}
\end{itemize}

In the next section, we present the dissertation outline.

\section{Dissertation Outline}  \label{sec:outline}

This dissertation is organized as follows.

Chapter~\ref{chap:relatedwork} briefly reviews related efforts in the literature along six axes. First, we discuss the methodology used by efforts that measure demographic aspects in Twitter. Second, we discuss the main efforts that have explored gender and race inequality in social networks. Third, we refer studies that combine linguistic with demographic status. Then, we introduce some efforts regarding algorithmic and data transparency. Moreover, we discuss recommendation diversity. Next, we discuss the main efforts regarding fairness.

In Chapter~\ref{chap:dataset}, we present our methodology to gather the demographic data of Twitter users. It consists of gathering tweets of U.S. users and using their profile picture in order to gather their demographic information. We also discuss how we infer the linguistic metrics of tweets using the Linguistic Inquiry and Word Count (LIWC). Then, we describe our methodology to gather the social connections and interactions between users. Finally, we discuss the potential limitations of our data.

Next, Chapter~\ref{chap:results_inequalities} characterizes gender and race inequalities in visibility by analyzing the association of demographic status with visibility and discovering possible inequalities.

In Chapter \ref{chap:linguistics}, we introduce linguistic patterns of users based on gender and race. We also analyze the differences in terms of topics interests.

Then, Chapter~\ref{chap:results_interconnections} discusses results in terms of connections and interactions in Twitter. In other words, we analyze the interconnection among groups of users separated by gender and race and the probability of male and female users to be friends with (i.e. to follow) other male and females user. 

Chapter~\ref{chap:system} presents a system design of demographic distribution of Twitter trending topics and describes its data collection and trending topic analysis steps.

Finally, Chapter~\ref{chap:conclusion} concludes the dissertation and offers directions for future work. 

 \chapter{Related Work} \label{chap:relatedwork}

In this section, we review the related literature along six axes. First, we discuss the methodology used by efforts that measure demographic aspects in Twitter. Second, we discuss the main efforts that have explored gender and race inequality in social networks. Third, we refer studies that combine linguistic with demographic status. Then, we introduce some efforts regarding algorithmic and data transparency. Moreover, we discuss recommendation diversity. Next, we discuss the main efforts regarding fairness. Finally, we present the concluding remarks.

\section{Demographics in Social Media} 

There are several studies that retrieve information from social media to analyze and predict real-world phenomena such as stock market~\cite{bollen2011twitter}, migration~\cite{Messias2016@asonam}, election~\cite{o2010tweets, tumasjan2010predicting}, and also political leaning inference of Twitter users \cite{juhi2017,pennacchiotti2011machine}. 

One of the first efforts to extract and analyze demographic information presents a comparative study between the demographic distribution of gender/race of Twitter users and U.S. population \cite{mislove2011understanding}.
After that, several efforts have arisen that investigate demographic information, in various social media, using different strategies for distinct purposes \cite{blevins2015jane, karimi2016, burger2011discriminating,hannak2017@icwsm,Hannak2017CSCW}.

Recent studies focused on demographics~\cite{blevins2015jane,karimi2016,liu2013s,nilizadeh2016twitter,vikatos2017} present methodologies to extract the necessary data through analysis and pattern matching of screen/full name as well as descriptions of user profiles and image in the profile status. Particularly, \cite{chen2015comparative} focus on demographic inference using profile self-descriptions and profile images. They analyze five signals in order to categorize users to their demographic status. They use users' names, self-descriptions, tweets, social networks, and profile images to infer demographic aspects (ethnicity, gender, age). \cite{culotta2015predicting} describe an alternative methodology assuming  that the demographic profiles  of visitors to a website are correlated with the demographic profiles of followers of that website on social network (e.g. Twitter), and  propose a regression model to determine demographic aspects (gender, age, ethnicity, education, income, and child status). In another study, \cite{reis2017} study demographic through news sharing in U.S. perspective and show that male and white users tend to be more active in terms of sharing news, biasing news audience to the interest of these demographic groups.

Another study, \cite{an2016greysanatomy} examine the correlation of hashtags that are used in different demographic groups, using Face++ and users' profile pictures. They show that this strategy is reliable and provides accurate demographic information for gender and race. More recently, \cite{chakraborty2017@icwsm} analyze the demographic of Twitter trending topics, using Face++ as well. They use an extensive dataset collected from Twitter in order to make the first attempt to quantify and explore the demographic biases in crowdsourced recommendations. They find that a large fraction of trends is promoted by crowds whose demographics are significantly different from the overall Twitter population. More worryingly, certain demographic groups are systematically under-represented among the promoters of the trending topics.

\section{Inequality in Twitter Visibility}

Homophily states that similarity breeds social connections~\cite{mcpherson2001birds}. In our society, people tend to characterize and categorize others in terms of demographic aspects, such as gender and race~\cite{willis2006first}, making them intuitively natural concepts through which people might account similarities. The existing literature is still limited in explaining the inequalities that arise from the establishments of these social connections in Twitter. 

Most of the existing efforts in this space attempt to quantify and understand other factors that are highly correlated with visibility (i.e. number of followers). Particularly, \cite{morales2014efficiency} show that user's reputation and her impact on society constitute significant factors of the online visibility growth. Other efforts show that the attention that users gain is correlated with his/her online behavior in terms of the frequency of their interactions~\cite{Comarela:2012:UFA:2309996.2310017,romero2011influence,wu2011says}. \cite{Freitas2015@asonam} created $120$ socialbots in Twitter, with different behaviors and demographics aspects. They show that users who post more, post about specific topics, reply, favorite or mention others frequently, may acquire more followers, and consequently, more visibility. 
Despite the extreme importance of these efforts, they do not investigate the correlation of demographic issues with visibility. Thus, our effort is complementary to theirs. 

There have been efforts quantifying differences and inequalities in many other social media, including Wikipedia~\cite{graells2015first,wagner2016women} and Pinterest~\cite{Gilbert:2013:INT:2470654.2481336}. 
However, the study that is closest related to ours explores gender inequalities in Twitter~\cite{nilizadeh2016twitter}. Authors show that gender may allow inequality to persist in terms of online visibility. This dissertation further explores race as a new demographic dimension. More important, unlike previous studies, our work focuses on investigating how different demographic groups (i.e. male/female, asian/black/white) connect with other.

\section{Demographics and Linguistic Analysis} 

In the field of demographics, most studies use linguistic analysis in order to extract useful features for predicting demographic information as gender, race, and age. \cite{burger2011discriminating} produce $ngrams$ from users' tweets, description, screen name, and full name, in order to predict Twitter user gender. They conclude that the training of an SVM classifier with the combination of all factors can create an efficient and accurate prediction scheme ($92\%$ accuracy) for gender classification. Also, \cite{chen2015comparative} introduce a similar methodology for predicting gender, ethnicity, and age. However, using ngrams from the social neighbors, including followers and friends, and the distribution of 100 generated topics of LDA algorithm as the input of SVM classifier. In their results, the performance of classification is much lower in terms of ethnicity and age. \cite{Gilbert:2013:INT:2470654.2481336} present an interesting statistical overview in Twitter and Pinterest using textual analysis and comparing what users write on Pinterest to what write text in Twitter. \cite{Cunha:2012:GBS:2309996.2310055} used Twitter data to analyze the difference between males and females in terms of generation of hashtags. Their results emphasize gender as a factor that influences the user's choice of specific hashtags to a specific topic.

We motivate this research topic based on \cite{DeChoudhury:2017:GCD:2998181.2998220} study which discovers gender and cultural differences in Twitter. They correlate several linguistic features to mental illness.
Our findings reinforce their observations about linguistic and topical differences among male and female users in Twitter and also contribute with a new analysis of race~\cite{vikatos2017}.

\section{Algorithmic and Data Transparency}
Increasingly, researchers and governments are recognizing the importance of making algorithms transparent. The White House recently released a report that concludes that the practitioners must ensure that AI-enabled systems are open, transparent, and understandable~\cite{white_house_transparency}. In news production, black-box automated writing bots or any other difficult-to-parse algorithmic systems used to produce or curate news may represent infringements on the basic ethics of journalism~\cite{diakopoulos2016algorithmic}.

Indeed, the controversy around Facebook using human editors on their trending topics teaches a lesson about the importance of transparency. On one hand, when humans were editing trending topics, they were accused to select and filter content~\cite{facebook_bias}. On the other hand, when humans were removed from the process, Facebook was accused of featuring fake news as trending~\cite{Ohlheiser2016}. Recently, Twitter announced the use of deep learning in their timeline algorithm and how it works in order to make it transparent to the public~\cite{twitter_timeline_deep}.

Demographics is also an important aspect of providing transparency. Therefore, demographic distribution of users may be leveraged to make other crowdsourced systems (e.g., social search~\cite{juhi2017} and Twitter trending topic analysis~\cite{chakraborty2017@icwsm}) transparent as well. For algorithms that operate on large-scale crowdsourced data, it is paramount to make the outputs of the algorithms (and the algorithm itself) transparent, and understand the non-uniformities in the inputs to the algorithms.

\section{Recommendation Diversity}
Diversity is increasingly recognized as an important metric for evaluating the effectiveness of online recommendations~\cite{Chen2016SIP}. There have been multiple attempts in the past to introduce diversity in content search and recommendation systems. \cite{gollapudi2009axiomatic} and \cite{agrawal2009diversifying} explored several approaches for search result diversification. \cite{ziegler2005improving} proposed algorithms to introduce \textit{Novelty and Serendipity} in collaborative filtering based recommendations. Basically, search result diversification suggests a trade-off between showing diverse results in the top position for a given query and have more appropriate results that approximate to the user intent~\cite{carbonell199use,chen2006less}. There is also an effort to study stereotypes in Google and Bing search engines. \cite{araujo2016identifying} query the search engines for beauty and ugly females. Their results show the existence of both negative stereotype for black females and positive stereotype for white females in terms of beauty. They also noticed that there are negative stereotypes about older females in U.S. while in hispanic countries share negative stereotypes about black females, positive for whites and neutral for asians. \cite{sheth2011towards} propose the ``Social Diversity'', a new approach to diversity in recommendations that uses social network information to diversify recommendation results.

In 2015, Google search engine suffered from a lack of diversity when their search algorithm labeled black people as ``gorillas''~\cite{guynn2015google,barr2015google}. It reinforces the importancy of using demographic information to provide diversity in recommendation systems. More recently, in 2016, a Microsoft Twitter bot that uses artificial intelligence quickly became a racist~\cite{victor2017nyt}. It also highlights the need for a better understanding of designing systems. Depending on which kind of output the algorithm gives to someone it could ruin her life.

\section{Fairness}
Some countries have anti-discrimination laws in order to prohibit unfair treatment of individuals based on sensitive attributes (e.g. race and gender). It can be hard to identify the source of the problem or to explain it to a court~\cite{barocas2016big}.  Moreover, in 2013 the New York Police Department (NYPD) started the Stop-question-and-frisk program (SQF\footnote{\url{https://en.wikipedia.org/wiki/Stop-and-frisk\_in\_New\_York\_City}}) to reduce criminality. It consists of temporally detaining, questioning, and at times searching civilians on the street for weapons and other contraband. ~\cite{goel2016precinct} found that blacks and hispanics make up more than $80\%$ of individuals stopped by SQF, even though they constitute approximately $50\%$ of the New York City population. 

Likewise, labor discrimination is also a difficult problem that has been studied for many years \cite{todisco2014share}. There are some studies that uncovered discrimination in traditional labour markets \cite{bertrand2004emily,carlsson2007evidence,dovidio2000aversive}. More recently, \cite{Hannak2017CSCW} explore the correlations between two freelance marketplaces, TaskRabbit\footnote{\url{https://www.taskrabbit.com}} and Fiverr\footnote{\url{https://www.fiverr.com}}, and quantify race- and gender-biases. They found that on TaskRabbit females, especially white females, receive $10\%$ fewer than males with equivalent work experience. Black workers, especially males, receive significantly lower feedback scores than others. On the other hand, on Fiverr, black workers, especially males, receive $~32\%$ fewer reviews than other males. Asian workers, also especially males, receive higher scores than others. Finally, their linguistic analysis shows that reviews for black females include fewer positive adjectives, while for blacks (in general) more negatives. Another, study shows that black tenants have much fewer chances of renting a place on Airbnb\footnote{\url{http://www.airbnb.com}}~\cite{edelman2017racial}. Also, some drivers for both UberX\footnote{\url{http://www.uber.com}} and Lyft\footnote{\url{http://www.lyft.com}} discriminate on the basis of the perceived race of the traveler~\cite{ge2016racial}. Google served ads that despised African-Americans~\cite{sweeney2013discrimination}, whilst Google did not show ads for high-paying jobs from females~\cite{datta2015automated}. Further, black and female sellers earn less than white and male sellers on eBay\footnote{\url{http://www.ebay.com}}~\cite{ayres2015race,kricheli2016many}.

A number of studies found racial disparities in automated~\cite{angwin2016machine} as well as human \cite{goel2016precinct,ridgeway2009doubly} decision making systems related to criminal justice. Recent studies focus on designing decision making systems that avoid unfairness \cite{Feldman2015KDD,Luong2011KDD,zafar2017AISTATS,zemel2013learning}, proposing methods for detecting \cite{Feldman2015KDD,Luong2011KDD,pedreshi2008discrimination,romei2014multidisciplinary} and removing \cite{dwork2012fairness,Feldman2015KDD,goh2016satisfying,kamiran2012data,kamishima2012fairness,pedreshi2008discrimination,zafar2017AISTATS,zemel2013learning} unfairness. Furthermore, \cite{Zafar2017WWW} propose fair classifier formulations in order to remove disparate only on false-positive and false-negative rates. They provide a flexible trade-off between disparate mistreatment-based fairness and accuracy. 

Demographics is key to providing more fairness algorithm and classifiers. Therefore, we hope our study may help developers and the research community to improve their systems in order to provide tools to decrease racial and gender disparities.

\section{Concluding Remarks}

In this chapter, we provided related work regarding the importance of demographic aspects of users in Online Social Networks (OSN). 

Our effort uses a similar strategy to gather demographic information as \cite{chakraborty2017@icwsm}, but we investigate very different research questions as we focus on inequalities, demographic aspects of users connections, and linguistic patterns of users. We note that our strategy to gather demographic aspects, applied to active users within the United States, allows us to take a step further in many existing studies on the field. It allows us to study race inequalities and also allows us to investigate how connections are established among demographic groups (i.e. male/female, asian/black/white). We also quantify to what extent one group follows and interacts with each other and the extent to which these connections and interactions reflect in inequalities in Twitter. 

In terms of data transparency, we hope that the methodology to compute demographic distribution of users can be leveraged to make other crowdsourced systems (e.g., social search~\cite{juhi2017} and Twitter trending topic analysis~\cite{chakraborty2017@icwsm}) transparent as well. More importantly, for algorithms that operate on large-scale crowdsourced data, is to make the outputs of the algorithms (and the algorithm itself) transparent, as it is also important to understand the non-uniformities in the inputs to the algorithms.

Regarding fairness, we hope our study may improve systems in order to provide tools to decrease racial and gender disparities. Demographic information is also an important factor to take in consideration to improve diversity in content search and recommendation system. Researchers could use age, gender, and race as attributes to provide diverse results to users. 

Demographics also reflect users' intent which may improve search engines results. All of the existing techniques of diversification have focused on bringing in diversity in the \textit{topical coverage of the contents}. Here, we propose a fundamentally new approach to introducing diversity. Instead of focusing on the diversity of content coverage, we attempt to increase diversity in the population who promotes the contents~\cite{chakraborty2017@icwsm}. Therefore we developed a system to show the demographic distribution of users in Twitter. We present this system in Chapter~\ref{chap:system}.

We hope our new and large-scale dataset may largely contribute to the research community. In the next chapter, we introduce the data collection of this dissertation.

 \chapter{Demographic Information Dataset} \label{chap:dataset}

This chapter describes the data collection and methods to infer demographic information of individual Twitter users. Our ultimate goal consists of gathering a dataset containing active U.S. Twitter users, their demographic information (gender and race), a sample of their connection graph (friends), and their tweets. Next, we describe our steps to create two distinct datasets, (i) inequalities in visibility and linguistic patterns of users; (ii) social connections and interactions of users; and also discuss their main limitations.

\section{Twitter Dataset Gathering}
To identify active Twitter users we gathered data from the $1\%$ random sample of all tweets, provided by the Twitter Stream API~\footnote{\url{https://dev.twitter.com/streaming/public}}, it offer samples of the public data flowing through Twitter. Our dataset covers a period of three complete months, from July to September 2016. In total, we collected $341,457,982$ tweets posted by around $50,270,310$ users.

We restricted our demographic studies to Twitter users located in the United States who posted during this period, at least, one tweet in English. As geographic coordinates are available in Twitter only for a limited number of users (i.e. $<$ $2\%$)~\cite{icwsm10cha}, our strategy to identify U.S. Twitter users was based on part of the methodology used in previous efforts~\cite{kulshrestha2012geographic,chakraborty2017@icwsm}. We have used the time zone information as a first filter and then we attempted to remove from this filtered dataset those users that provided free text location indicating they are not U.S. (i.e. Montreal, Vancouver, Canada). We end up with a dataset containing $6,286,477$ users likely located in the United States.

\section{Crawling Demographic Information} 
\label{sec:crawling_demographic}

Most of the existing studies related to demographics of users in Twitter have looked into gender and age. Some efforts attempt to infer the user's gender from the user name~\cite{blevins2015jane,karimi2016,liu2013s,mislove2011understanding}. However, some users may not use their real names, consequently, their gender could not be inferred properly~\cite{liu2013s}. Others have attempted to identify patterns like \textit{`$25$ yr old'} or \textit{`born in $1990$'} in Twitter profile description to identify the user age~\cite{sloan2015tweets}. 

Here, we use a different strategy that allows us to study another demographic aspect: the user's race. To do that, we crawl the profile picture Web link of all Twitter users identified as located within the United States. 
In December $2016$, we crawled the profile picture's URLs of about $6$ million users, discarding $4,317,834$ ($68.68\%$) of them.  We discarded users in two situations: first, when the user does not have a profile picture and second, when the user has changed her picture since our first crawl. When users change their picture, their profile picture URL changes as well, making it impossible for us to gather these users in a second crawl via Twitter Stream API.   

From the remaining $1,968,643$ users, we submitted the profile picture Web links into the \textit{Face++ API}. Face++ is a face recognition platform based on deep learning~\cite{fan2014learning,yin2015learning} able to identify the gender (i.e. male and female), age, race (limited to asian, black, and white), and more attributes related to smiling, face positions, glasses information from recognized faces in images. In this dissertation, we focus on gender and race attributes.

We have also discarded those users whose profile pictures do not have a recognizable face or have more than one recognizable face, according to Face++. Our baseline dataset, also used by \cite{chakraborty2017@icwsm}, contains $1,670,863$ users located in U.S. with identified demographic information. Table~\ref{table:expected_baseline} shows the demographic distribution of users in our baseline dataset across the different demographic groups. The phases of our data crawling and the amount of data on each step are summarized in Table~\ref{table:dataset}. 


\begin{table*}[!htb]
\centering
\caption{Absolute demographic distribution of $1.6$ million users in baseline dataset.}
\label{table:expected_baseline}
\begin{tabular}{|c|c|c|c|}
\hline
\multirow{2}{*}{\textbf{Race}} & \multicolumn{2}{c|}{\textbf{Gender}} & \multirow{2}{*}{\textbf{Total}} \\ \cline{2-3}
                               & \textbf{Male}     & \textbf{Female}  &                                 \\ \hline
Asian                          & $120,950$ ($7.24\%$)   & $177,205$ ($10.61\%$) & $298,155$ ($17.85\%$)                \\ \hline
Black                          & $130,954$ ($7.84\%$)   & $107,827$ ($6.45\%$)  & $238,781$ ($14.29\%$)                \\ \hline
White                          & $538,625$ ($32.23\%$)  & $595,302$ ($35.63\%$) & $1,133,927$ ($67.86\%$)               \\ \hline
\textbf{Total}                 & $790,529$ ($47.31\%$)  & $880,334$ ($52.69\%$) & $1,670,863$ ($100\%$)                 \\ \hline
\end{tabular}
\end{table*}

\begin{table}[h]
 \centering
\caption{Dataset construction}
\label{table:dataset}
\begin{tabular}{| c | c | p{3.3cm} |}
\hline
\textbf{Phase} & \textbf{Number of Users}\\
\hline
Crawling $3$ months of Tweets & $50$ million\\
\hline
Filtering U.S. users &  $6$ million\\
\hline
U.S. users with profile image &  $1.9$ million\\
\hline
Baseline: U.S. users with one face &  $1.6$ million\\
\hline
Dataset 1: Recognized U.S. users with linguistic attributes &  $304,477$ \\
\hline
Dataset 2: Social connections and interactions & $428,697$\\
\hline
\end{tabular}
\end{table}

\section{Baseline Dataset} \label{sec:baseline}

In this section, we use a null model as a method to estimate the statistical significance of the observed trend in the given data. We compare the distribution of random samples created by the null model with that of the original dataset and we measure the statistical significance. This step is important due to some limitations to gather all tweets and linguistic attributes for $1.6$ million users. Therefore, to circumvent this limitation, we prefer to randomly select $304,477$ users from the $1.6$ million dataset and show the statistical significance of this sub-dataset which we use to show results regarding race and gender inequalities and linguistic patterns of Twitter users.

Table \ref{table:expected_baseline} shows the distribution of gender and race in the baseline dataset of the $\approx 1.6$  million Twitter users between July and September 2016. Additionally, Table~\ref{table:expected} shows the demographic breakdown of the $304,477$ randomly selected users across the different demographic groups. To construct a null model, we create $k$ random samples from the whole population (our crawled dataset with demographic aspects), where each sample has exactly $304,477$ users. For each sample, we count how many asians, blacks, whites, males, and females are included. Then, the $Z_{White}$ is computed as follows:

\begin{equation}
    Z_{White}=\frac{|U_{White}|-mean(|S_{White}|)}{std(|S_{White}|)}
\end{equation}

where $mean(White)$ is the mean and $std(White)$ is the standard deviation of the values from multiple samples of white users. We use the same equation for the other gender and race aspects. Table \ref{table:null_model} presents a detailed description of the comparison with null model. Additionally, Table \ref{table:expected} presents the demographic distribution of $304,477$ users with linguistic attributes. The numbers in the parenthesis correspond to the $Z$-values.

Intuitively, when the absolute value of $Z$-value becomes bigger (either positive or negative), the trend (more or less number, respectively) is less likely observed by chance. In this work, the size of population is $304,477$, and $k$=100.

\begin{table*}[t]
\centering
\caption{Demographic distribution of $304,477$ users with linguistic attributes. The numbers in the parenthesis correspond to the $Z$-values.}
\label{table:expected}
\begin{tabular}{|c|c|c|c|}
\hline
\multirow{2}{*}{\textbf{Race (\%)}} & \multicolumn{2}{c|}{\textbf{Gender (\%)}} & \multirow{2}{*}{\textbf{Total (\%)}} \\ \cline{2-3}
                               & \textbf{Male}     & \textbf{Female}  &                                 \\ \hline
Asian                          & $7.07$ ($-3.85$)   & $10.05$ ($-11.28$) & $17.12$ ($-10.90$)                \\ \hline
Black                          & $8.17$ ($8.53$)   & $6.74$ ($7.68$)  & $14.91$ ($11.69$)               \\ \hline
White                          & $32.88$ ($8.49$)  & $35.09$ ($-7.69$) & $67.97$ ($1.20$)              \\ \hline
\textbf{Total}                 & $48.12$ ($10.91$)  & $51.88$ ($-10.91$) & $100.00$                 \\ \hline
\end{tabular}
\end{table*}

\begin{table*}[]
\centering
\scriptsize
\caption{Basic statistical description of null models. $k=100$ samples with a population of $304,477$ randomly selected users. We use confidence intervals of $95\%$ confidence level. }
\label{table:null_model}
\begin{tabular}{@{}crrrrrrrr@{}}
\toprule
Demographic & \multicolumn{1}{c}{Mean} & \multicolumn{1}{c}{$Z$-value} & \multicolumn{1}{c}{S.D.} & \multicolumn{1}{c}{Min} & \multicolumn{1}{c}{25-perc} & \multicolumn{1}{c}{Median} & \multicolumn{1}{c}{75-perc} & \multicolumn{1}{c}{Max} \\ \midrule
Male         & $144,035.1 \pm 44.86$        & 10.91                       & 228.88                   & 143,544                  & 143,883.00                      & 144,054.5                   & 144,156.50                    & 144,680                  \\
Female       & $160,441.9 \pm 44.86$        & -10.91                      & 228.88                   & 159,797                  & 160,320.50                    & 160,422.5                   & 160,594.00                      & 160,933                  \\
Asian       & $54,311.5 \pm 39.17$         & -10.90                      & 199.87                   & 53,907                   & 54,177.25                    & 54,296.5                    & 54,444.00                     & 54,803                   \\
Black       & $43,514.01 \pm 31.72$        & 11.69                       & 161.85                   & 43,196                   & 43,380.75                    & 43,503.5                    & 43,633.50                     & 43,887                   \\
White       & $206,651.49 \pm 46.82$       & 1.20                        & 238.91                   & 205,921                  & 206,490.25                   & 206,666.5                   & 206,789.25                   & 207,110                  \\
Asian Male   & $22,043.64 \pm 26.24$        & -3.85                       & 133.88                   & 21,674                   & 21,958.75                    & 22,040.5                    & 22,115.50                     & 22,429                   \\
Asian Female & $32,267.86 \pm 28.92$        & -11.28                      & 147.56                   & 31,900                   & 32,153.50                     & 32,262.0                    & 32,371.75                    & 32,667                   \\
Black Male   & $23,857.98 \pm 23.81$        & 8.53                        & 121.48                   & 23,634                   & 23,777.75                    & 23,858.0                    & 23,930.00                     & 24,197                   \\
Black Female & $19,656.03 \pm 21.82$        & 7.68                        & 111.34                   & 19,342                   & 19,585.25                    & 19,660.5                    & 19,737.75                    & 19,944                   \\
White Male   & $98,133.48 \pm 45.61$        & 8.49                        & 232.73                   & 97,538                   & 97,995.25                    & 98,130.5                    & 98,297.50                     & 98,623                   \\
White Female & $108,518.01 \pm 43.04$       & -7.69                       & 219.62                   & 108,025                  & 108,348.25                   & 108,501.5                   & 108,688.00                    & 109,015                  \\ \bottomrule
\end{tabular}
\end{table*}

\section{Gathering Tweets} \label{sec:tweets}

We are interested in correlating linguistic features of Twitter users with demographic information. We crawled the recent $3,200$ tweets of $304,477$ users for the purpose of linguistic analysis. Table~\ref{table:expected} shows the demographic breakdown of users in this dataset across the different demographic groups. We can note a prevalence of females ($51.88\%$) in comparison to males ($48.12\%$) and a predominance of whites ($67.97\%$) in comparison to blacks ($14.91\%$) and asians ($17.12\%$). This means if we pick users randomly in our dataset, we would expect demographic groups with these proportions. Table~\ref{table:tweets_stats} shows the statistical descriptions of the number of tweets with confidence intervals of $95\%$ confidence level for each demographic groups.

\begin{table}[]
\centering
\caption{Basic statistical descriptions of number of tweets with confidence intervals of $95\%$ confidence level.}
\label{table:tweets_stats}
\begin{tabular}{@{}crrr@{}}
\toprule
Demographic & \multicolumn{1}{c}{Mean} & \multicolumn{1}{c}{Median} & \multicolumn{1}{c}{Max} \\ \midrule
Male         & $11,624.76 \pm 109.40$                    & $3,874$      & $1,683,948$                 \\
Female       & $12,933.40 \pm 105.89$                    & $4,885$      & $1,132,964$                 \\
Asian       & $14,020.92 \pm 183.73$                    & $5,544$      & $1,108,525$                 \\
Black       & $18,949.91 \pm 248.46$                    & $8,245$      & $973,225$                 \\
White       & $10,432.49 \pm 85.28$                     & $3,637$      & $1,683,948$                 \\ \bottomrule
\end{tabular}
\end{table}

\section{Extraction of Topics}

We extracted the information about topics of interests for active users using the Who Likes What\footnote{\url{http://twitter-app.mpi-sws.org/who-likes-what}} Web-based service~\cite{Bhattacharya2014}. The produced topics are derived from the list of friends (other users the user is following) of each user. Then, we sort the produced topics based on their frequency to obtain the $20$ most common topics from Twitter users, including them as binary variables (1 if the user is interested in this topic, 0 otherwise). We manually cleaned several top topic labels following the same procedure as \cite{nilizadeh2016twitter}. Therefore, we merge topics like \textit{businesses} and \textit{biz}, group topics into similarity (e.g. \textit{celebrities} and \textit{famous}, \textit{actors} and \textit{actor}), and remove some topics like \textit{best}, \textit{br}, \textit{bro}, \textit{new}. Table \ref{table:topics} presents a list of the 20-top topics and the merged sub-topics in each one as well as the number of users that are interested in.

\begin{table*}[]
\centering
\caption{20-top topics of user's interests}
\label{table:topics}
\begin{tabulary}{1\textwidth}{lLl}

\textbf{Topic} & \multicolumn{1}{c}{\textbf{Sub-Topics}}                                                                                                              & \textbf{\#Users} \\ \midrule
World          & world, earth, hollywood, usa, canada, texas, international, nyc, country, city, boston, san francisco, france, america, los angeles, brasil, london, india           & $290,030$         \\
Celebrities    & celebrities, famous, stars, celebs, celebrity, star, celeb                                                                                                & $245,125$        \\
Entertainment  & entertainment                                                                                                                                       & $244,956$         \\
Music          & music, pop, hip hop, rap, gospel, hiphop                                                                                                                 & $227,986$         \\
TV             & tv, television                                                                                                                                       & $225,682$         \\
Life           & life, lifestyle, health, healthcare, fitness, food, style, smile, drink                                                                                     & $157,032$         \\
Fun            & fun, funny, humor, lol, laugh                                                                                                                           & $154,058$         \\
Info           & info, information                                                                                                                                    & $147,567$         \\
Artists        & musicians, singers, artist, singer, musician, rappers, bands                                                                                              & $141,519$         \\
Actors         & actors, actresses, actress, actor                                                                                                                      & $140,647$         \\
Media          & sports news, tech news, newspapers, music news, breaking news, world news, news media, radio, internet, social media, youtube, sports media, magazines, magazine & $135,849$         \\
Writers        & writers                                                                                                                                             & $126,051$         \\
Bloggers       & bloggers, blogs, blog                                                                                                                                 & $110,699$         \\
Business       & business, biz, businesses                                                                                                                             & $107,361$         \\
Sports         & sports, football, basketball, baseball, soccer, futbol, basket, martial arts, sport, mma, golf, cricket, boxing, motorsports, f1, racing                           & $93,611$         \\
Movie          & movie, movies, film, films                                                                                                                             & $88,863$         \\
Organizations  & organizations, nfl, nba, mlb, nhl, ufc, lfc, lgbt                                                                                                          & $82,568$         \\
Technology     & technology, tech, iphone, digital, geek, software, computer, electronic, android, xbox, mac, gadgets, programming, geeks                                         & $72,137$         \\
Politics       & politics, government, political, politicians, politician                                                                                                & $64,735$         \\
Companies      & companies, apple, company, microsoft, google                                                                                                            & $53,128$          \\ \bottomrule
\end{tabulary}
\end{table*}

\section{Linguistic Measures}
To quantify gender and race dimensions in the language of Twitter users, we use the psycholinguistic lexicon Linguistic Inquiry and Word Count (LIWC)~\cite{tausczik2010psychological}. The crawled tweets of a user were gathered in a supertext (the concatenation of all tweets) in order to extract a variety of linguistic metrics. The features are categorized into $3$ main categories, (1) affective attributes; (2) cognitive attributes; and (3) linguistic style attributes, as \cite{DeChoudhury:2017:GCD:2998181.2998220} propose. For this work, we considered $36$ features from LIWC categorized in six groups (affective attributes, cognitive attributes, lexical density and awareness, temporal references, social/personal concerns, and interpersonal focus) in order to find the main differences across each demographic group. 

The affective attributes contemplate features that show how strong is the expression of feelings like anger, anxiety, sadness, and swear. Cognitive attributes are related to the process of knowledge acquisition through perception. The lexical density and awareness group gather features related to the language itself and its structure. Temporal references are related to the tense expressed in the writing, while interpersonal focus presents features related to the speech. The social/personal concerns group comprises features that express characteristics inherent to the individual as well his/her relation to the environment where he/she lives.

\section{Gathering Social Connections and Interactions}

Ideally, to study how different demographic groups are connected and interact with each other, we would like to have at our disposal the followers and friends of all users from our dataset. However, the number of followers and friends of the $1.6$ million users in our dataset surpasses $6.4$ billion users, making it unfeasible to crawl all profiles through the Twitter and the Face++ API. Next, we discuss a sampling procedure we used to create a dataset of users to study their connections and their interactions with demographic information.  

We randomly select a total of $6,000$ users from our dataset, $1,000$ users of each demographic group (i.e. asian male, asian female, black male, black female, white male, and white female). Then, we gathered their friends. As some of these users have a prohibitively large number of friends, we limited our crawl to gather only the most recent $5,000$ friends of a given user, as this is the maximum number of user IDs the Twitter API returns per request. However, this strategy gathered the entire friends' list for $98.51\%$ of the users. 

Then, we follow the same methodology we discussed before to gather the demographic information of users. First, we remove users that are not located in U.S., and then we attempt to identify the demographic aspects of each user using Face++ API. 
It is undeniable that the aggregated number of friends is extremely large and it is difficult to gather due to the limitation on the number of requests that Face++ API allows us to make. Thus we were limited to gather demographics of at least $5\%$ of each user's friends.

Our social connections and interactions dataset contains $428,697$ users with the proper demographic information identified. Table~\ref{table:number_friends} presents the total number of friends gathered for each demographic group that we examined. The average and median percentage of friends of the $6,000$ users for which we were able to gather demographic information are $10.15\%$ and $9.40\%$, respectively. We note that these fractions are usually higher than $5\%$ as some extra users were previously gathered in our $1.6$ million demographic dataset.

Also, we worked in a similar way for the analysis of interactions between demographic groups. More specifically, we select the users that retweet and mention the tweets of the randomly selected users. To do that, we gathered all tweets (limited to a maximum of $3,200$ tweets due to Twitter REST API limitation\footnote{\url{https://dev.twitter.com/rest/reference/get/statuses/user\_timeline}}), it returns a collection of the most recent tweets posted by the user, of our $6,000$  users. We then identified which of these users were mentioned or retweeted and we limited to gather the demographic information of only $5\%$ of retweeters and mentioned users for our analysis. Table~\ref{table:number_interactions} summarizes the amount of crawled users for each demographic group.

Our study about the connections among demographic groups (Chapter~\ref{chap:results_interconnections}) is based on this specific dataset.

\begin{table}[!t]
 \centering
\caption{Number of Friends in each Group}
\label{table:number_friends}
\begin{tabular}{| c | c | c | c | c |}
\hline
 & \textbf {White} & \textbf{Black}& \textbf{Asian}& \textbf{Total} \\
\hline
\textbf{Male} &  $151,840$ & $52,437$ & $24,299$  & $228,576$  \\
\hline
\textbf{Female} & $137,010$  & $31,011$ & $32,100$  & $200,121$ \\
\hline
\textbf{Total} & $288,850$  & $83,448$  & $56,399$  & $428,697$ \\
\hline
\end{tabular}
\end{table}

\begin{table}[!t]
 \centering
\caption{Number of Interactions in each Group}
\label{table:number_interactions}
\begin{tabular}{| c | c | c | c | c |}
\hline
 & \textbf {White} & \textbf{Black}& \textbf{Asian}& \textbf{Total} \\
\hline
\textbf{Male} & $246,879$  & $109,744$ & $51,370$ & $407,993$ \\
\hline
\textbf{Female} & $202,338$  & $60,108$ & $71,137$ & $333,583$ \\
\hline
\textbf{Total} & $449,217$ & $169,852$ & $122,507$ & $741,576$ \\
\hline
\end{tabular}
\end{table}

\section{Potential Limitations}

The gender and race inference are challenge tasks, and as other existing strategies have limitations and the accuracy of Face++ inferences is an obvious concern in our effort. The limitations of our data is discussed next.

\quad \quad \textbf{Accuracy of the inference by Face++:} Face++ itself returns the confidence levels for the inferred gender and race attributes, and it returns an error range for inferred age. In our data, the average confidence level reported by Face++ is $95.22\pm0.015\%$ for gender and $85.97\pm0.024\%$ for race, with a confidence interval of $95\%$. Recent efforts have used Face++ for similar tasks and reported high confidence in manual inspections~\cite{an2016greysanatomy,Zagheni2014,chakraborty2017@icwsm}. In addition, in recent scientific effort~\cite{chakraborty2017@icwsm} they evaluated the effectiveness of the inference made by Face++, using human annotators to label randomly selected profile images from Twitter. They measured the inter-annotator agreement in terms of the Fleiss’ $\kappa$ score which was $1.0$ and $0.865$ for gender and race, respectively.

Our dataset may also include fake accounts and bots. Previous studies provide evidence for an significant fraction of fake accounts \cite{Freitas2015@asonam,messias13@firstmonday} in Twitter. 

\quad \quad \textbf{Data:} Finally, we note that our approach to identify users located in U.S. may bring together some users located in the same U.S. time zone, but from different countries. We, however, believe that these users might represent a small fraction of the users, given the predominance of active U.S. users in Twitter~\cite{sysomos}. Also, we are using the $1\%$ random sample of all tweets. However, the $1\%$ random sample is not the best data to capture all the dynamics happening in Twitter, its limitations are known~\cite{morstatter2014biased} and it is the best available option at our disposal.

\section{Concluding Remarks}

In this chapter, we presented our methodology to gather the demographic data of Twitter users. First, we crawled the English tweets from U.S. in order to get the users who posted them. Second, we submitted their profile picture Web links into the Face++ API. Face++ is a face recognition platform based on deep learning able to identify the gender, race, and age attributes. We also discussed how we inferred the linguistic metrics of tweets using the LIWC. Then, we described our methodology to gather the social connections and interactions between users. Finally, we discussed our potential limitations. In the next chapter, we analyze the association of demographic status with visibility and discovering possible inequalities in Twitter. In other words, we analyze whether the gender and race affect users' number of followers and how many lists they are added to.

 \chapter{Inequality in Visibility} \label{chap:results_inequalities}

We are interested in analyzing the relationship between demographic aspects and visibility, and discovering possible inequalities. These asymmetries can be derived from the prejudices and stereotypes in the selection of which user to follow based on gender or race. \cite{nilizadeh2016twitter} provide evidence of inequalities and asymmetries in terms of visibility between males and females. 

We focus only on the visibility in the social network and not in user's influence of the audience to which he/she is exposed. For our purpose, we use two different features to measure a user visibility: its follower count and its listed count. In other words, we test whether the gender and race affect users' number of followers and how many lists they are added to.

The number of followers measures the real size of the audience that someone is exposed~\cite{icwsm10cha}. Finally, listed count represents the number of times a user was added to a specific list by others in Twitter. Twitter Lists allow users to group and organize Twitter accounts that tweet on a topic that is of interest to her and follow their collective tweets. Many users carefully create lists to include other Twitter users who they consider experts on a given topic. Previous research efforts~\cite{sharma2012inferring} attempt to gather lists in a large scale to find experts in Twitter.

As a baseline for comparisons, Table~\ref{table:expected} shows the demographic breakdown of users in our dataset across the different demographic groups. We can note a prevalence of females ($51.88\%$) in comparison to males ($48.12\%$) and a predominance of whites ($67.97\%$) in comparison to blacks ($14.91\%$) and asians ($17.12\%$). This means that if we pick users randomly in our dataset, we would expect demographic groups with these proportions. We used these proportions next as a baseline for characterizing inequalities in Twitter.

\section{Gender Inequality}

\begin{figure}[!htb]
  \centering
    \includegraphics[width=1\textwidth]{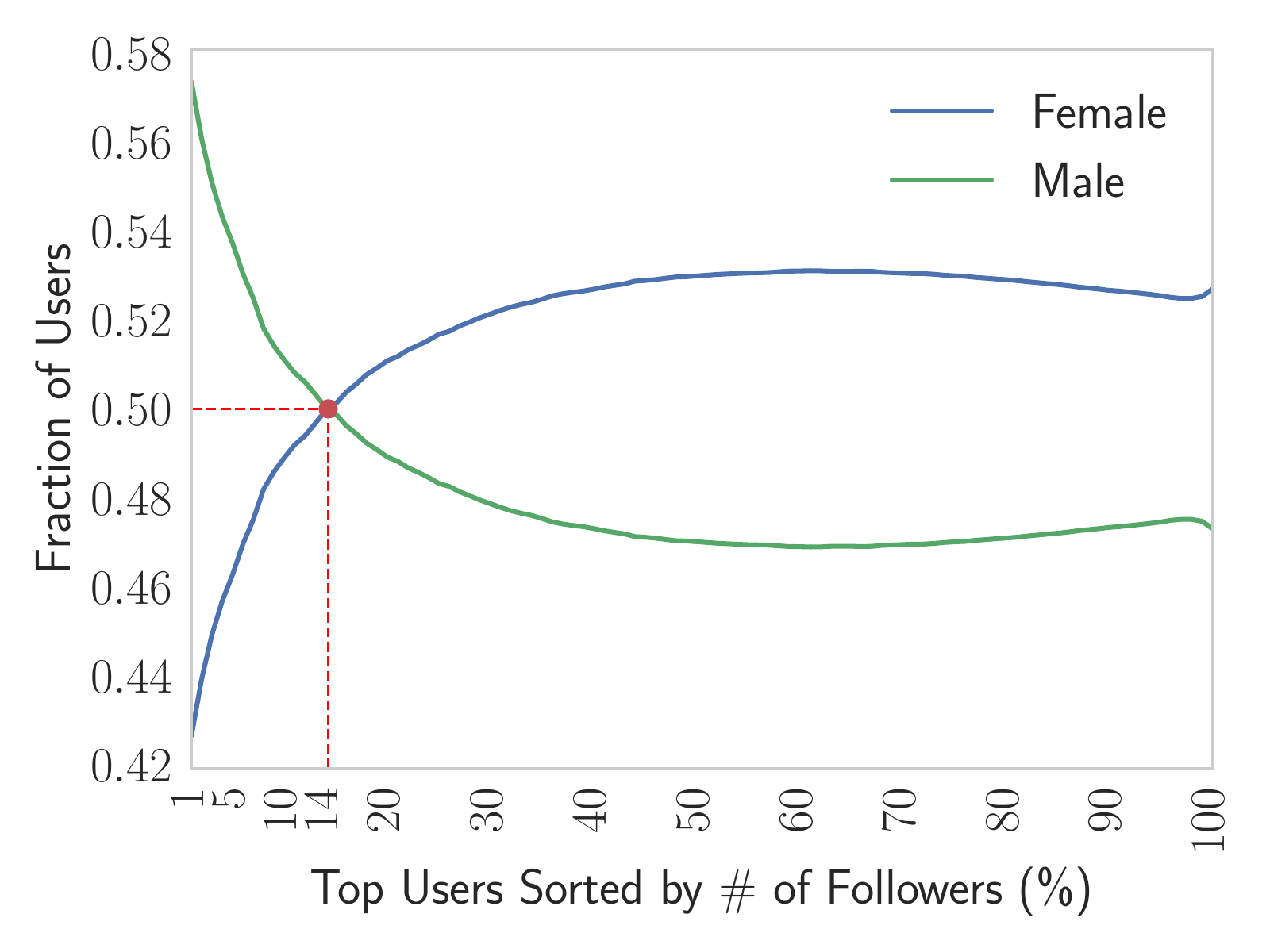}
  \caption{Distribution of the fraction of females (in blue color) and males (in green color) in the rank of users with most followers. The intersection point happens in the top $14\%$, meaning the fraction of females became higher than males from this point.}
  \label{fig:top_gender}
\end{figure}

We begin our analysis by sorting all users with respect to the number of followers in our dataset and then we computing the fraction of males and females among the top followed users. Figure~\ref{fig:top_gender} provides some insight on this point. Although, the number of females ($51.88\%$) is larger than the males ($48.12\%$) as shown in Table~\ref{table:expected}, males tend to be among the top followed users. The most significant difference exists in top $1\%$ with a fraction of $57.34\%$ of males and $42.65\%$ of females, which is almost $15\%$ and this difference is decreasing until the top $14\%$, where the fraction of females become higher than the fraction of males.

\begin{figure}[!htb]
  \centering
    \includegraphics[width=1\textwidth]{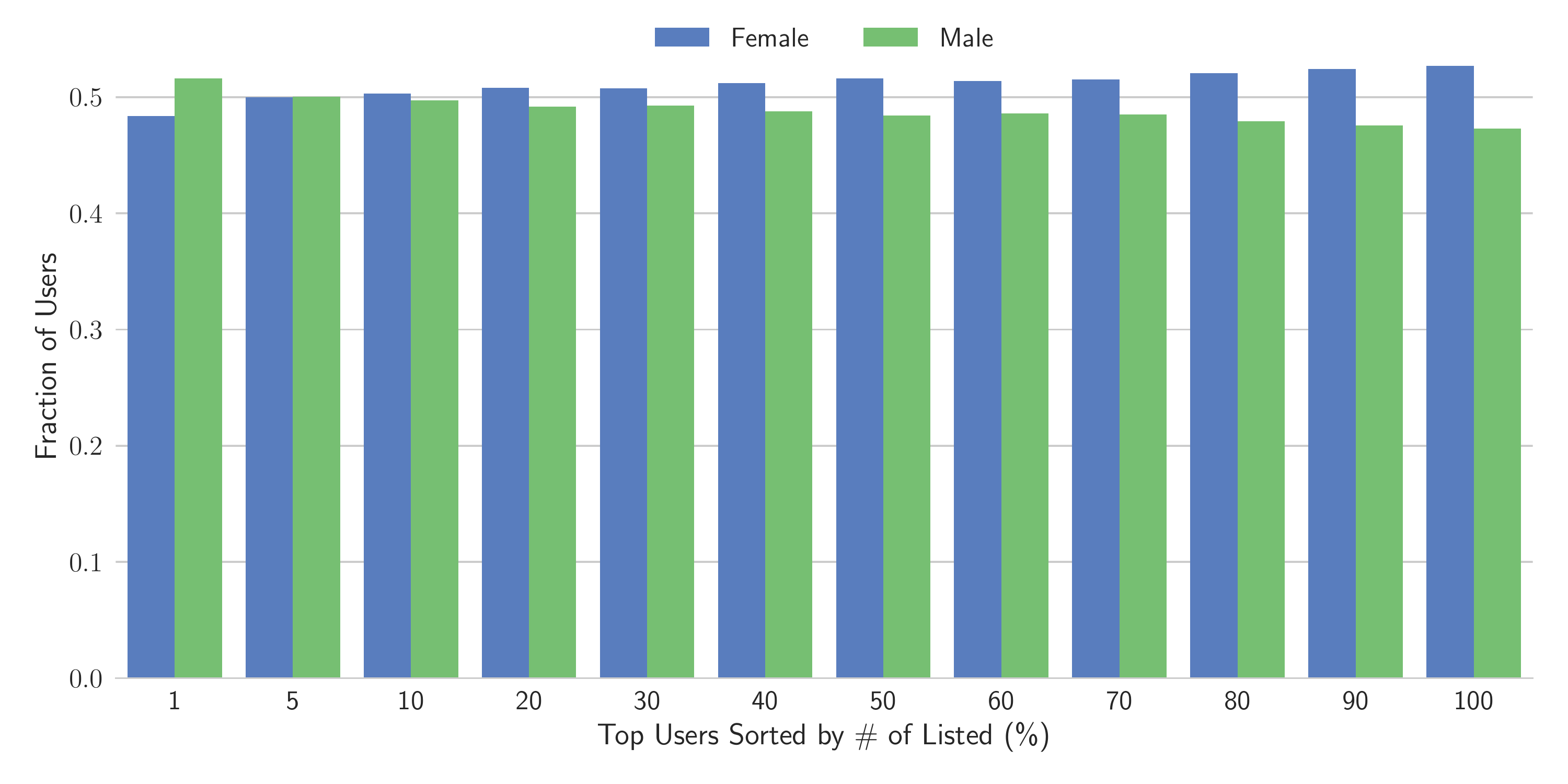}
  \caption{Distribution of the fraction of females (in blue color) and males (in green color) in the rank of most listed users.}
  \label{fig:top_gender_list}
\end{figure}

The same kind of observation can be made for the ranking of listed counts, as shown in Figure~\ref{fig:top_gender_list}. However, the discrepancy perceived is smaller in the lists. Basically, it is restricted to the top $1\%$ of the most listed users with $51.23\%$ of males and $48.64\%$ of females. In the top $6\%$, the fraction of females become higher than males. The same occurs in the top $8\%$.

The results of this analysis reflect the general idea that higher positions are usually taken by males \cite{sugimoto2013global}. 

Our observations also reinforce previous findings related to gender in Twitter~\cite{nilizadeh2016twitter}, suggesting the existing of the so called 'glass ceiling effect', which posits that females face an invisible barrier at the highest levels of an organization~\cite{cotter2001glass}. One concern raised by authors on that work is that disparity may be driven by a small number of 'elite' users. Our findings diminish such concerns as our dataset is at least two orders of magnitudes larger and we noted such discrepancies even in the top $10\%$ positions.

\section{Race Inequality}

\begin{figure}[!htb]
  \centering
    \includegraphics[width=1\textwidth]{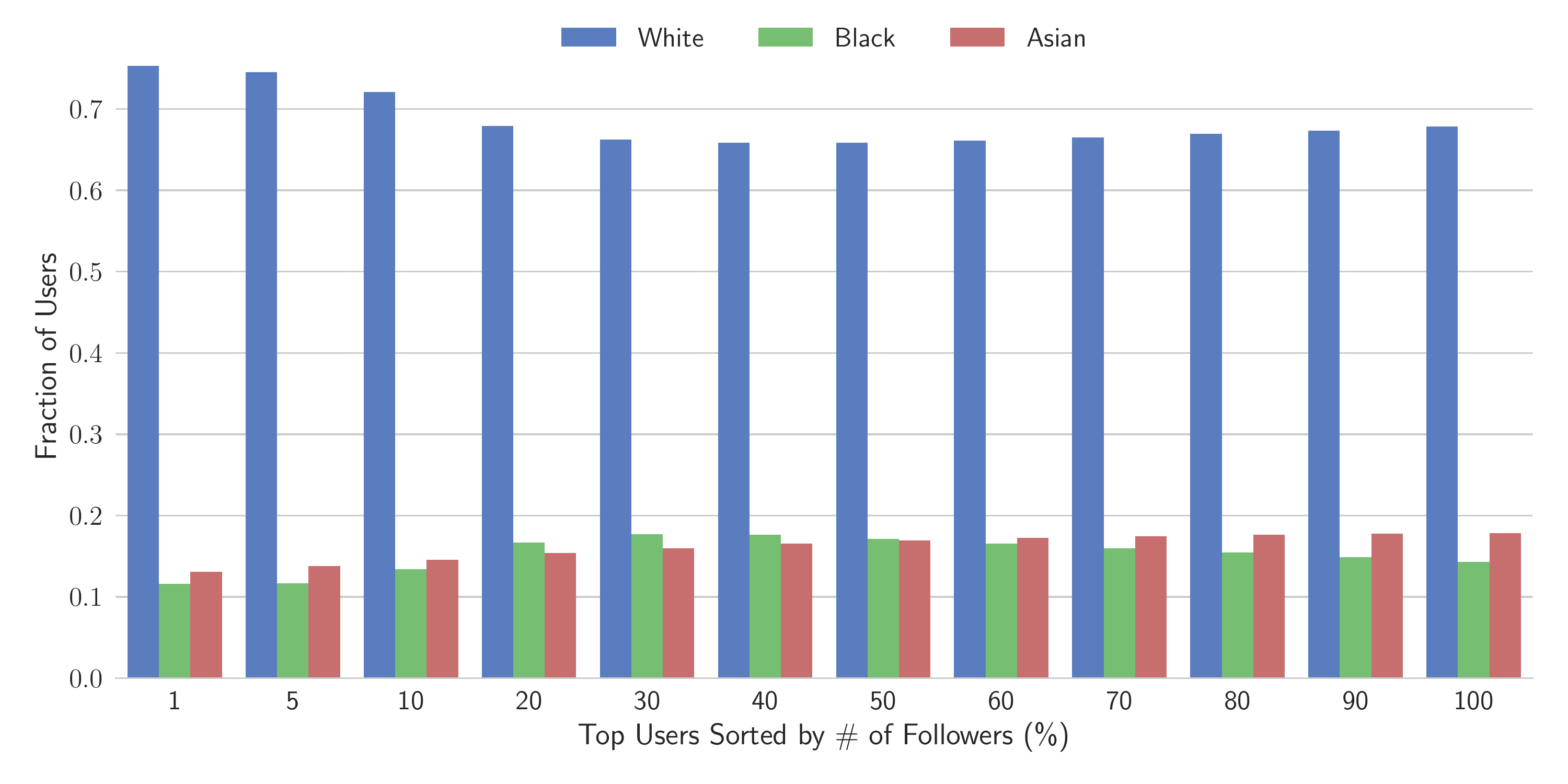}
  \caption{Amount of whites, blacks, and asians in the rank of users with most followers}
  \label{fig:top_race}
\end{figure}

\begin{figure}[!htb]
  \centering
    \includegraphics[width=1\textwidth]{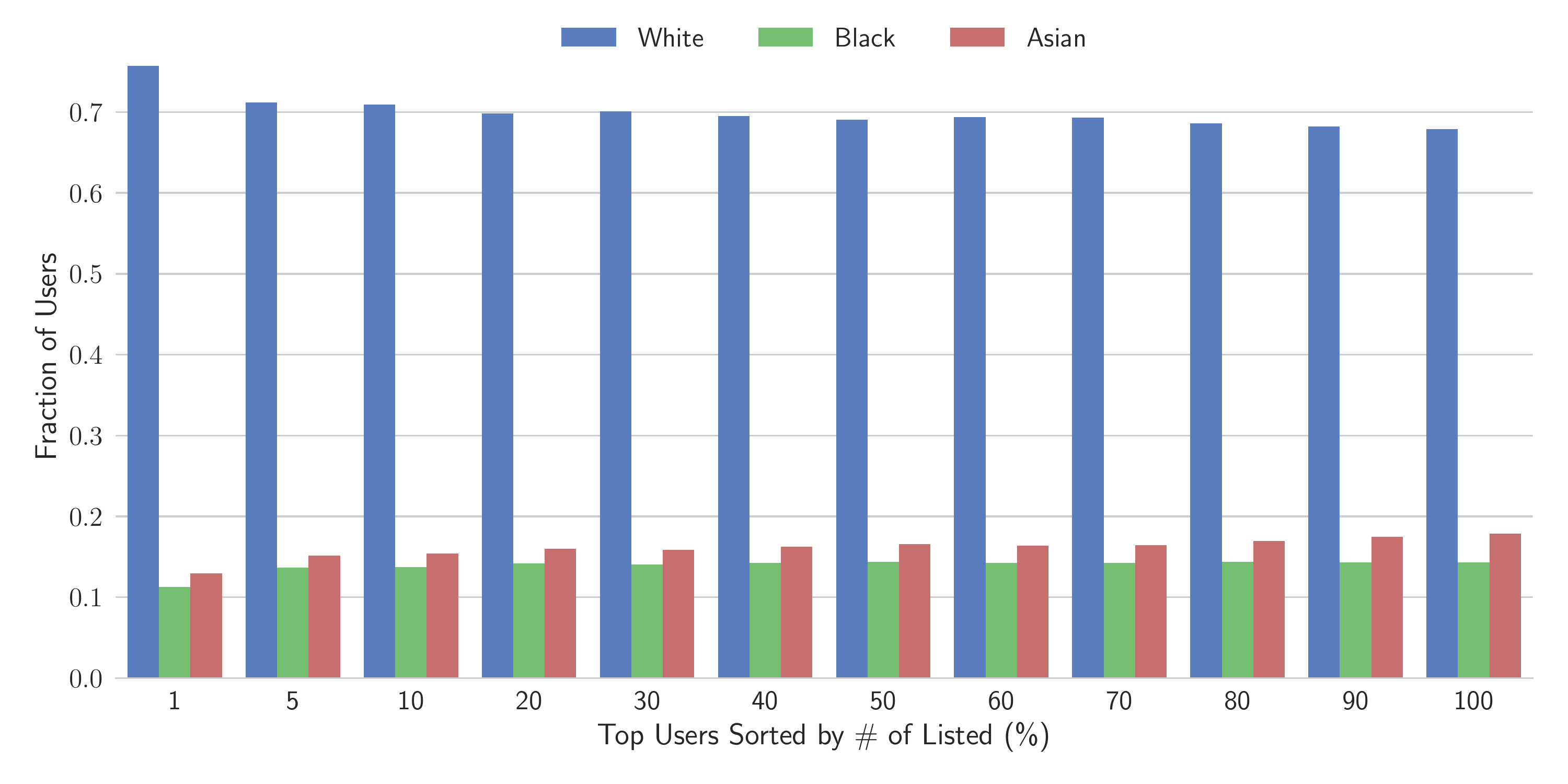}
  \caption{Amount of whites, blacks, and asians in the rank of the most listed users}
  \label{fig:top_race_list}
\end{figure}

We now turn our attention to the analysis of race inequalities. 
Similarly to what we have done in the previous section, we examine here the presence of each user race within the top positions in a rank of top followed users and top listed users. Figure~\ref{fig:top_race} and Figure~\ref{fig:top_race_list} provide the necessary information of the top followed and the top listed users for each race. We note that the number of white users tends to be higher in the both top positions (in the rank of users with more followers and in the rank of most listed users).

Our results suggest that race disparity in Twitter visibility occurs, meaning that at the highest levels of visibility, users perceived to be white come out on top position. This observation reinforces many previous observations related to race inequality in United States~\cite{bonilla2006racism,oliver2006black}.

\section{Taking Together Gender and Race Inequality}

\begin{figure}[!htb]
  \centering
    \includegraphics[width=1\textwidth]{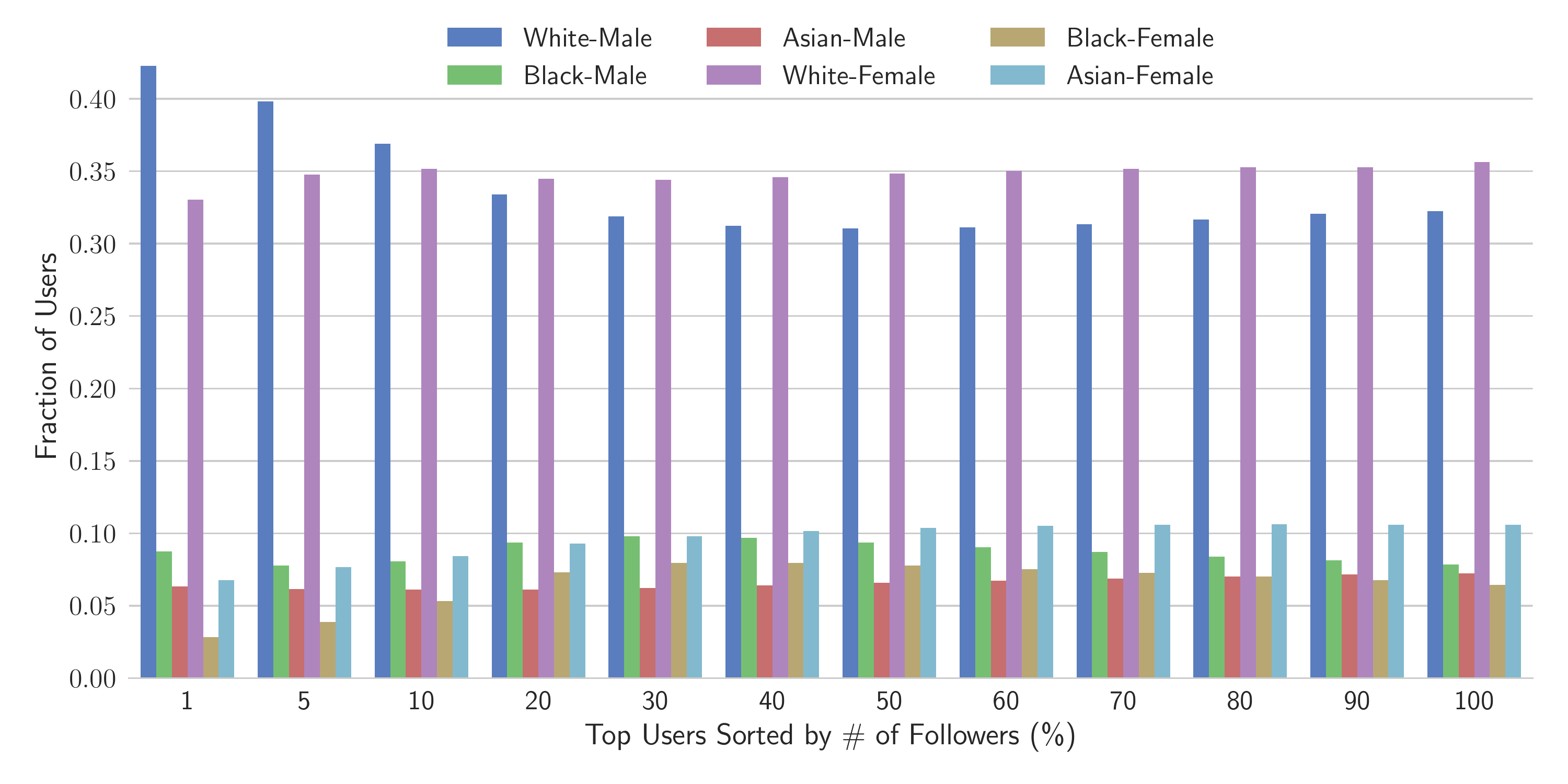}
  \caption{Demographic groups in terms of gender and race as a function of the rank of the most listed users}
  \label{fig:top_categories_followers}
\end{figure}

\begin{figure}[!htb]
  \centering
    \includegraphics[width=1\textwidth]{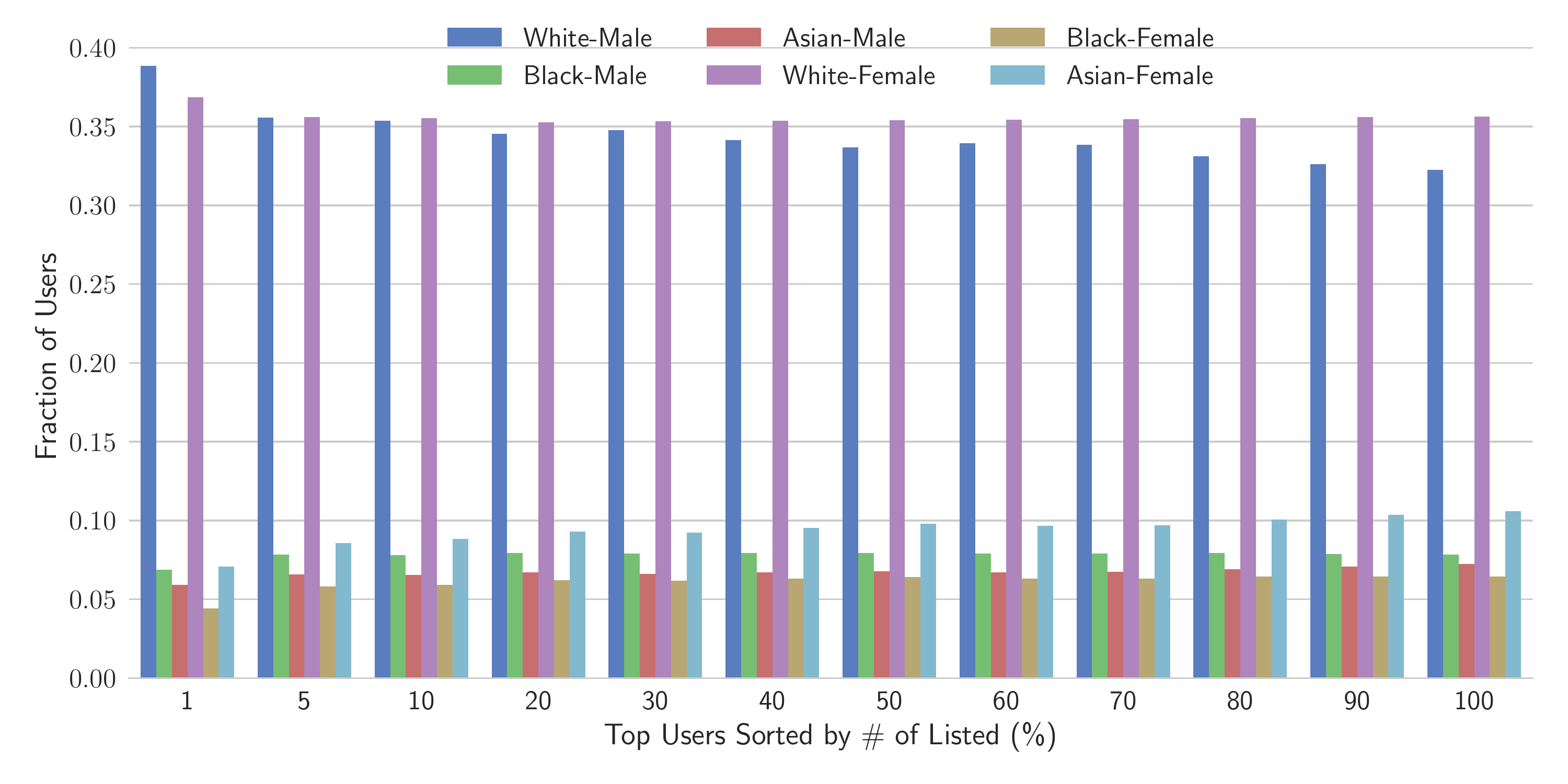}
  \caption{Demographic groups in terms of gender and race as a function of the rank of the most listed users}
  \label{fig:top_categories_listed}
\end{figure}

Finally, we attempt to quantify the existing inequality when we look into gender and race aspects at the same time. We note from Figure~\ref{fig:top_categories_followers} that the amount of white males among the top $1\%$ users with more followers is $42.27\%$. This represents an increase of $9.39\%$ in the fraction of white males in comparison to the expected proportion of white males in our dataset, which is $32.88\%$. We can note that the population of black females and asian females experience a reduction of $3.9\%$ and $3.29\%$ in the top $1\%$ most followed users, from $6.74\%$ and $10.05\%$ to $2.84\%$ and $6.76\%$, respectively. We note similar trends in terms of lists as we can see Figure~\ref{fig:top_categories_listed}.

To better analyze to what extent demographic groups in the top $1\%$ is substantially higher or lower than what one would expect based on our demographic information dataset, Table~\ref{table:top_followers_listed_diff} shows the increase or decrease in the proportion of each demographic group in comparison with such baseline expected population. Overall, these results quantify the perceived inequalities that six demographic groups experience at the highest levels of visibility (i.e. top $1\%$ most followed and listed). The most privileged demographic group in the rank of followers is the group of white males, with $28.56\%$ higher than we expected for the most followed users. The number of white females is also higher than we expected for the most listed users, but much smaller amount (only $5.04\%$) in comparison to males. The most unprivileged groups are asian females and black females, under-represented in the top $1\%$ of the most followed users by more than $32\%$. 

\begin{table}[]
\centering
\caption{Relative proportion of each demographic group in the top $1\%$ rank of users with more followers (followers column) and the most listed (listed column) in comparison to the baseline expected population.}
\label{table:top_followers_listed_diff}
\begin{tabular}{|l|c|c|c|c|}
\hline
\multirow{2}{*}{\textbf{Race}} & \multicolumn{2}{c|}{\textbf{Followers}}                    & \multicolumn{2}{c|}{\textbf{Listed}}                       \\ \cline{2-5} 
                               & Male                          & Female                       & Male                          & Female                       \\ \hline
Asian                          & \textcolor{red}{$-10.60$}  & \textcolor{red}{$-32.70$} & \textcolor{red}{$-16.36$}  & \textcolor{red}{$-29.61$} \\ \hline
Black                          & \textcolor{blue}{$+7.17$} & \textcolor{red}{$-57.73$} & \textcolor{red}{$-15.90$}  & \textcolor{red}{$-34.20$} \\ \hline
White                          & \textcolor{blue}{$+28.56$} & \textcolor{red}{$-5.84$}  & \textcolor{blue}{$+18.15$} & \textcolor{blue}{$+5.04$} \\ \hline
\end{tabular}
\end{table}

\section{Concluding Remarks}

In this chapter, we analyzed the association of demographic status with visibility and discovering possible inequalities. We showed that the Twitter glass ceiling effect, typically applied to females, also occurs in Twitter for males, if they are black or asian. Our analysis reinforces evidence about gender inequality in terms of visibility and introduces race as a key demographic aspect, which reveals hiding prejudices between demographic groups. In the next chapter, we investigate how demographic groups differ from each other in terms of linguistic and topic interests.

 \chapter{Linguistic Patterns} \label{chap:linguistics}

In this chapter, we want to investigate how demographic groups differ from each other in terms of linguistic and topic interests. In other words, we want to find out the extent to which male differs from female in terms of the way they post content in Twitter and also the extent to which they have interest for some specific topics. This chapter also presents information regarding race, and combinations of gender and race.

We believe that studying the differences in how demographic groups write content about specific topics would help systems developers to provide more transparency to users. Therefore, first, we present results regarding the linguistic differences of demographic groups. After words, we show their interests in Twitter.

\section{Linguistic Differences} 

In order to show how demographic groups differ from each other in both gender and race domains, this section presents the absolute difference between groups with respect to various categories of linguistic measures. Table \ref{table:liwc_gender} shows the linguistic features extracted from LIWC into six categories (affective attributes, cognitive attributes, lexical density and awareness, temporal references, social and personal concerns, and interpersonal focus).

\begin{figure}[!htb]
  \centering
    \includegraphics[width=0.9\textwidth]{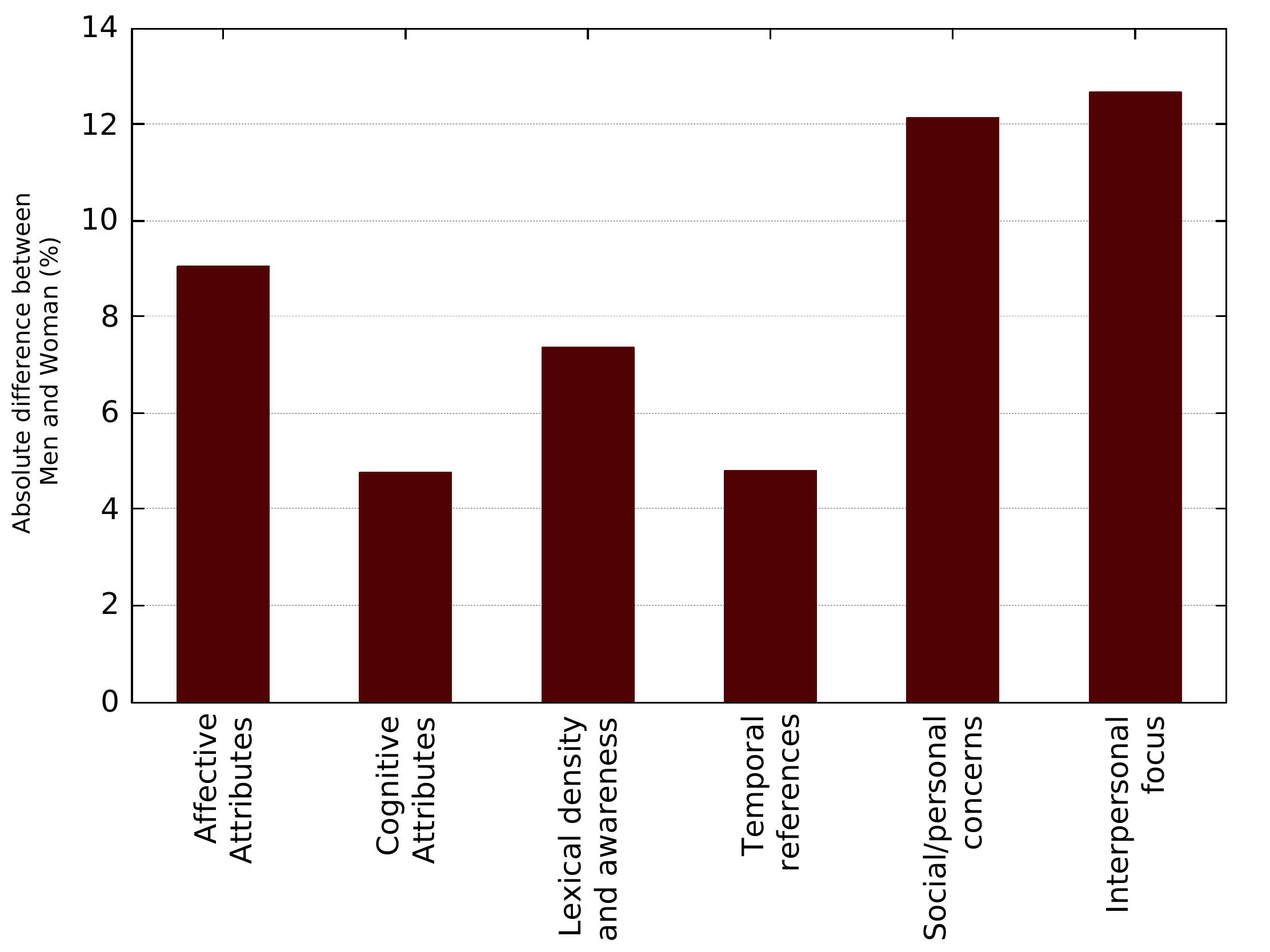}
  \caption{Mean absolute differences between male and female users per various categories of linguistic measures}
  \label{fig:gender_abs_diff}
\end{figure}

Figure \ref{fig:gender_abs_diff} shows the mean absolute differences between male and female users across each linguistic category. The difference for a specific group of features is calculated by taking the average ratio of the difference between the values for male and female to the values of the measure among male. The mean difference in the first group (affective attributes), for example, is calculated as the average of the absolute difference of each feature that comprises this group. It shows in which linguistic category the analyzed users differ the most. The number of users considered in each group were the same.

Figure \ref{fig:gender_abs_diff} also shows that interpersonal focus, which contemplates features like family, friends, health, religion, body, achievement, home, and sexual as the most prominent linguistic difference among males and females.

\begin{figure}[!htb]
  \centering
    \includegraphics[width=0.9\textwidth]{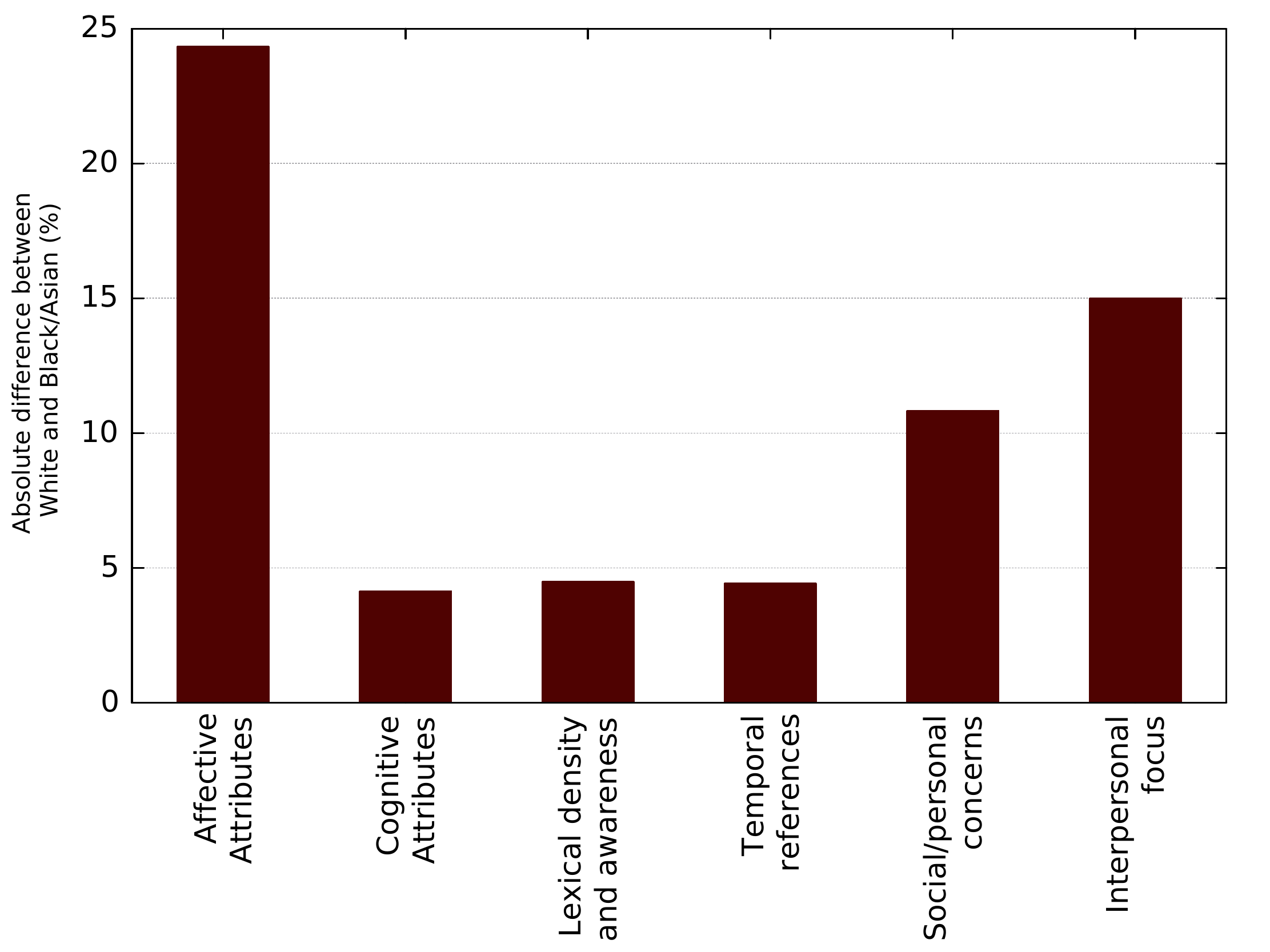}
  \caption{Mean absolute differences between white and black/asian users combined per various categories of linguistic measures}
  \label{fig:white_abs_diff}
\end{figure}

In the race domain, the analysis of the linguistic difference for each race was performed in the same way as gender, but considering the other two races combined. Figure \ref{fig:white_abs_diff} shows the mean absolute differences between white and black/asians combined. As we can see, there is a stronger difference in affective attributes, which comprises the expression of anger, anxiety, sadness, and swear. Other linguistic aspects such as social/personal concerns and interpersonal focus showed to be relevant when comparing the writing of white users against the black and asian group.

Respectively, the linguistic difference among black users was compared against white and asian users combined. Again, affective attributes are the linguistic group with the features that most differ from one ethnicity to the others. 

\begin{figure}[!htb]
  \centering
    \includegraphics[width=0.9\textwidth]{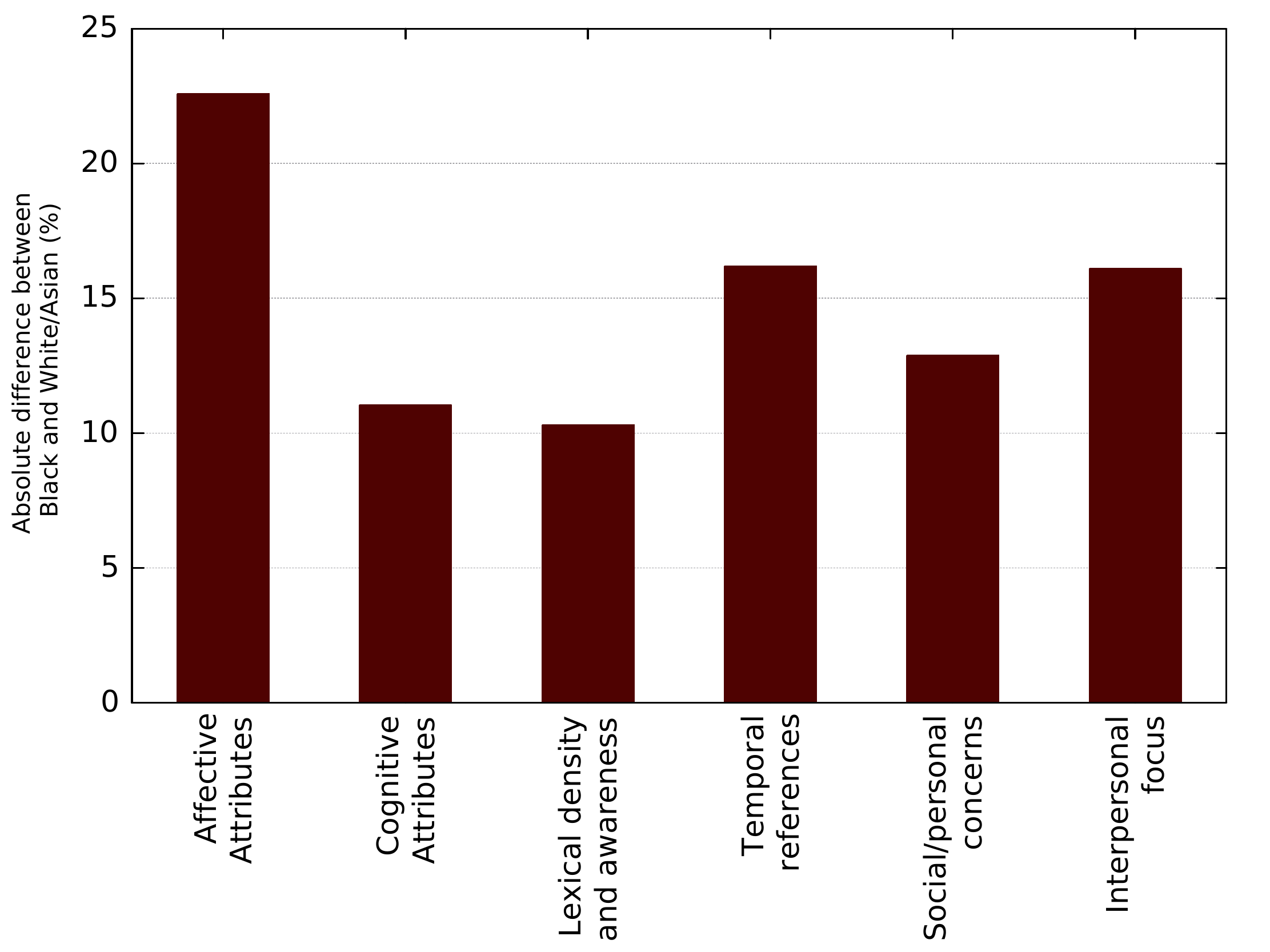}
  \caption{Mean absolute differences between black and white/asian users combined per various categories of linguistic measures}
  \label{fig:black_abs_diff}
\end{figure}

When it comes to comparing the asian linguistic to that in white and black users, some group of features that did not present higher absolute differences when comparing black and white groups, now tend to be higher such as lexical density and awareness and temporal references, which reveal some differences reflected by such different cultures especially in their way of writing.

\begin{figure}[!htb]
  \centering
    \includegraphics[width=0.9\textwidth]{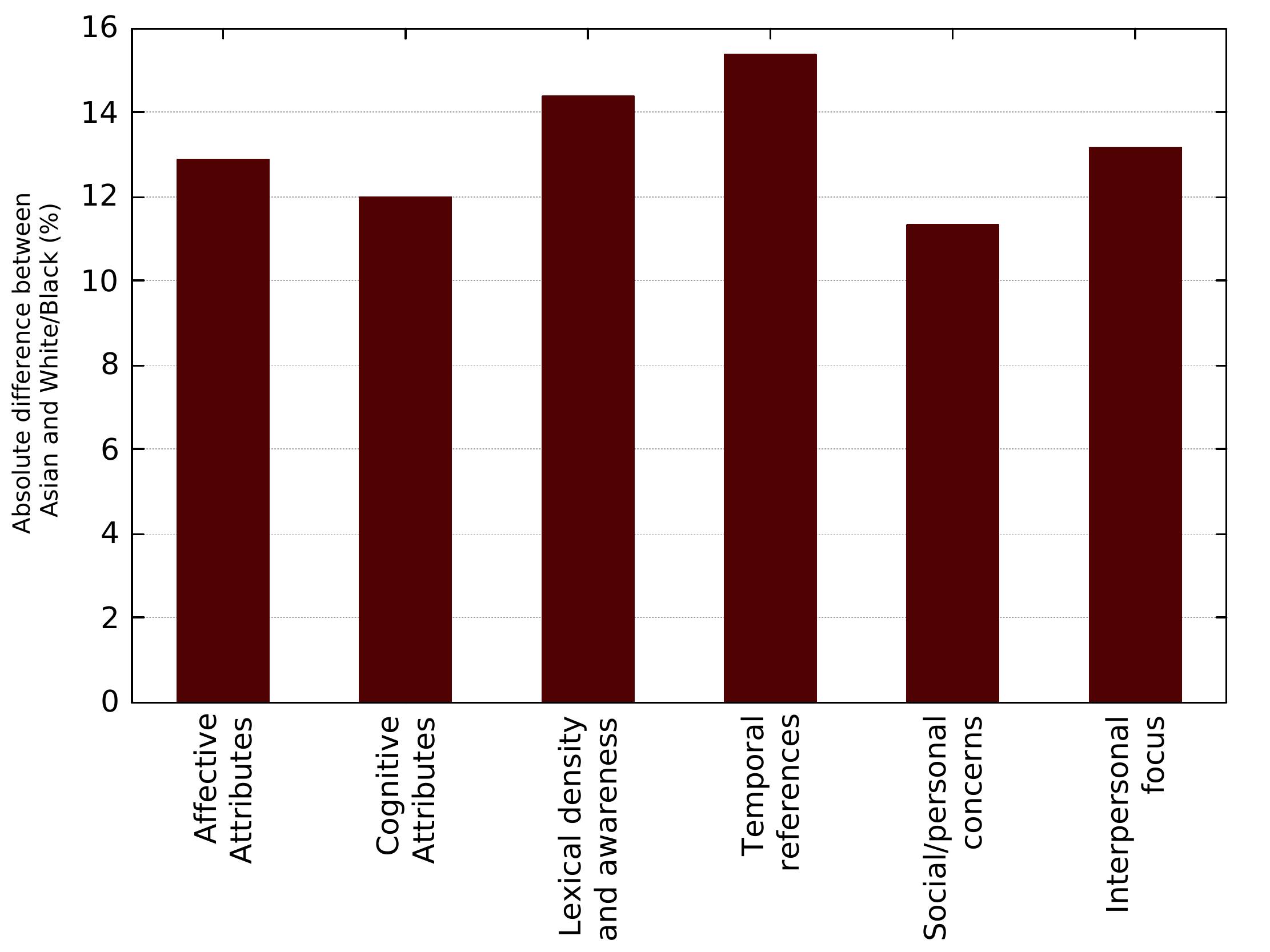}
  \caption{Mean absolute differences between asian and white/black users combined per various categories of linguistic measures}
  \label{fig:asian_abs_diff}
\end{figure}

Additionally, we correlate the produced linguistic features with gender using Wilcoxon rank sum significance tests.
As Table \ref{table:liwc_gender} presents, females tend to use anxiety ($z=-74.534$) and sadness ($z=-74.394$) terms and phrases. On the other hand, males express anger ($z=4.733$) in their tweets.

In terms of cognitive attributes, females are more likely to write phrases that express cognition and perception. From this group of features, two stand out: certainty ($z=-60.593$) and feel ($z=-70.766$) showing how females express more confidence and feelings in their writing.

In lexical density and awareness, we can see that females make more use of verbs ($z=-45.808$), auxiliary verbs ($z=-46.441$), conjunctions ($z=-72.098$), and adverbs ($z=-66.915$), while males use more articles ($z=77.303$) and prepositions ($z=32.596$).

The temporal references attributes are more present in the females writing, as we can see from the values for present tense ($z=-62.110$) and future tense ($z=-15.118$)

From social/personal Concerns perspective there is a clear trend on the usage of these features by females more than by males. Among the most noticeable values shown in Table \ref{table:liwc_gender}, are family ($z=-93.252$), bio ($z=-102.681$). Also, the predominance of features like friends, social, health, and body show that females express more social and personal concerns in their writing than males. The only feature in this group that is more present in males' writing is achievement ($z=65.265$)

Noticeably, females also have a higher tendency to write in the first person singular ($z=-97.329$) and in the second person ($z=-88.482$) than males, while there is a slight trend towards males using the first person plural in detriment of females ($z=4.309$).

Also, from the race perspective, the difference of values between each race shows some particularities in the way of writing for each race. In this analysis, one race is compared with the other two combined (e.g. white users are compared with blacks and asians).
From affective attributes, it is possible to see that black users tend to express more anger ($z=94.610$) and swear ($z=107.344$) than white/asian.

From cognitive attributes, almost all features were more present in black users texts than in the other races, with higher values for certainty ($z=62.239$), hear ($z=62.137$), and feel ($z=63.963$).

In terms of lexical density and awareness, black users have more presence in features like verbs, auxiliary verbs, conjunctions, and adverbs, while prepositions are more present among white users. 

In regards to social/personal concerns, there is a higher presence of black people, noticeably in family ($z=86.721$), social ($z=90.830$), religion ($z=85.163$), and body ($z=86.903$). 

The interpersonal focus feature set reveals that there is a predominance in the use of first person plural for white ($z=77.425$) while first person singular ($z=63.492$), second person ($z=95.495$) and third person ($z=87.717$) are more prominent in the black group.

\begin{table}[!htb]
\centering
\small
\caption{Differences between tweets from male and female users based on linguistic measures.
 $\mu(male)$ and  $\mu(female)$ are the median value of the feature for male and female respectively.
 Statistical significance is count based on \textit{Wilcoxon rank} sum tests. Linguistic features are extremely significant except from $hear$ (*:=significant) based on $p$-values.}
\label{table:liwc_gender}
\begin{tabular}{lllll}
\hline
\multicolumn{1}{l|}{}                 & $\mu(male)$ & $\mu(female)$ & z \\ \hline
\multicolumn{4}{l}{\textbf{Affective attributes}}         \\ \hline
\multicolumn{1}{l|}{anger}   &     0.0055   &    0.0056   &    4.733  \\
\multicolumn{1}{l|}{anxiety} &    0.0016   &    0.0019   &    -74.534  \\
\multicolumn{1}{l|}{sadness}    &    0.0029   &    0.0034   &    -74.394 \\ 
\multicolumn{1}{l|}{swear}   &    0.0023   &    0.0026   &    -7.411 \\ \hline
\multicolumn{4}{l}{\textbf{Cognitive attributes}}         \\ \hline
\multicolumn{4}{l}{Cognition}                             \\ \hline
\multicolumn{1}{l|}{causation} &    0.0101   &    0.0104   &    -18.627 \\
\multicolumn{1}{l|}{certainty}   &    0.0101   &    0.0111   &    -60.593 \\
\multicolumn{1}{l|}{tentativeness}   &    0.0136   &    0.0141   &    -14.641  \\ \hline
\multicolumn{4}{l}{Perception}                            \\ \hline
\multicolumn{1}{l|}{see}      &    0.00957   &    0.0099   &    -24.538   \\
\multicolumn{1}{l|}{hear}         &    0.0055   &    0.0056   &    -0.033$^{*}$ \\
\multicolumn{1}{l|}{feel}    &    0.0035   &    0.0041  &    -70.766   \\
\multicolumn{1}{l|}{percepts}   &    0.0207   &    0.0218   &    -41.373     \\
\multicolumn{1}{l|}{insight}     &    0.0115   &    0.0125   &    -46.806    \\
\multicolumn{1}{l|}{relative}    &    0.1014   &    0.0999   &    18.026  \\ \hline
\multicolumn{4}{l}{\textbf{Lexical Density and Awareness}} \\ \hline
\multicolumn{1}{l|}{verbs}  &    0.1103   &    0.1170   &    -45.808    \\
\multicolumn{1}{l|}{auxiliary verbs}  &    0.0539   &    0.0583   &    -46.441    \\
\multicolumn{1}{l|}{articles}   &    0.0370   &    0.0340   &    77.303       \\
\multicolumn{1}{l|}{prepositions}    &    0.0843   &    0.0817   &    32.596       \\
\multicolumn{1}{l|}{conjunctions}  &    0.0279   &    0.0314   &    -72.098        \\
\multicolumn{1}{l|}{adverbs}    &    0.0317   &    0.0355   &    -66.915       \\ \hline
\multicolumn{4}{l}{\textbf{Temporal references}}          \\ \hline
\multicolumn{1}{l|}{present tense}  &    0.0802   &    0.0871   &    -62.110        \\
\multicolumn{1}{l|}{future tense}    &    0.0103   &    0.0106   &    -15.118     \\ \hline
\multicolumn{4}{l}{\textbf{Social/Personal Concerns}}     \\ \hline
\multicolumn{1}{l|}{family}        &    0.0026   &    0.0034   &    -93.252    \\
\multicolumn{1}{l|}{friends}       &    0.0028   &    0.0033   &    -66.168    \\
\multicolumn{1}{l|}{social}      &    0.0938   &    0.1021   &    -77.896   \\
\multicolumn{1}{l|}{health}     &    0.0037   &    0.0044   &    -76.446     \\
\multicolumn{1}{l|}{religion}   &    0.0024   &    0.0025   &    -26.485       \\
\multicolumn{1}{l|}{bio}       &    0.0157   &    0.0203   &    -102.681    \\
\multicolumn{1}{l|}{body}       &    0.0045   &    0.0056   &    -58.386    \\
\multicolumn{1}{l|}{achievement}   &    0.0116   &    0.0105   &    65.265     \\
\multicolumn{1}{l|}{home}    &    0.0022   &    0.0026   &    -74.049      \\
\multicolumn{1}{l|}{sexual}   &    0.0011   &    0.0012   &    -18.691     \\
\multicolumn{1}{l|}{death}  &    0.0014   &    0.0013   &    29.463     \\ \hline
\multicolumn{4}{l}{\textbf{Interpersonal focus}}          \\ \hline
\multicolumn{1}{l|}{1st p. singular}  &    0.0245   &    0.0340   &    -97.329   \\
\multicolumn{1}{l|}{1st p. plural}   &    0.0046   &    0.0045   &    4.309    \\
\multicolumn{1}{l|}{2nd p.}   &    0.0160   &    0.0198   &    -88.482    \\
\multicolumn{1}{l|}{3rd p.}   &    0.0030   &    0.0031   &    -3.371$^{***}$  \\ \hline
\end{tabular}
\end{table}

\begin{table*}[!htb]
\centering
\small
\caption{Differences between tweets from white, black, and asian users based on linguistic measures.
$\mu(White)$, $\mu(Black)$ and $\mu(Black)$ is the median value of features for each demographic group respectively.
Statistical significance is count based on Wilcoxon rank sum tests. The $p$-values are extremely significant for all linguistic features. We test the correlation of each unique demographic group with the others.}
\label{table:liwc_race}
\begin{tabular}{llllllll}
\hline
\multicolumn{1}{l|}{}                 & $\mu(White)$ & $\mu(Black)$ & $\mu(Asian)$ & $z_{W/B-A}$ & $z_{B/W-A}$ & $z_{A/W-B}$\\ \hline
\multicolumn{5}{l}{\textbf{Affective attributes}}         &    &    &    \\ \hline
\multicolumn{1}{l|}{anger}   &   0.0051   &   0.0081   &   0.0056   &   -67.261   &   94.610   &   -5.236  \\
\multicolumn{1}{l|}{anxiety}   &   0.0017   &   0.0019   &   0.0016   &   -0.696   &   33.789   &   -30.517  \\
\multicolumn{1}{l|}{sadness}   &   0.0031   &   0.0034   &   0.0032   &   -20.814   &   28.205   &   -0.625  \\
\multicolumn{1}{l|}{swear}   &   0.0021   &   0.0064   &   0.0027   &   -90.375   &   107.344   &   11.329  \\ \hline
\multicolumn{5}{l}{\textbf{Cognitive attributes}}         &    &    &    \\ \hline
\multicolumn{5}{l}{Cognition}                             &    &    &    \\ \hline
\multicolumn{1}{l|}{causation}   &   0.0104   &   0.0105   &   0.0096   &   29.931   &   19.465   &   -54.832  \\
\multicolumn{1}{l|}{certainty}   &   0.0105   &   0.0116   &   0.0101   &   -19.404   &   62.239   &   -33.955  \\
\multicolumn{1}{l|}{tentativeness}   &   0.0138   &   0.0152   &   0.0130   &   -8.958   &   55.174   &   -40.226  \\ \hline
\multicolumn{5}{l}{Perception}                            &    &        \\ \hline
\multicolumn{1}{l|}{see}   &   0.0098   &   0.0098   &   0.0095   &   18.756   &   6.970   &   -29.506  \\
\multicolumn{1}{l|}{hear}   &   0.0055   &   0.0062   &   0.0054   &   -26.349   &   62.137   &   -25.331  \\
\multicolumn{1}{l|}{feel}   &   0.0037   &   0.0044   &   0.0039   &   -44.180   &   63.963   &   -5.128  \\
\multicolumn{1}{l|}{percepts}   &   0.0212   &   0.0223   &   0.0210   &   -14.067   &   43.711   &   -23.308  \\
\multicolumn{1}{l|}{insight}   &   0.0122   &   0.0128   &   0.0112   &   11.133   &   40.420   &   -51.201  \\
\multicolumn{1}{l|}{relative}   &   0.1020   &   0.1012   &   0.0936   &   50.614   &   15.841   &   -76.870   \\ \hline
\multicolumn{5}{l}{\textbf{Lexical Density and Awareness}} &    &    &    \\ \hline
\multicolumn{1}{l|}{verbs}   &   0.1125   &   0.1222   &   0.1082   &   -16.435   &   64.214   &   -39.436  \\
\multicolumn{1}{l|}{auxiliary verbs}   &   0.0554   &   0.0612   &   0.0529   &   -12.202   &   58.285   &   -39.130  \\
\multicolumn{1}{l|}{articles}   &   0.0366   &   0.0339   &   0.0314   &   96.532   &   -26.056   &   -94.363  \\
\multicolumn{1}{l|}{prepositions}   &   0.0851   &   0.0817   &   0.0743   &   77.024   &   1.032   &   -95.556  \\
\multicolumn{1}{l|}{conjunctions}   &   0.0291   &   0.0319   &   0.0286   &   -11.852   &   43.571   &   -25.898  \\
\multicolumn{1}{l|}{adverbs}   &   0.0329   &   0.0363   &   0.0325   &   -17.239   &   48.159   &   -23.542  \\ \hline
\multicolumn{5}{l}{\textbf{Temporal references}}          &    &    &    \\ \hline
\multicolumn{1}{l|}{present tense}   &   0.0825   &   0.0912   &   0.0798   &   -21.972   &   69.126   &   -37.196  \\
\multicolumn{1}{l|}{future tense}   &   0.0103   &   0.0119   &   0.0099   &   -28.333   &   79.181   &   -38.719     \\ \hline
\multicolumn{5}{l}{\textbf{Social/Personal Concerns}}     &    &    &    \\ \hline
\multicolumn{1}{l|}{family}   &   0.0029   &   0.0040   &   0.0032   &   -74.318   &   86.721   &   10.755  \\
\multicolumn{1}{l|}{friend}   &   0.0031   &   0.0033   &   0.0033   &   -26.248   &   25.332   &   8.717  \\
\multicolumn{1}{l|}{social}   &   0.0956   &   0.1101   &   0.0971   &   -60.389   &   90.830   &   -10.166  \\
\multicolumn{1}{l|}{health}   &   0.0040   &   0.0044   &   0.0039   &   -9.579   &   45.973   &   -30.920  \\
\multicolumn{1}{l|}{religion}   &   0.0024   &   0.0031   &   0.0024   &   -53.672   &   85.163   &   -13.154  \\
\multicolumn{1}{l|}{bio}   &   0.0176   &   0.0204   &   0.0179   &   -32.215   &   53.914   &   -10.492  \\
\multicolumn{1}{l|}{body}   &   0.0048   &   0.0067   &   0.0052   &   -62.906   &   86.903   &   -3.428  \\
\multicolumn{1}{l|}{achievement}   &   0.0114   &   0.0109   &   0.0097   &   69.227   &   -1.632   &   -83.506  \\
\multicolumn{1}{l|}{home}   &   0.0025   &   0.0024   &   0.0022   &   50.362   &   -4.554   &   -57.624  \\
\multicolumn{1}{l|}{sexual}   &   0.0011   &   0.0019   &   0.0012   &   -51.768   &   71.799   &   -3.084  \\
\multicolumn{1}{l|}{death}  &   0.0014   &   0.0015   &   0.0013   &   4.356   &   31.454   &   -34.554  \\\hline
\multicolumn{5}{l}{\textbf{Interpersonal focus}}          &    &    &    \\ \hline
\multicolumn{1}{l|}{1st p. singular}  &   0.0268   &   0.0355   &   0.0296   &   -51.874   &   63.492   &   4.760  \\
\multicolumn{1}{l|}{1st p. plural}   &   0.0048   &   0.0042   &   0.0039   &   77.425   &   -28.107   &   -68.994  \\
\multicolumn{1}{l|}{2nd p.}      &   0.0169   &   0.0227   &   0.0177   &   -63.930   &   95.495   &   -10.148  \\
\multicolumn{1}{l|}{3rd p.}   &   0.0030   &   0.0039   &   0.0028   &   -36.070   &   87.717   &   -37.143  \\ \hline
\end{tabular}
\end{table*}

Table \ref{table:ranking-top-20-gender} and Table \ref{table:ranking-top-20-race} present the ranking difference for the 20 most common phrases for gender and races respectively. To find these differences, we randomly selected $1,000$ users from each group (male, female, asian, black, and white). Their tweets were used to create ngrams for each group. With this subset of our dataset, we extracted the top $100$ phrases for each demographic group and the top $20$ are shown in these Tables.

\begin{table}[!htb]
\centering
\caption{Ranking differences of gender top phrases. We use $ne$ for no existing phrases in a group.}
\label{table:ranking-top-20-gender}
\begin{tabular}{rccc}
\hline
\textbf{}                            & \textbf{Rank(female)} & \textbf{Rank(male)} & \textbf{Diff(F-M)} \\ \hline
\multicolumn{1}{r|}{i do n't}        & 1                    & 1                  & 0                  \\
\multicolumn{1}{r|}{i ca n't}        & 2                    & 2                  & 0                  \\
\multicolumn{1}{r|}{you do n't}      & 3                    & 3                  & 0                  \\
\multicolumn{1}{r|}{i 'm not}        & 4                    & 4                  & 0                  \\
\multicolumn{1}{r|}{ca n't wait}     & 5                    & 8                  & 3                  \\
\multicolumn{1}{r|}{i 'm so}         & 6                    & 19                 & 13                 \\
\multicolumn{1}{r|}{i love you}      & 7                    & 15                 & 8                  \\
\multicolumn{1}{r|}{do n't know}     & 8                    & 11                 & 3                  \\
\multicolumn{1}{r|}{i want to}       & 9                    & 24                 & 15                 \\
\multicolumn{1}{r|}{more for virgo}  & 10                   & 55                 & 45                 \\
\multicolumn{1}{r|}{more for cancer} & 11                   & 29                 & 18                 \\
\multicolumn{1}{r|}{i wan na}        & 12                   & 28                 & 16                 \\
\multicolumn{1}{r|}{! i 'm}          & 13                   & 25                 & 12                 \\
\multicolumn{1}{r|}{you ca n't}      & 14                   & 16                 & 2                  \\
\multicolumn{1}{r|}{more for libra}  & 15                   & 39                 & 24                 \\
\multicolumn{1}{r|}{it 's a}         & 16                   & 10                 & 6                  \\
\multicolumn{1}{r|}{and i 'm}        & 17                   & 33                 & 16                 \\
\multicolumn{1}{r|}{more for pisces} & 18                   & ne                 & -                  \\
\multicolumn{1}{r|}{i need to}       & 19                   & 34                 & 15                 \\
\multicolumn{1}{r|}{do n't have}     & 20                   & 27                 & 7                  \\ \hline
\end{tabular}
\end{table}

As we can see in Table \ref{table:ranking-top-20-gender} phrases expressing negation are in the top 4-positions for both males and females. It is also clear to see that females are more into signs than males since phrases with this kind of content present higher differences in the gender ranking. 

Due to the informal nature of Twitter, the top phrases also reveal that it is common the usage of slangs like ``do n't'', ``ca n't'' and ``wan na'' for both genders. 

\begin{table*}[!htb]
\centering
\scriptsize
\caption{Ranking differences of race top phrases. We use $ne$ for no existing phrases in a group.}
\label{table:ranking-top-20-race}
\begin{tabular}{r|cccccc}
\textbf{}     & \textbf{Rank(White)} & \textbf{Rank(Black)} & \textbf{Rank(Asian)} & \textbf{Diff(W-B)} & \textbf{Diff(W-A)} & \textbf{Diff(B-A)} \\ \hline
i do n't      & 1                    & 1                    & 1                    & 0         & 0         & 0         \\
i ca n't      & 2                    & 2                    & 2                    & 0         & 0         & 0         \\
ca n't wait   & 3                    & 18                   & 7                    & 15        & 4         & 11        \\
you do n't    & 4                    & 4                    & 3                    & 0         & 1         & 1         \\
i 'm not      & 5                    & 8                    & 6                    & 3         & 1         & 2         \\
i love you    & 6                    & 33                   & 4                    & 27        & 2         & 29        \\
i 'm so       & 7                    & 16                   & 6                    & 9         & 1         & 10        \\
do n't know   & 8                    & 19                   & 11                   & 11        & 3         & 8         \\
it 's a       & 9                    & 26                   & 16                   & 17        & 7         & 10        \\
one of the    & 10                   & 48                   & 20                   & 38        & 10        & 28        \\
i want to     & 11                   & 47                   & 10                   & 36        & 1         & 37        \\
! i 'm        & 12                   & 46                   & 29                   & 34        & 17        & 17        \\
if you 're    & 13                   & 28                   & 19                   & 15        & 6         & 9         \\
thank you for & 14                   & 126                  & 28                   & 112       & 14        & 98        \\
it 's not     & 15                   & 34                   & 32                   & 19        & 17        & 2         \\
and i 'm      & 16                   & 58                   & 21                   & 42        & 5         & 37        \\
you ca n't    & 17                   & 17                   & 17                   & 0         & 0         & 0         \\
i 'm at       & 18                   & 53                   & 26                   & 35        & 8         & 27        \\
n't wait to   & 19                   & 100                  & 51                   & 81        & 32        & 49        \\
i liked a     & 20                   & 7                    & ne                   & 13        & -         & -         \\ \hline
\end{tabular}
\end{table*}

When analyzing the ranking of race top phrases in Table \ref{table:ranking-top-20-race}, the trend of using negation phrases also happens here. Phrases containing expressions like ``i don't'', ``i can't'' and ``i'm not'' appear in the top positions for all the racial groups. Another interesting result is the position of the expression ``i love you'' in the writing of different races. White and asian users seem to be more likely to tweet contents with this expression than black users. Also, the expression ``i want to'' appears more often in the writing of white and asian users than in the blacks. These tables show differences regarding the way of writing of each demographic group and reveal interesting characteristics about the difference from one to another.

\section{Differences in Topic Interests}

\begin{figure}[!htb]
  \centering
    \includegraphics[width=0.9\textwidth]{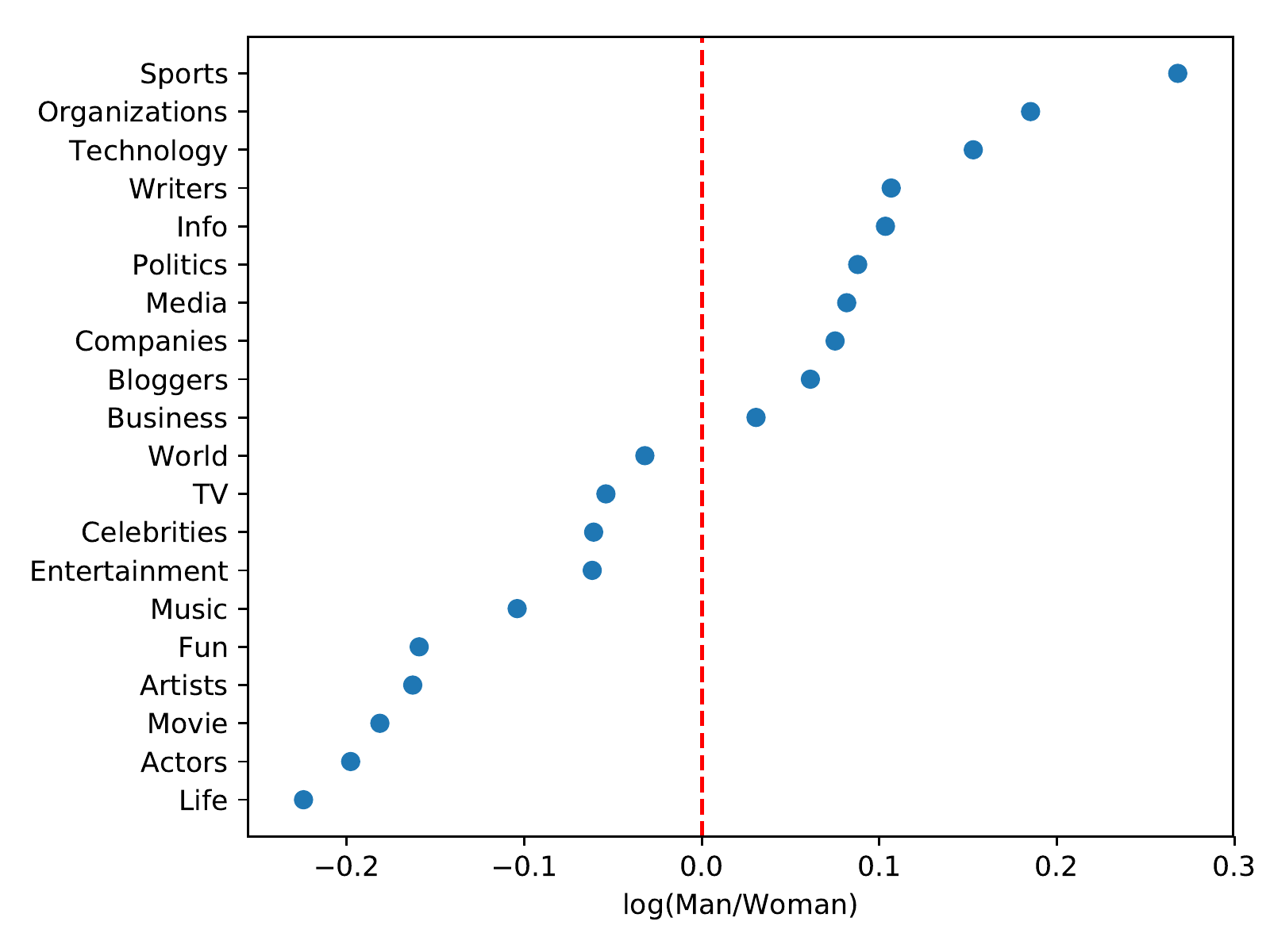}
  \caption{Gender interests: dots represent the gender interests for the 20-top popular topics.}
  \label{fig:topics_gender}
\end{figure}

Males and females may have differences in preferences and interests in digest information. In order to understand which topic is preferable to females than males, we analyze the differences in the topic interest of users in our dataset. The Figure \ref{fig:topics_gender} shows the gender distribution for the 20-top topics that we extracted, with log-ratio of perceived male to female. It shows the topic interest for users based on gender in our dataset. On the right side, we see topics related to males' interests while on the left side we see the topics that females are more interested than males. The 3-top topics for males are sports, organizations, and technology. In other words, males tend to interest more in these topics than females. However, females interest more for life, actors, and movie than males. More specifically,  the gender difference between topics varies among males and females.

\begin{figure*}%
    \centering
    \subfloat[White vs Asian]{{\includegraphics[width=.6\textwidth]{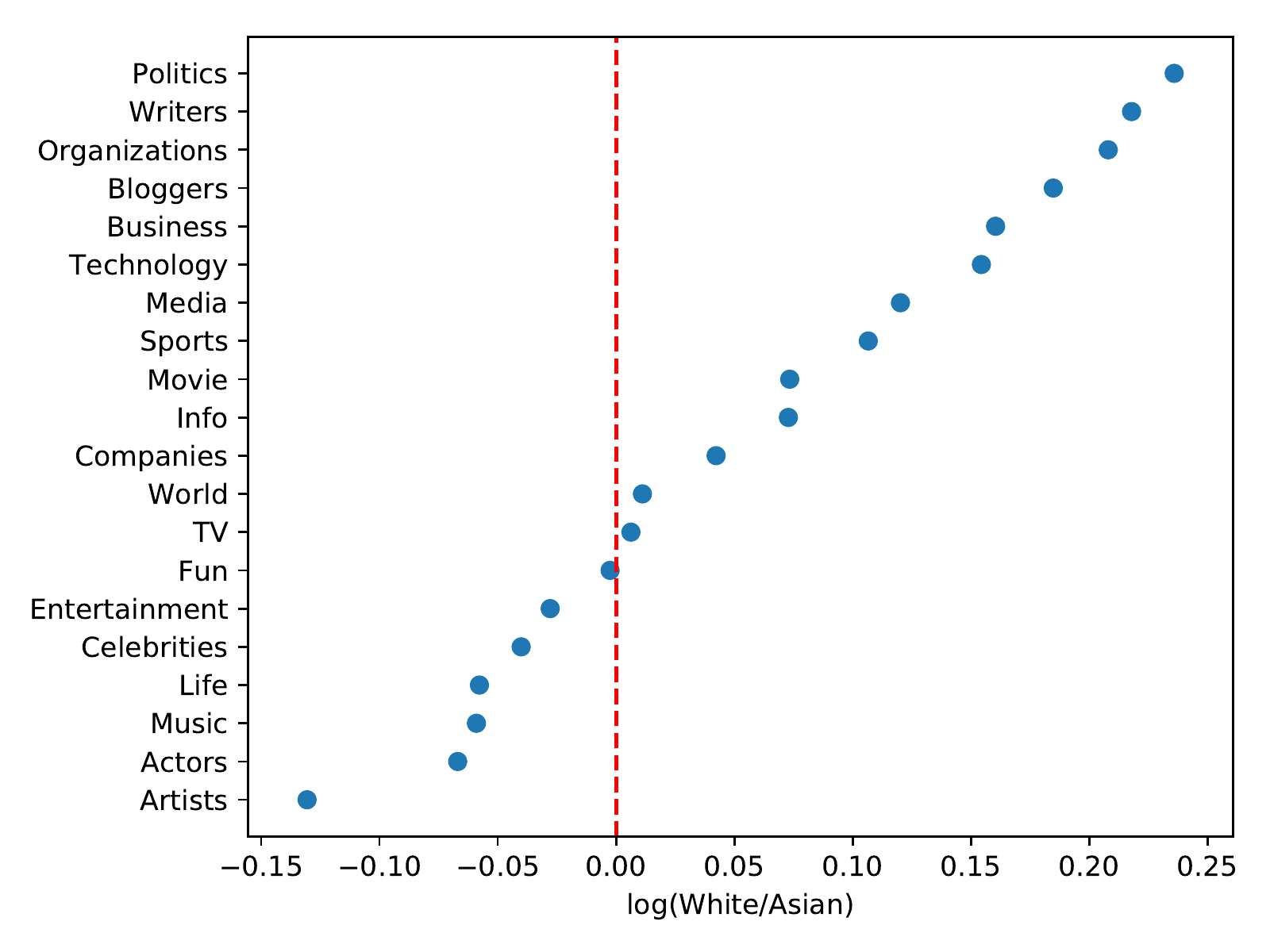} }}%
   
    \subfloat[White vs Black]{{\includegraphics[width=.6\textwidth]{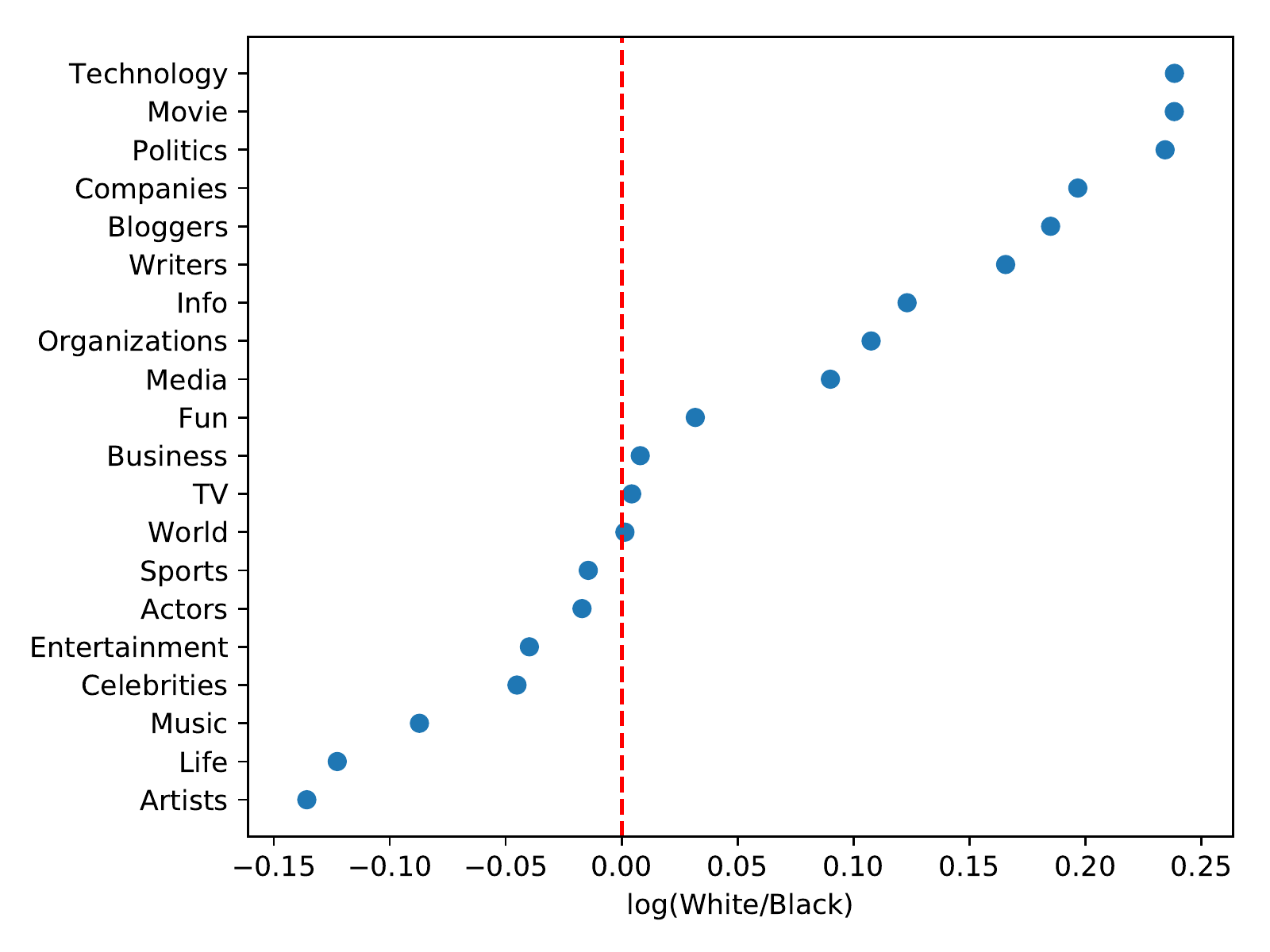} }}%
    \qquad
    \subfloat[Asian vs Black]{{\includegraphics[width=.6\textwidth]{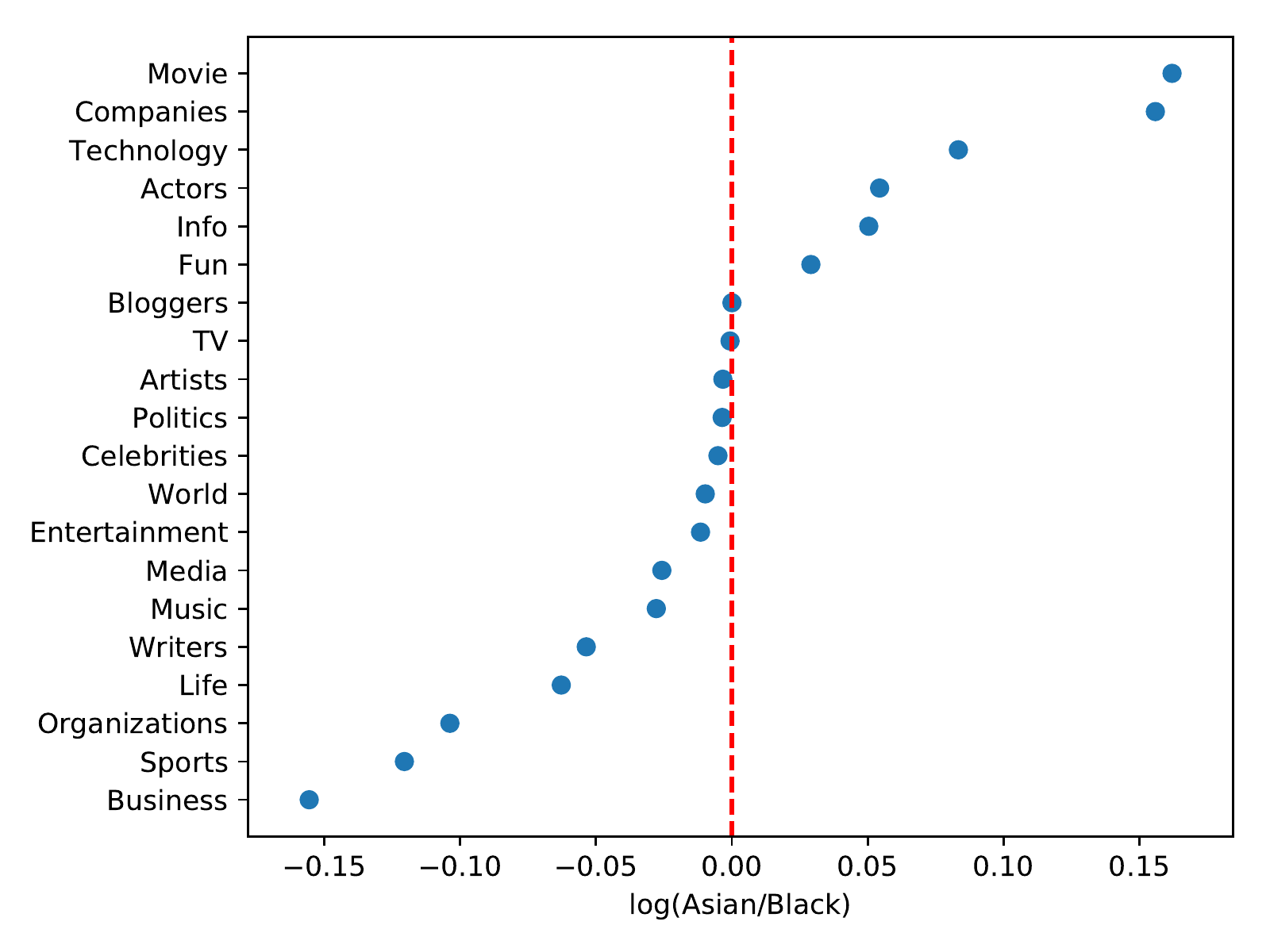} }}%
    \caption{Race interests: dots represent the race interests of (a) white against asians, (b) white against blacks, and (c) asian against blacks for the 20-top popular topics. The dataset is normalized by the number of blacks as shown in Table \ref{table:expected}.}%
    \label{fig:race_topics_against}%
\end{figure*}

In a similar way, we present the race distribution for the 20-top topics of asian, black, and white users in Figure \ref{fig:race_topics_against}. In order to show results regarding race, for this specific analysis, we have normalized the dataset by the number of black users once they are the minority amount of users in our dataset, as shown in Table \ref{table:expected}. Therefore, we have randomly selected $45,406$ users for each race to study their topic interests. Users from different races may also vary in interests and preferences. Figure \ref{fig:race_topics_against}-a shows that white users have more interest in politics, writers, and organizations than asians. However, asians prefer more artists, actors, and music topics than whites. Figure \ref{fig:race_topics_against}-b compares the differences in topic interests for white and blacks. We see that white users are interested in technology, movie, and politics more than blacks. Nonetheless, blacks prefer more artists, life, and music topics. Finally, from Figure \ref{fig:race_topics_against}-c, we see that asians have more interest for movie, companies, and technology topics than blacks. On the other hand, blacks prefer more business, sports, and organizations than asians.

\section{Concluding Remarks}

In this chapter, we presented the linguistic differences between gender and race. Further, we also analyzed the differences in terms of interests. We showed that interpersonal focus, which contemplates features like family, friends, health, religion, body, achievement, home, and sexual is the most prominent linguistic differences among males and females. There is a stronger difference in affective attributes for white and black users in comparison to other users, which comprises the expression of anger, anxiety, sadness, and swear. The differences in lexical density and awareness and temporal references attributes are the most significant between asians and other groups. This reveals some differences reflected by these different ethnicity especially in their way of writing.

Other interesting findings are that females tend to use anxiety and sadness terms while males express more with anger in their tweets. Females are more likely to write phrases that express cognition, perception, confidence and feelings. Additionally, females make more use of verbs, auxiliary verbs, conjunctions, and adverbs while males use more articles and prepositions. The temporal references and social/personal concerns attributes are more present in the females. Also, they have a higher tendency to write in the first person singular and in the second person while males use the first person plural and express more concern of achievement. In a similar analysis, black users tend to express more anger, swear, and cognitive attributes than white/asian. They also have more presence in features like verbs, auxiliary verbs, conjunctions, and adverbs. Blacks, tend more to use terms related to family, social, religion, and body. On the other hand, prepositions and the use of first person plural are more present among white users while the third person is more prominent for blacks.

Then, in terms of interests, we showed that the 3-top topics for males are sports, organizations, and technology. On the other hand, females have more interest for life, actors, and movie. Regarding race, we showed that white users have more interest in politics, writers, and organizations than asians. Asians prefer to write about artists, actors, and music topics more than whites. White users are more interested in technology, movie, and politics than blacks while blacks prefer more artists, life, and music topics. Asians have more interest for movie, companies, and technology topics than blacks. Finally, blacks prefer more business, sports, and organizations than asians.

Finally, phrases expressing negation are in the top positions for both males and females. It is also clear to see that
females are more into zodiac signs than males since phrases with this kind of content present higher differences in the gender ranking. Further, white and asian users seem to be more likely to tweet contents ``I love you'' and  ``I want to'' than blacks.

In the next chapter we discuss how different demographic groups are connected and interact with each other. For doing so, we dispose of the followers and friends of all users from our dataset. Our ultimate goal is to examine the proportion of connections and interactions among each demographic group.
 \chapter{Demographic Group Interconnections} \label{chap:results_interconnections}

Next, we analyze how demographic groups are connected with each other. Our ultimate goal is to examine the proportion of connections and interactions among each demographic group. Thus, for this analysis we used the dataset of $428,697$ users related to friends, followers and interactions discussed previously.

In order to represent connection probabilities among demographic groups we built a probabilistic graph and we compute the probability of a connection/interaction for every pair of users. To gain a holistic view, we aggregate all pairs of users making the nodes to represent demographic groups and each direct edge to represent the probability of a relationship happened, like friendship or interaction. The sum of all outgoing probabilities for each demographic group is $1.0$. 

\section{Gender and its Interconnections}

\begin{figure}[!htb]
  \centering
    \includegraphics[width=0.6\textwidth]{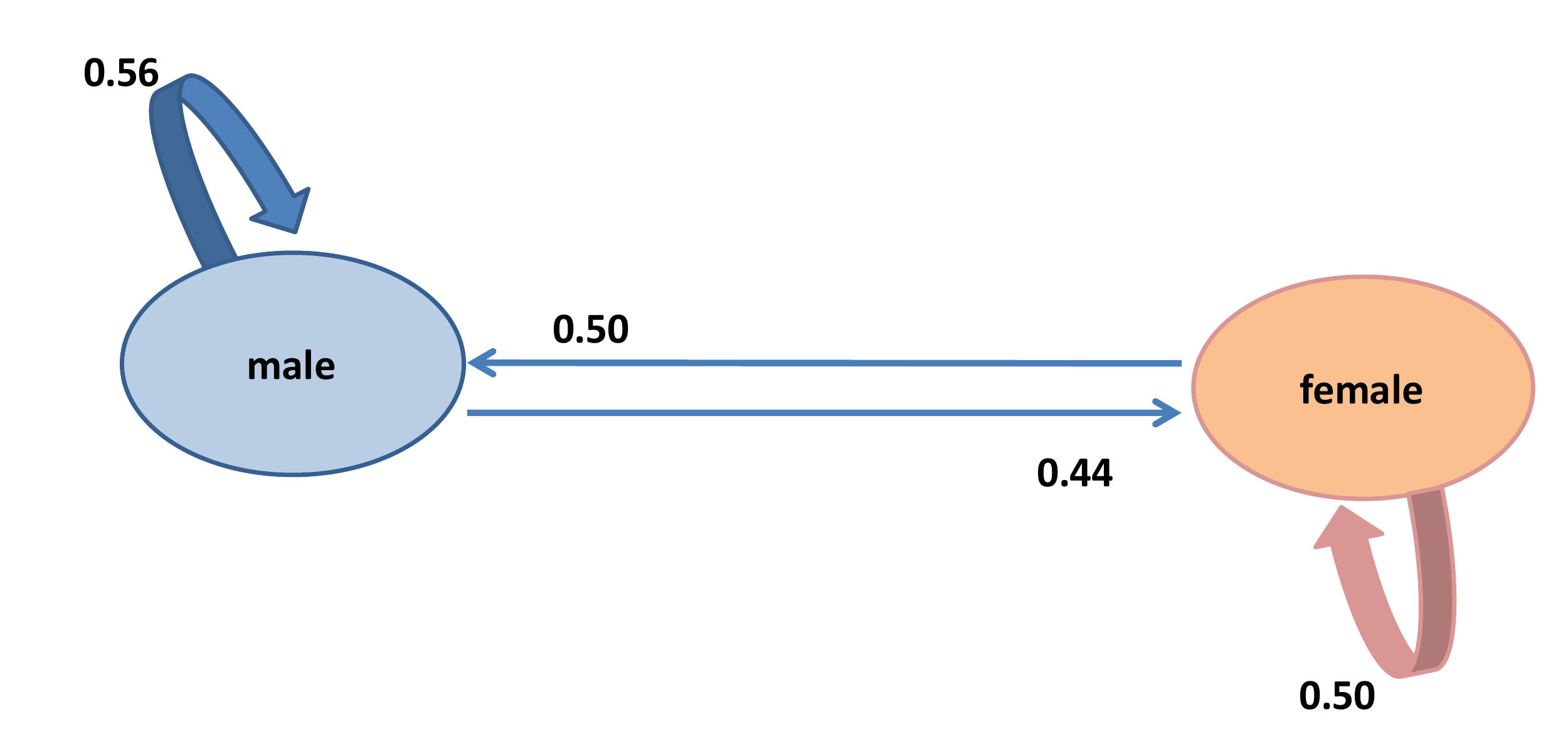}
  \caption{Probability of friendship for groups of users separated by gender}
  \label{fig:gender_graph_friends}
\end{figure}

We begin by analyzing interconnection among groups of users separated by gender. 
We analyze the probability of male and female users to be friends with (i.e. to follow) other male and female users. In our analysis, we note in Figure \ref{fig:gender_graph_friends} that females are equally likely to follow males and females in $50\%$, while males tend to follow $56\%$ of males versus $44\%$ of females. If we take the numbers in Table~\ref{table:expected} as a proxy for the expected population of males and females in a randomly created list of friends, we note that even the equally distributed list of friends of females is actually disproportional as the expected percentage of males and females are $48.12\%$ and $51.88\%$, respectively. This means that males tend to follow $16.37\%$ more other males than expected, whereas females tend to follow $3.9\%$ more males than expected. This suggests that both, males and females connections, take their part of the responsibility on these gender inequalities, but it also shows that males tend to take the larger part of the responsibility.

\begin{figure}[!htb]
  \centering
    \includegraphics[width=0.6\textwidth]{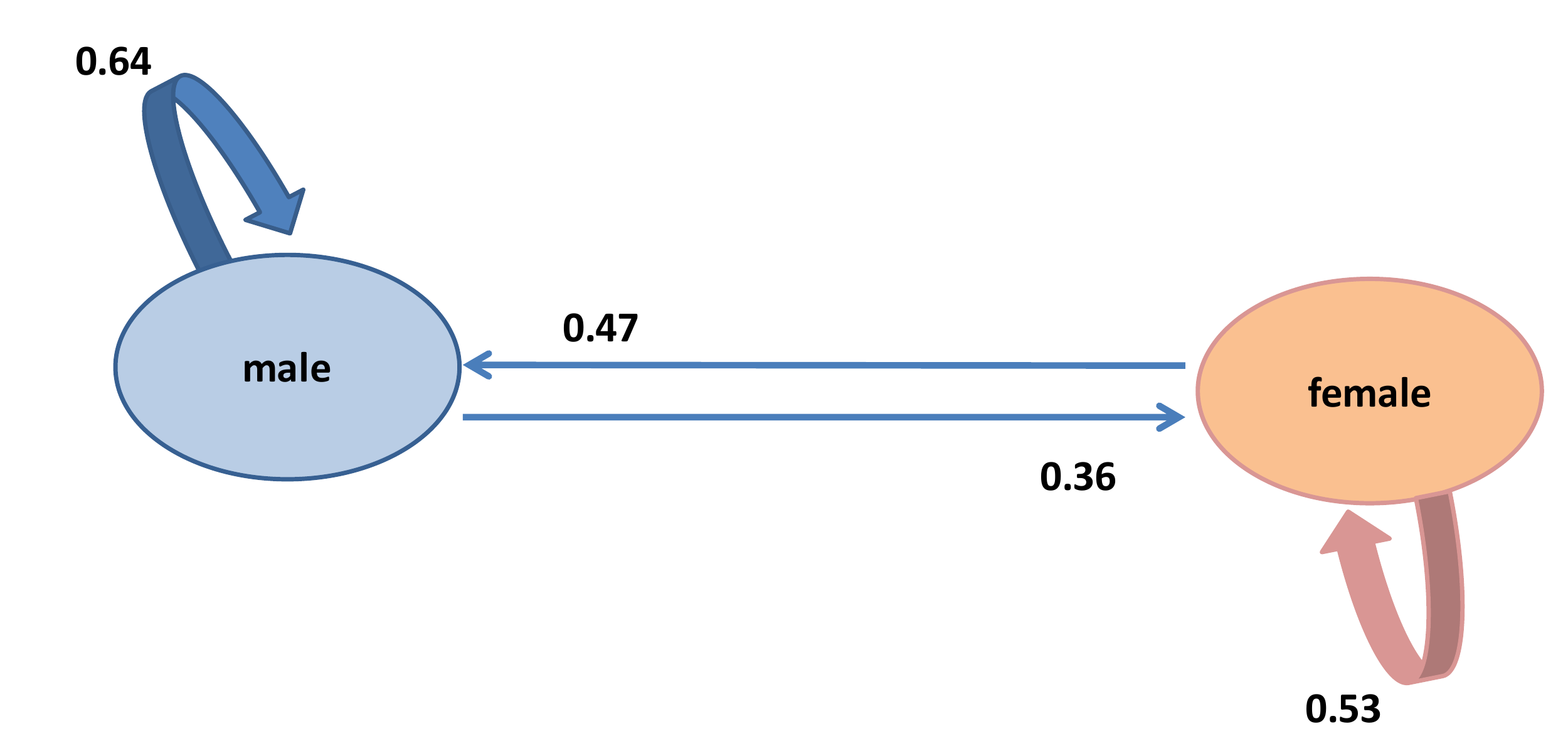}
  \caption{Probability of interactions for groups of users separated by gender}
  \label{fig:act_gender_graph}
\end{figure}

In terms of interactions, similar observations can be made. However, as we can see in Figure \ref{fig:act_gender_graph}, the proportion of males and females retweet/mentioned by females is quite close to the expected population ($53\%$ of females against $47\%$ for males). For the group of males, we noted a much higher disparity in terms of interactions in comparison to friendship. Males tend to interact with $64\%$ of other males and only $36\%$ of females. 

This result quantifies the perceived inequality in Twitter. These observations are in line with perceived inequalities in the offline world, including in academic job hiring, earnings, funding and academic rating~\cite{sugimoto2013global,sugimoto2015relationship} as well as income differences between males and females in the population of high earners \cite{merluzzi2015unequal}.

\section{Race and its Interconnections}

\begin{figure}%
    \centering
    \subfloat[Friendship]{{\includegraphics[width=7.5cm]{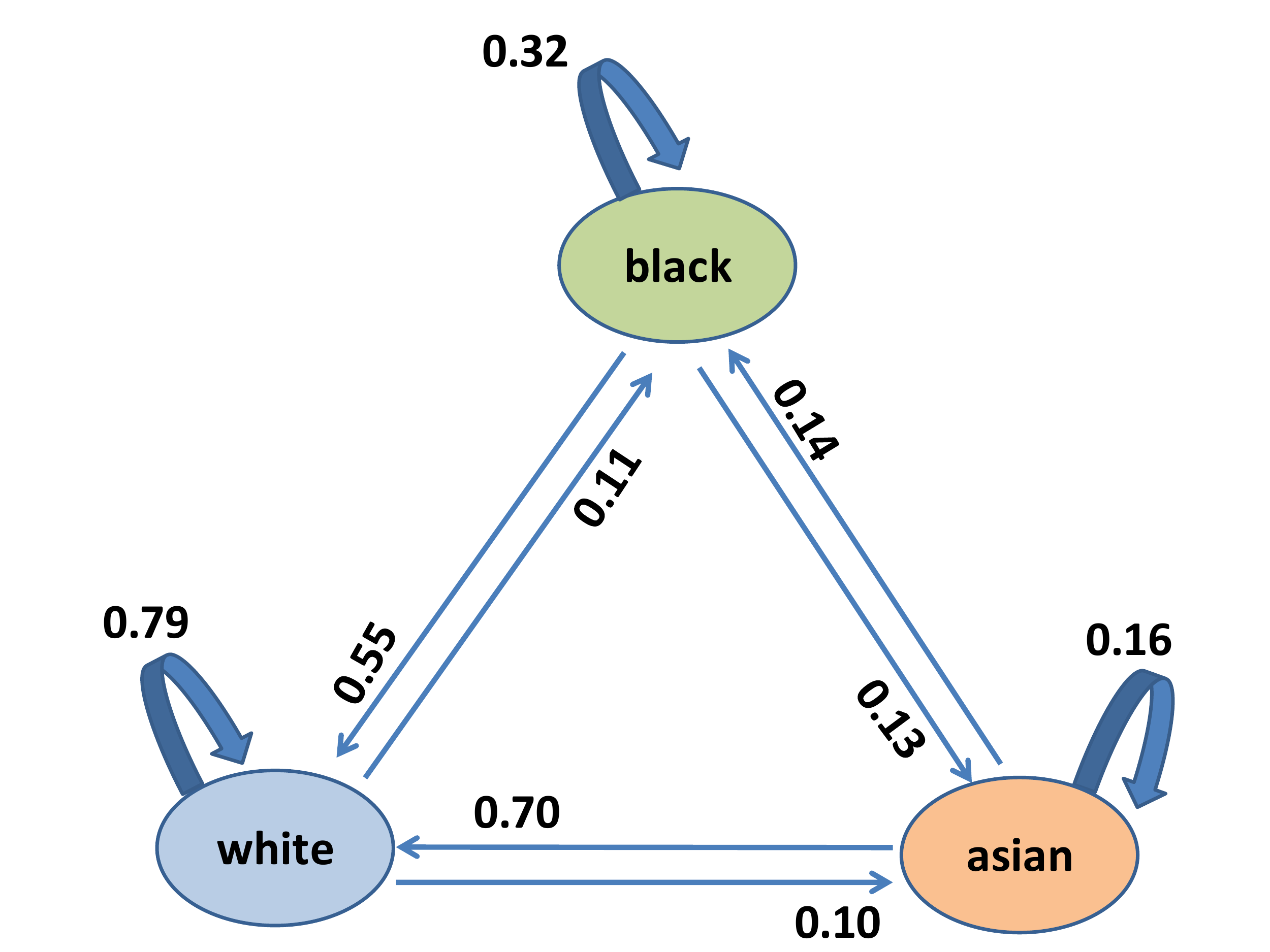} }}%
    \subfloat[Interaction]{{\includegraphics[width=7.5cm]{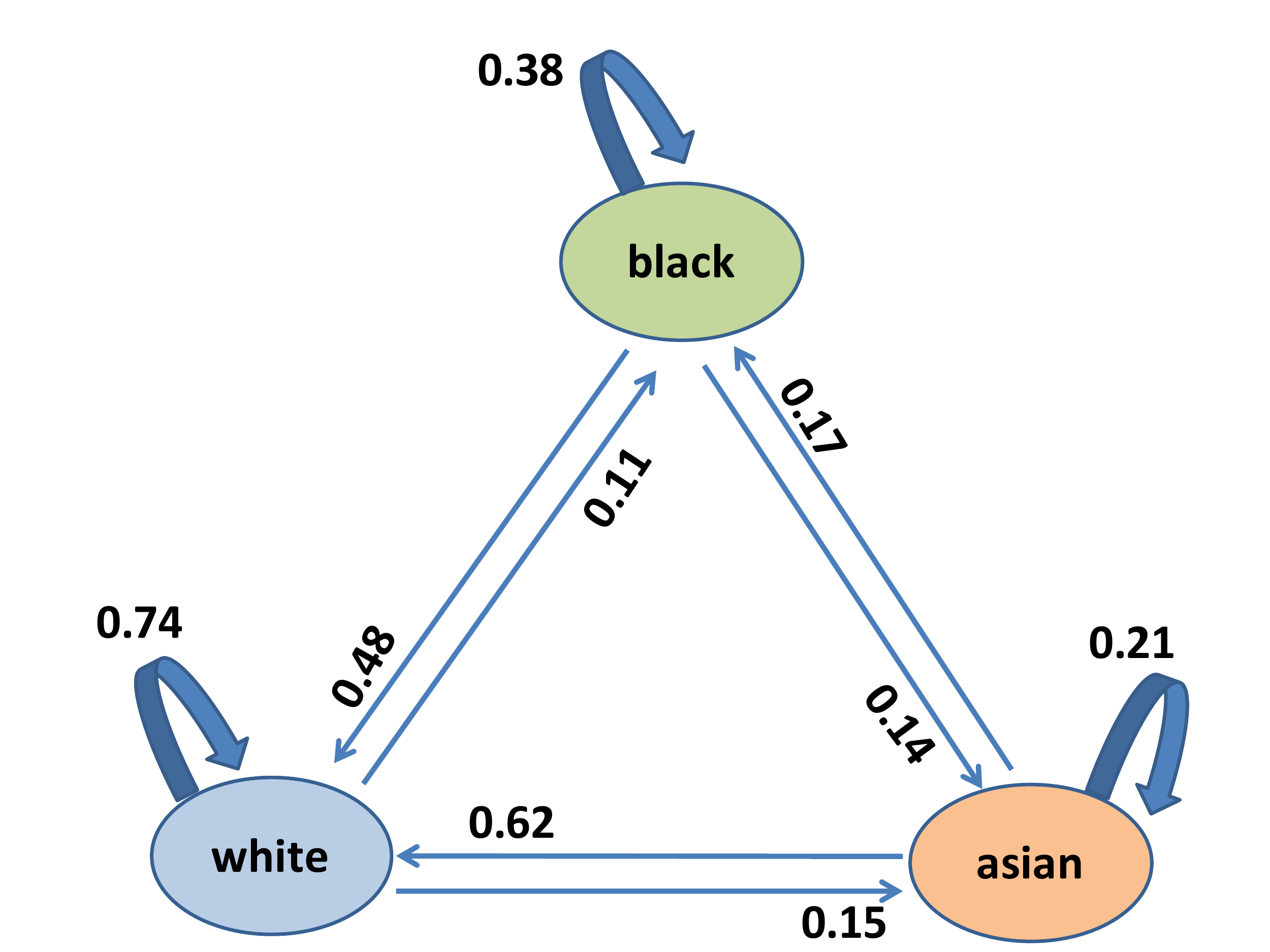} }}%
    \caption{Probability of friendship and interaction for groups of users separated by race}%
    \label{fig:race_graph_friendship_inter}%
\end{figure}

Next, we turn our attention to the analysis of interconnection among groups of users separated by race. Figure~\ref{fig:race_graph_friendship_inter} shows these interconnections in terms of friendship and interactions, respectively. 

We start by analyzing the self-loops on each node.  We note that white users tend to follow about $79\%$ of other white users, black users tend to follow $32\%$ of black users and asians tend to follow $16\%$ of asians. The key for analyzing these figures is to recall the reference distribution of users in our dataset according to their race, in which whites represent $67.97\%$, black accounts to $14.91\%$, and asian accounts to $17.12\%$. Thus, we can note that, in comparison to the expected distribution of users, whites tend to follow $16.22\%$ more whites than expect, blacks tend to follow impressive $2.14$ times more blacks than expect, and asians end up following less ($6.54\%$) asians than expected. In other words, the presence of homophily was not clear for the case of asians.

Similar observations can be made for interactions, the only exception is among asians, where the fraction of interactions is higher than the reference distribution. Although asians tend to follow more white users (even more than the reference distribution), the interactions with white users are below the reference fraction. Overall, whites and blacks showed quite high numbers of endogenous connections and interactions as both groups seem to avoid following a different race. The stronger mutual interconnection is between black users. 

\section{Demography of Interconnections}

\begin{figure}%
    \centering
    \subfloat[Friendship]{{\includegraphics[width=0.65\textwidth]{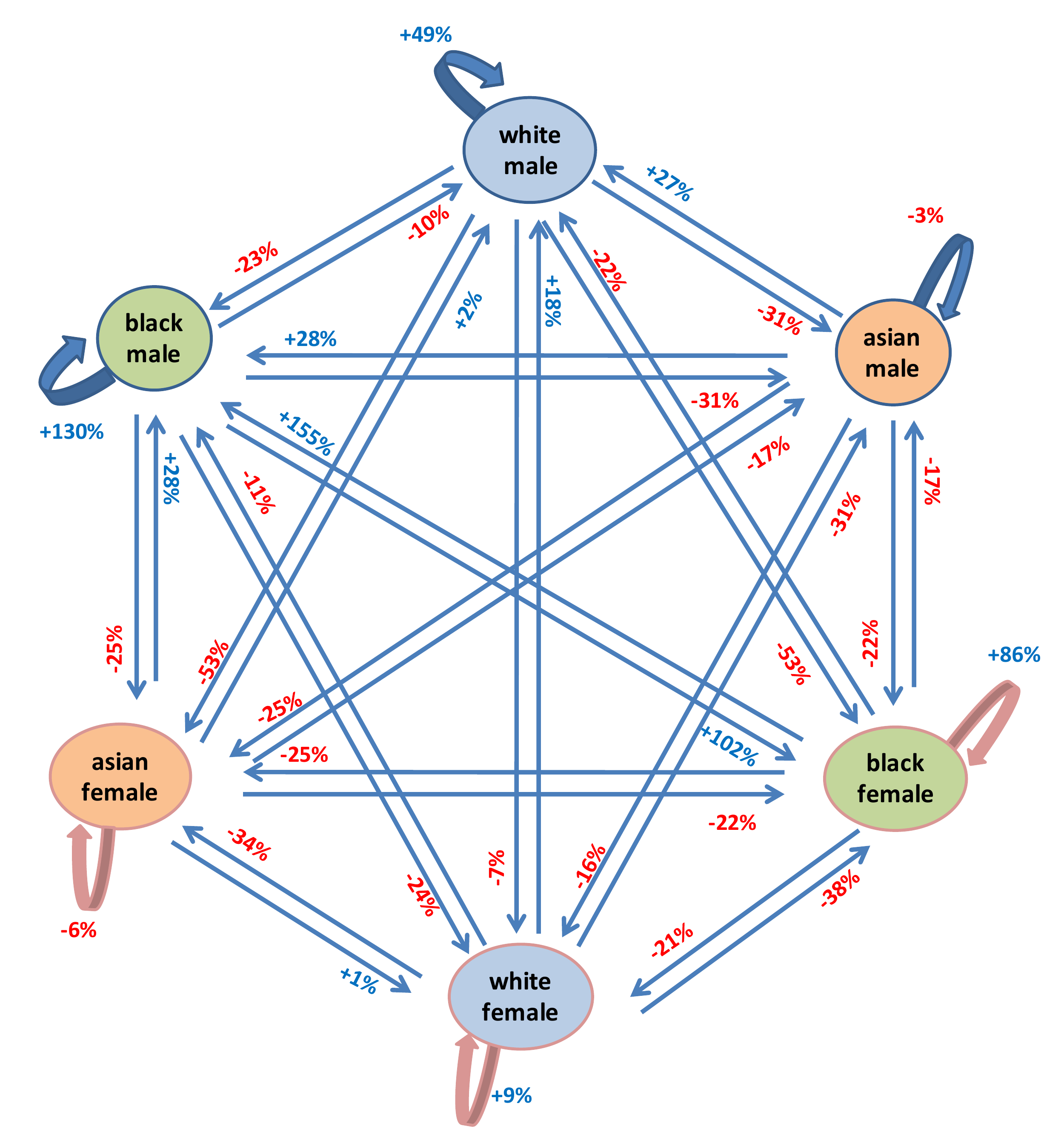} }}%
    \qquad
    \subfloat[Interaction]{{\includegraphics[width=0.65\textwidth]{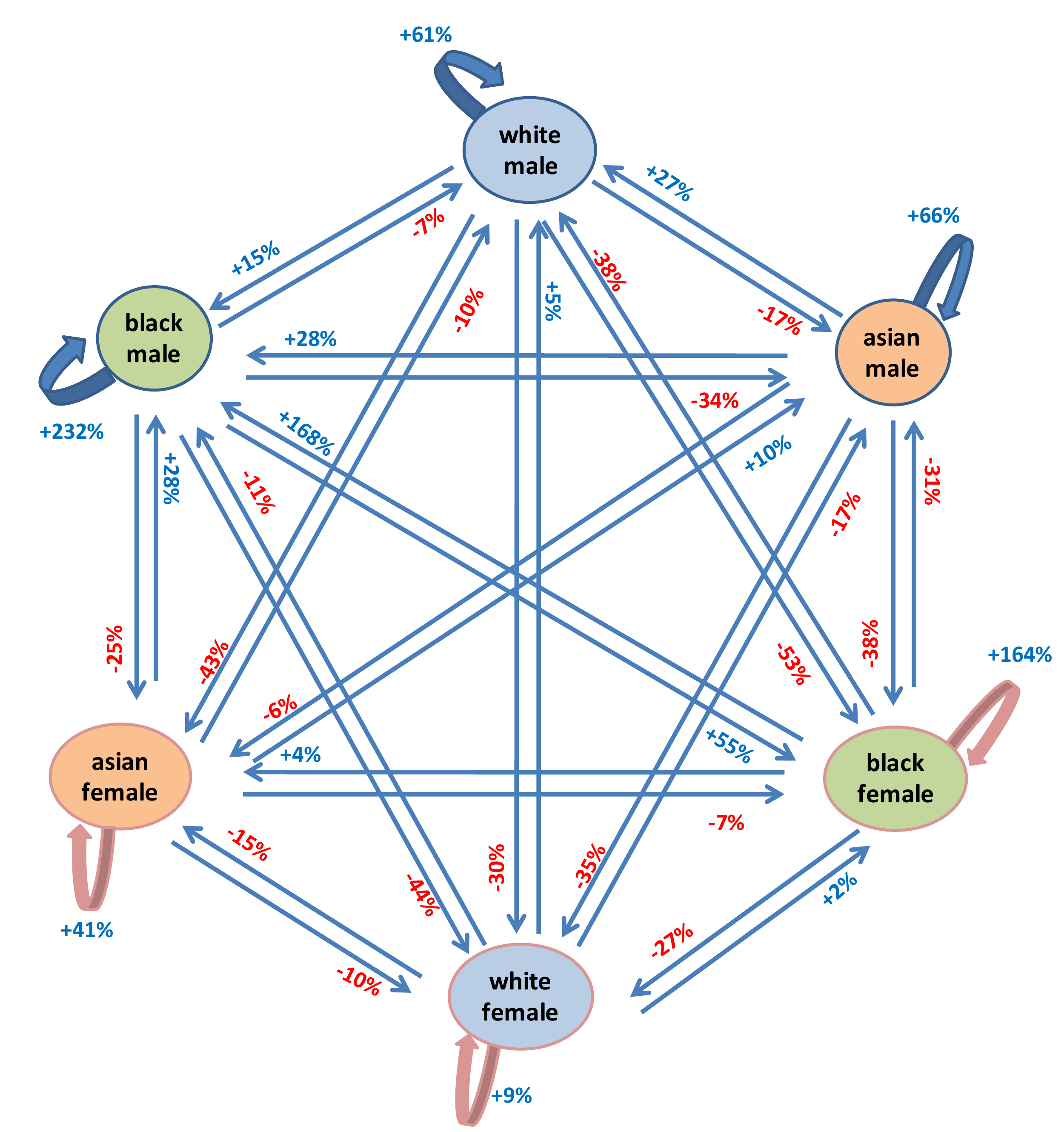} }}%
    \caption{Relative probability of friendship and interaction for groups of users separated by race and gender. The value declares the relative increase (\textcolor{blue}{blue} color) or decrease (\textcolor{red}{red} color) in comparison to the expected demographic population.}%
    \label{fig:all_graph_friendship_inter}%
\end{figure}

Finally, we characterize the interconnection of users grouped by race and gender. Figure~\ref{fig:all_graph_friendship_inter} shows these interconnections in terms of friendship and interactions, respectively. However, given the high number of edges in these figures, instead of providing the transition probabilities, we computed their relative increase or decrease in comparison to the expected demographic population.

In addition, to reinforce previous findings from the analysis of gender and race separately presented before, these results show interesting trends. First, we note that white males tend to receive only positive edges from white and asian groups, meaning that only black users (both males and females) tend to have a fraction of white males users as friends that are relatively lower than the expected demographic distribution in Twitter. This suggests that race plays a more important role for connections with white males. In terms of interactions, similar observations hold, except that asian females tend to have a relatively smaller number of interactions with white males. Interestingly, white females appear over-represented only in the list of friends of asian females.  

Another interesting pattern is that the highest values for self-loops are for black males ($130\%$) and black females ($86\%$). These numbers are higher for interactions, $232\%$ and $164\%$, respectively. The interconnections between these two groups are much larger than expected. In contrast, the lower values for self-loops are for asians, with negative values for friendship, but positive for interactions, as discussed before. Interestingly, all incoming links to the two asian groups are negative, meaning that the other two demographic groups tend to have an under-represented proportion of asians as friends and in terms of interactions.

\section{Concluding Remarks}

In this chapter, we analyzed the interconnection among groups of users separated by gender and race. We analyzed the probability of males and females users to be friends with (i.e. to follow) other males and females users. We found that males and females connections take their part of the responsibility on gender inequality. However, males tend to take the larger part of the responsibility. Further, race plays a more important role for connections with white males. In terms of interactions, similar observations hold, except that asian females tend to have a relatively smaller number of interactions with white males. Our analysis of connections and interactions among these groups explain part of causes of such inequalities and offer hints for the promotion of equality in the online space. The next chapter present the system design of the \textit{Who Makes Trends?} Web-based service in order to make the demographic biases in Twitter trends more transparent.

 \chapter{Leveraging Demographic Aspects to Design Transparent Systems} \label{chap:system}

Demographics aspects have a valuable power to provide transparency of algorithms and systems that use user-generated data. In this dissertation, we focus on providing transparency on the Twitter trending topics in order to show how powerful are demographic aspects for providing transparency. We believe this strategy can also be applied in different systems such as Google Suggestions. The ability to understand how Google suggestions algorithm works, by using demographic aspects, would allow us to analyze which demographic groups have created these suggestions and would help us to study stereotype, reappropriation, and many other concepts.

Increasingly, researchers and governments are recognizing the importance of making algorithms transparent. The White House recently released a report that concludes that the practitioners must ensure that AI-enabled systems are open, transparent, and understandable~\cite{white_house_transparency}. Hence, there is a concern in promoting more transparent data to users.  

Users of social media sites like Facebook and Twitter rely on crowdsourced content recommendation systems (such as Trending Topics) to retrieve important and useful information. Once the contents are chosen for the recommendation, they reach a large population, actually giving the initial users of the contents an opportunity to promote their messages to the wider public. Consequently, it is important to understand the demographics of users who make a content worthy of recommendation, and explore whether they are representative of the media site's overall population. Therefore, to make demographic bias more transparent in Twitter we developed and deployed the \textit{Who Makes Trends?}\footnote{\url{http://twitter-app.mpi-sws.org/who-makes-trends/}}~\cite{chakraborty2017@icwsm} Web-based service. It infers the demographic of Twitter trending topics in U.S.

\section{Who Makes Trends?}
To make the demographic biases in Twitter trends more transparent, we developed and deployed a real-time Web-based service \textit{Who Makes Trends?} at \url{http://twitter-app.mpi-sws.org/who-makes-trends}. It shows information of the Trend Promoters and Trend Adopters. Trend Promoters is the distribution (or combination) of demographic groups (such as middle-aged white males, young asian females, adolescent black males) in the crowd promoting (or posting about) a topic before the topic becomes trending in Twitter. On the other hand, Trend Adopters is the distribution (or combination) of demographic groups in the crowd adopting a topic after the topic becomes trending in Twitter. Here, we are only considering US-based Twitter users whose tweets on the trends appear in the $1\%$ random sample we obtained from Twitter.

\begin{figure}[!htb]
  \centering
    \includegraphics[width=1\textwidth]{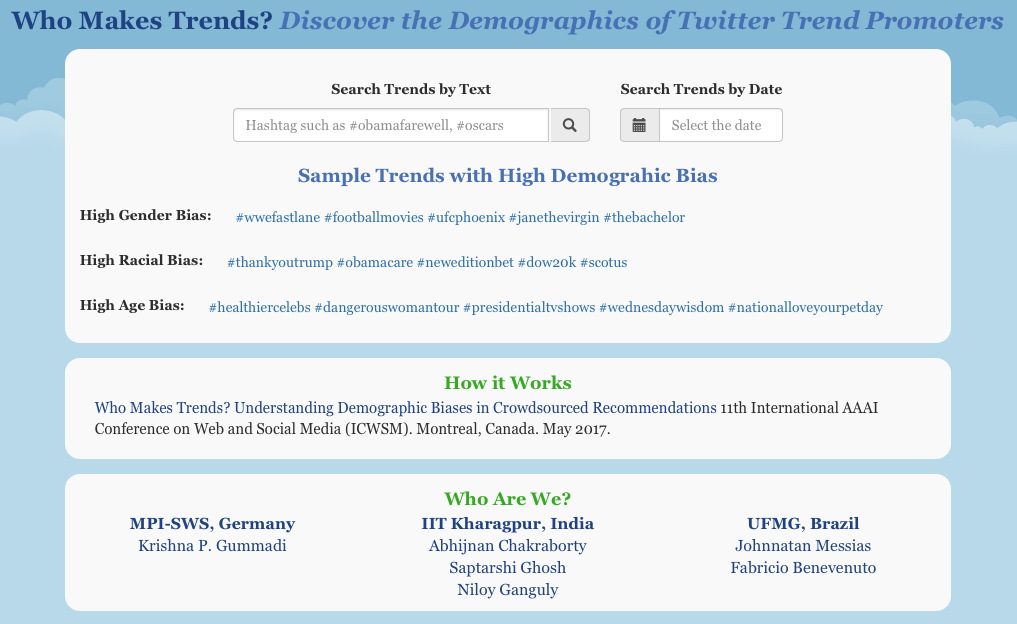}
  \caption{Main page of \textit{Who Makes Trends?} web-based service. It allows users to search trends by text or by date. On this page, there are some manually selected hashtags which contain high gender bias, high racial bias, and high age bias as an example of data transparency.}
  \label{fig:who-makes-trends-main}
\end{figure}

\begin{figure}[!htb]
  \centering
    \includegraphics[width=0.6\textwidth]{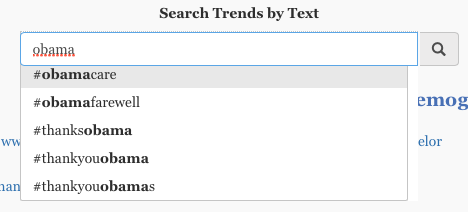}
  \caption{Search Trends by Text box with auto-complete functionality. It allows users to search for a specific query and it also suggests some trending hashtags based on the input provided by users.}
  \label{fig:who-makes-trends-search-by-text}
\end{figure}

\begin{figure}[!htb]
  \centering
    \includegraphics[width=0.5\textwidth]{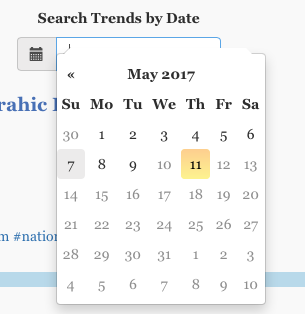}
  \caption{Search Trends By Date box which allows users to see which trends (we have data) appear on a specific day.}
  \label{fig:who-makes-trends-search-by-date}
\end{figure}

Figure \ref{fig:who-makes-trends-main} shows the main page (index page) of the service. It contains three sections: (i) the search trends and sample trends with high demographic bias; (ii) how it works; and (iii) who are we?. The first section allows users to search trends by text (see Figure \ref{fig:who-makes-trends-search-by-text}) or even by date (see Figure \ref{fig:who-makes-trends-search-by-date}). It also contains some trending hashtags manually selected with high gender bias, high racial bias, and high age bias as an example of data transparency. The second basically points to the \cite{chakraborty2017@icwsm} work and explains more details about gender, race, and age bias in Twitter trending topics. Finally, the last section shows the people who worked on this project.

This section describes necessary procedures to develop this application. Basically, it describes the data collection and methods to provide the distribution of Promoters and Adopters Trends of Twitter trending topics. Detailed information is available in \cite{chakraborty2017@icwsm} work.

\subsection{Data collection}

For this work, we gather and store, every day, the $1\%$ random sample of all English tweets posted in U.S, through the Twitter Stream API\footnote{\url{https://dev.twitter.com/streaming/overview/request-parameters}}, since January 1st, 2017. The filtering process is configurable via parameter setting in the Twitter Stream API. Therefore, we use as a filter the U.S. bounding box to infer tweets from U.S. and also the language as English. After this process, we are able to get all English tweets from U.S. provided by the Twitter Stream API. We prefer to use the Stream Filter because it returns more tweets from U.S. than the worldwide stream since $1\%$ of the total tweets posted in the U.S. is higher than the total $1\%$ posted worldwide. Simultaneously, and along the same period, by querying the Twitter REST API\footnote{\url{http://dev.twitter.com/rest/reference/get/trends/place}} at every $5$-minutes, we collected all topics which became trending in the US. It is important to highlight, once we gather data from U.S., we use the Eastern Standard Time (EST)\footnote{\url{https://www.timeanddate.com/time/zones/est}} as a default time zone in order to control the day (12 am to 11:59:59 pm) we gather the data. Additionally, some hashtags may appear as trending in consecutive days. Hence, in order to reflect this restriction in our service, we decided to update the service every 2 days. Therefore, we process the data for the day $D$ as being the data for the day $D$ - $12$ hours and also day $D$ + $12$ hours. This step is necessary because some hashtags may have been posted the day before ($D$ - $1$ day), became trending on day $D$ and/or even become trending one day after ($D$ + $1$ day). 

Once we have the daily Twitter data with all English tweets posted in U.S., the next step is to filter those tweets that contain at least one hashtag which appeared in the Twitter Trending Topic on that specific day. This step is necessary because the Web service shows demographic information of users who promote (post a tweet before it becomes trending) or adopt (post a tweet after it becomes trending) this trending hashtag. To do that, for every tweet in our daily dataset, we look at the tweet field \textit{entities/hashtags} which provides all hashtags available in a tweet and we check if the hashtag appeared in the Twitter Trending Topic. Then, in order to show the demographic information regarding Trend Promoters and Trend Adopters we create three set of users: (i) who posted a tweet before it becomes trending; (ii) who posted a tweet after it becomes trending; and (iii) the union of all users we could infer the demographic information since January 1st. This is possible because each trending topic gathered from Twitter has the precise time it became trending. Also, we would like to show the expected demographic distribution for each trending in order to compare the Trend Promoters and Trend Adopters as being under- or over-represented with respect to the Twitter users population for which we have demographic data.

Now, once we have the three set of users, (i) Trend Promoters; (ii) Trend Adopters; and (iii) Expected Distribution, similarly to the methodology in Section~\ref{sec:crawling_demographic}, we submit the profile picture Web links into the \textit{Face++ API}. We also discard those users whose profile pictures do not have a recognizable face or have more than one recognizable face, according to Face++. Finally, the service is able to show the demographic distribution of Twitter Trending Topics.

\subsection{Trending Topic Analysis}

To compare the demographics of Twitter users with the demographics of the \textit{offline} population, we collect the demographics of U.S. residents from the U.S. Census Bureau\footnote{\url{https://www.census.gov/topics/population/age-and-sex.html}}
\footnote{\url{http://census.gov/prod/cen2010/briefs/c2010br-02.pdf}} and present in Table~\ref{tab:reference_dist}.
We see that some demographic groups appear more in Twitter compared to their share of U.S. population. For instance, the presence of asians in Twitter is about $3.06$ times larger than in the overall U.S. population. Similarly, the adolescent and young people are much more in Twitter. However, mid-aged and old population have comparatively significantly less presence in Twitter. Our findings corroborate a recent survey on social media population conducted by Pew Research\footnote{\url{http://pewinternet.org/2016/11/11/social-media-update-2016}}.

\begin{table}[!htb]
\footnotesize
\centering
\caption{Comparing the demographics of the population in U.S., and the demographics of U.S.-based Twitter users, whose tweets were included in the $1\%$ random sample during January -- May 2017, and whose demographic information could be inferred.}
\label{tab:reference_dist}
\begin{tabular}{|c|c|c|c|c|c|c|c|c|c|}
\hline
\multirow{2}{*}{\textit{\textbf{Baseline}}}                                    & \multicolumn{2}{c|}{\textbf{Gender (\%)}} & \multicolumn{3}{c|}{\textbf{Race (\%)}} & \multicolumn{4}{c|}{\textbf{Age Group (\%)}} \\ \cline{2-10} 
                                                                               & Male                & Female              & White       & Black       & Asian       & Adolescent   & Young   & Mid-aged   & Old    \\ \hline
\textbf{\textbf{\begin{tabular}[c]{@{}c@{}}U.S.\\ Population\end{tabular}}}    & 49.20               & 50.80               & 72.40       & 12.60       & 4.80        & 13.60        & 26.70   & 33.20      & 13.50  \\ \hline
\textbf{\textbf{\begin{tabular}[c]{@{}c@{}}Twitter\\ Population\end{tabular}}} & 45.97               & 54.03               & 73.05       & 12.25       & 14.70       & 26.37        & 62.58   & 10.80      & 0.25   \\ \hline
\end{tabular}
\end{table}

Table~\ref{table:top_31_days_hashtags}, shows the amount of promoters, adopters, and the total users with inferred demographic information and also the total of users we could not infer the demographics who used the specific trending hashtag from April 2nd to May 2nd, 2017, sorted in a descending order based on the date. As an interesting finding, some trending hashtags appear to be promoted by Twitter even when there are not so many users posting tweets about this topic. We believe it may happen due to some important and well-known events. Therefore, to give more visibility to those contents, Twitter chooses them as trending. For example, the trending hashtags \textit{\#fyrefestival}, \textit{\#unicornfrappuccino}, and \textit{\#eastersunday}.

\textit{\#fyrefestival}\footnote{\url{http://www.fyrefestival.com}}\footnote{\url{http://www.nytimes.com/2017/04/28/arts/music/fyre-festival-ja-rule-bahamas.html}} (Fyre Festival): is regarding the music festival \textit{Fyre Festival} scheduled to happen in Bahamian island of Great Exuma. However, the festival was a disaster, grounds that were woefully lacking the promised amenities, replaced instead by dirt fields, soggy tents, and folding chairs. Hence, people got upset and started to post in Twitter what was going on in the event. The organizers had to postpone the festival.

\textit{\#unicornfrappuccino}\footnote{\url{http://en.wikipedia.org/wiki/Unicorn_Frappuccino}}\footnote{\url{https://www.cnet.com/news/starbucks-unicorn-frappuccino-comparisons-twitter-jokes/}} (Unicorn Frappuccino): It is a drink created by Starbucks in 2017. However, it got viral because users did not like it and started to post jokes about this new drink.

\textit{\#eastersunday}\footnote{\url{http://en.wikipedia.org/wiki/Easter}} (Easter Sunday): It celebrates the resurrection of Jesus. It is an event celebrated by Christian culture around the world which made it appears as trending in Twitter.

\begin{sidewaystable}[]
\centering
\footnotesize
\caption{Top hashtags during 31 days from April 2nd to May 2nd, 2017. They are sorted in a descending order based on the date. This Table shows the number of promoters, adopters, and the total users with inferred demographic information. Also, it shows the number of total users who we could not infer the demographic information. The hashtags are clickable links that point to their demographic distribution on \textit{Who Makes Trends?} service.}
\label{table:top_31_days_hashtags}
\begin{tabular}{@{}lccccccc@{}}
\toprule
\multicolumn{1}{c}{\textbf{Hashtag}}   & \textbf{Date} & \textbf{\begin{tabular}[c]{@{}c@{}}\#Promoters\\ With\\ Demographic\\ Inference\end{tabular}} & \textbf{\begin{tabular}[c]{@{}c@{}}\#Promoters\\ Without\\ Demographic\\ Inference\end{tabular}} & \textbf{\begin{tabular}[c]{@{}c@{}}\#Adopters\\ With\\ Demographic\\ Inference\end{tabular}} & \textbf{\begin{tabular}[c]{@{}c@{}}\#Adopters\\ Without\\ Demographic\\ Inference\end{tabular}} & \textbf{\begin{tabular}[c]{@{}c@{}}\#Total\\ With\\ Demographic\\ Inference\end{tabular}} & \textbf{\begin{tabular}[c]{@{}c@{}}\#Total\\ Without\\ Demographic\\ Inference\end{tabular}} \\ \midrule

\textit{\href{http://twitter-app.mpi-sws.org/who-makes-trends/app.php?query=mayday2017\&date=2017-05-02}{\textbf{\#mayday2017}}}         & 02-05-2017    & 609                                                                                           & 532                                                                                              & 162                                                                                          & 165                                                                                             & 736                                                                                       & 660                                                                                          \\
\textit{\href{http://twitter-app.mpi-sws.org/who-makes-trends/app.php?query=metgala\&date=2017-05-01}{\textbf{\#metgala}}}            & 01-05-2017    & 1563                                                                                          & 616                                                                                              & 491                                                                                          & 257                                                                                             & 1988                                                                                      & 830                                                                                          \\
\textit{\href{http://twitter-app.mpi-sws.org/who-makes-trends/app.php?query=wwepayback\&date=2017-04-30}{\textbf{\#wwepayback}}}         & 30-04-2017    & 862                                                                                           & 660                                                                                              & 88                                                                                           & 70                                                                                              & 912                                                                                       & 695                                                                                          \\
\textit{\href{http://twitter-app.mpi-sws.org/who-makes-trends/app.php?query=climatemarch\&date=2017-04-29}{\textbf{\#climatemarch}}}       & 29-04-2017    & 637                                                                                           & 582                                                                                              & 155                                                                                          & 142                                                                                             & 753                                                                                       & 694                                                                                          \\
\textit{\href{http://twitter-app.mpi-sws.org/who-makes-trends/app.php?query=fyrefestival\&date=2017-04-28}{\textbf{\#fyrefestival}}}       & 28-04-2017    & 32                                                                                            & 15                                                                                               & 1363                                                                                         & 925                                                                                             & 1384                                                                                      & 936                                                                                          \\
\textit{\href{http://twitter-app.mpi-sws.org/who-makes-trends/app.php?query=nfldraft\&date=2017-04-27}{\textbf{\#nfldraft}}}           & 27-04-2017    & 4846                                                                                          & 3635                                                                                             & 2318                                                                                         & 1873                                                                                            & 6183                                                                                      & 4813                                                                                         \\
\textit{\href{http://twitter-app.mpi-sws.org/who-makes-trends/app.php?query=wednesdaywisdom\&date=2017-04-26}{\textbf{\#wednesdaywisdom}}}    & 26-04-2017    & 317                                                                                           & 341                                                                                              & 77                                                                                           & 57                                                                                              & 383                                                                                       & 392                                                                                          \\
\textit{\href{http://twitter-app.mpi-sws.org/who-makes-trends/app.php?query=dwts\&date=2017-04-25}{\textbf{\#dwts}}}               & 25-04-2017    & 360                                                                                           & 188                                                                                              & 89                                                                                           & 51                                                                                              & 435                                                                                       & 232                                                                                          \\
\textit{\href{http://twitter-app.mpi-sws.org/who-makes-trends/app.php?query=mondaymotivation\&date=2017-04-24}{\textbf{\#mondaymotivation}}}       & 24-04-2017    & 673                                                                                           & 717                                                                                              & 141                                                                                          & 131                                                                                             & 803                                                                                       & 840
\\
\textit{\href{http://twitter-app.mpi-sws.org/who-makes-trends/app.php?query=sundayfunday\&date=2017-04-23}{\textbf{\#sundayfunday}}}           & 23-04-2017               & 678                          & 613                        & 153                         & 112                       & 812                      & 712                    \\
\textit{\href{http://twitter-app.mpi-sws.org/who-makes-trends/app.php?query=earthday\&date=2017-04-22}{\textbf{\#earthday}}}               & 22-04-2017               & 2636                         & 2719                       & 3105                        & 2618                      & 5531                     & 5123                   \\
\textit{\href{http://twitter-app.mpi-sws.org/who-makes-trends/app.php?query=ripprince\&date=2017-04-21}{\textbf{\#ripprince}}}              & 21-04-2017               & 453                          & 300                        & 171                         & 104                       & 603                      & 393                    \\
\textit{\href{http://twitter-app.mpi-sws.org/who-makes-trends/app.php?query=happy420\&date=2017-04-20}{\textbf{\#happy420}}}               & 20-04-2017               & 805                          & 567                        & 132                         & 121                       & 918                      & 661                    \\
\textit{\href{http://twitter-app.mpi-sws.org/who-makes-trends/app.php?query=bostonmarathon\&date=2017-04-19}{\textbf{\#bostonmarathon}}}         & 19-04-2017               & 789                          & 598                        & 282                         & 225                       & 1005                     & 778                    \\
\textit{\href{http://twitter-app.mpi-sws.org/who-makes-trends/app.php?query=unicornfrappuccino\&date=2017-04-18}{\textbf{\#unicornfrappuccino}}}     & 18-04-2017               & 36                           & 13                         & 1712                        & 996                       & 1737                     & 1005                   \\
\textit{\href{http://twitter-app.mpi-sws.org/who-makes-trends/app.php?query=cleveland\&date=2017-04-17}{\textbf{\#cleveland}}}              & 17-04-2017               & 709                          & 442                        & 406                         & 355                       & 1049                     & 711                    \\
\textit{\href{http://twitter-app.mpi-sws.org/who-makes-trends/app.php?query=eastersunday\&date=2017-04-16}{\textbf{\#eastersunday}}}           & 16-04-2017               & 64                           & 77                         & 1570                        & 1233                      & 1616                     & 1300                   \\
\textit{\href{http://twitter-app.mpi-sws.org/who-makes-trends/app.php?query=aprilthegiraffe\&date=2017-04-15}{\textbf{\#aprilthegiraffe}}}        & 15-04-2017               & 810                          & 421                        & 76                          & 56                        & 872                      & 467                    \\
\textit{\href{http://twitter-app.mpi-sws.org/who-makes-trends/app.php?query=goodfriday\&date=2017-04-14}{\textbf{\#goodfriday}}}             & 14-04-2017               & 1674                         & 1422                       & 862                         & 640                       & 2452                     & 1996                   \\
\textit{\href{http://twitter-app.mpi-sws.org/who-makes-trends/app.php?query=stanleycup\&date=2017-04-13}{\textbf{\#stanleycup}}}             & 13-04-2017               & 369                          & 303                        & 600                         & 493                       & 842                      & 709                    \\
\textit{\href{http://twitter-app.mpi-sws.org/who-makes-trends/app.php?query=bucciovertimechallenge\&date=2017-04-12}{\textbf{\#bucciovertimechallenge}}} & 12-04-2017               & 171                          & 244                        & 770                         & 992                       & 867                      & 1137                   \\
\textit{\href{http://twitter-app.mpi-sws.org/who-makes-trends/app.php?query=nationalpetday\&date=2017-04-11}{\textbf{\#nationalpetday}}}         & 11-04-2017               & 2637                         & 1887                       & 1455                        & 891                       & 4056                     & 2744                   \\
\textit{\href{http://twitter-app.mpi-sws.org/who-makes-trends/app.php?query=nationalsiblingsday\&date=2017-04-10}{\textbf{\#nationalsiblingsday}}}    & 10-04-2017               & 3828                         & 1837                       & 2512                        & 1202                      & 6296                     & 3023                   \\
\textit{\href{http://twitter-app.mpi-sws.org/who-makes-trends/app.php?query=sundayfunday\&date=2017-04-09}{\textbf{\#sundayfunday}}}           & 09-04-2017               & 775                          & 585                        & 181                         & 130                       & 939                      & 705                    \\
\textit{\href{http://twitter-app.mpi-sws.org/who-makes-trends/app.php?query=nationalbeerday\&date=2017-04-08}{\textbf{\#nationalbeerday}}}        & 08-04-2017               & 1188                         & 1581                       & 368                         & 333                       & 1529                     & 1876                   \\
\textit{\href{http://twitter-app.mpi-sws.org/who-makes-trends/app.php?query=syria\&date=2017-04-07}{\textbf{\#syria}}}                  & 07-04-2017               & 1263                         & 856                        & 654                         & 472                       & 1771                     & 1217                   \\
\textit{\href{http://twitter-app.mpi-sws.org/who-makes-trends/app.php?query=themasters\&date=2017-04-06}{\textbf{\#themasters}}}             & 06-04-2017               & 420                          & 452                        & 3015                        & 2513                      & 3225                     & 2771                   \\
\textit{\href{http://twitter-app.mpi-sws.org/who-makes-trends/app.php?query=13reasonswhy\&date=2017-04-05}{\textbf{\#13reasonswhy}}}           & 05-04-2017               & 280                          & 87                         & 985                         & 363                       & 1225                     & 439                    \\
\textit{\href{http://twitter-app.mpi-sws.org/who-makes-trends/app.php?query=nationalchampionship\&date=2017-04-04}{\textbf{\#nationalchampionship}}}   & 04-04-2017               & 3376                         & 2561                       & 238                         & 174                       & 3533                     & 2682                   \\
\textit{\href{http://twitter-app.mpi-sws.org/who-makes-trends/app.php?query=finalfour\&date=2017-04-03}{\textbf{\#finalfour}}}              & 03-04-2017               & 4146                         & 3448                       & 851                         & 618                       & 4739                     & 3887                   \\
\textit{\href{http://twitter-app.mpi-sws.org/who-makes-trends/app.php?query=openingday\&date=2017-04-02}{\textbf{\#openingday}}}             & 02-04-2017               & 1732                         & 1342                       & 4129  & 3437 & 5461 & 4506
\\ \bottomrule
\end{tabular}
\end{sidewaystable}

\subsection{Disparate Demographics}

While analyzing the demographics of the promoters of different trends, we observed that different trends are promoted by user-groups having highly disparate demographics. We classify these disparate demographics into three categories: (i) high gender bias; (ii) high racial bias; and (iii) high age bias.

High gender bias contains trending hashtags that are highly under- or over-expected for gender. As examples, we have the hashtags \#footballmovies, \#ufcphoenix, and \#thebachelor. 
On the other hand, high racial bias contains trending hashtags that are highly under- or over-expected for race. As examples, we have the hashtags \#thankyoutrump, \#obamacare, and \#neweditionbet.
Finally, for the high age bias, we have the hashtags \#dangerouswomantour, \#presidentialtvshows, and \#nationalloveyourpetday which are highly under- or over-expected for age.
Table \ref{table:promoters} shows the demographic distribution of promoters of different trending hashtags. We can see that trending hashtags like \#thankyoutrump was promoted by mid-aged white users, similarly, \#obamacare was promoted by mid-aged white males. On the other hand, trending hashtags like \#neweditionbet were promoted by adolescent black females.

\begin{table}[!htb]
\tiny
\centering
\caption{Demographics of promoters of Twitter trends. Demographic groups shown in bold blue are represented more (over-expected), and groups in red italics are represented less (under-expected) among the promoters. We consider differences of $5\%$ as the threshold. The hashtags are clickable links that point to their demographic distribution on \textit{Who Makes Trends?} service.}
\label{table:promoters}
\begin{tabular}{|l|c|c|c|c|c|c|c|c|}
\hline
                                   & \multicolumn{8}{c|}{\textbf{Demographics of Promoters}}                                                                                                                                                                                                                                                                     \\ \cline{2-9} 
                                   & \multicolumn{2}{c|}{\textbf{Gender (\%)}}                                     & \multicolumn{3}{c|}{\textbf{Race (\%)}}                                                                             & \multicolumn{3}{c|}{\textbf{Age Group (\%)}}                                                                          \\ \cline{2-9} 
\multirow{-3}{*}{\textbf{Hashtag}} & \textbf{Male}                         & \textbf{Female}                       & \textbf{White}                        & \textbf{Black}                       & \textbf{Asian}                       & \textbf{Adolescent}                   & \textbf{Young}                        & \textbf{Mid-aged}                     \\ \hline
\href{http://twitter-app.mpi-sws.org/who-makes-trends/app.php?query=footballmovies\&date=2017-02-02}{\textit{\#footballmovies}}            & {\color[HTML]{3531FF} \textbf{65.82}} & {\color[HTML]{FE0000} \textit{34.18}} & {\color[HTML]{3531FF} \textbf{83.55}} & {\color[HTML]{FE0000} \textit{5.06}} & 11.39                                & {\color[HTML]{FE0000} \textit{10.13}} & {\color[HTML]{3531FF} \textbf{70.88}} & {\color[HTML]{3531FF} \textbf{18.99}} \\ \hline
\href{http://twitter-app.mpi-sws.org/who-makes-trends/app.php?query=ufcphoenix\&date=2017-01-15}{\textit{\#ufcphoenix}}                & {\color[HTML]{3531FF} \textbf{77.03}} & {\color[HTML]{FE0000} \textit{22.97}} & 73.65                                 & 10.81                                & 15.54                                & {\color[HTML]{FE0000} \textit{16.89}} & {\color[HTML]{3531FF} \textbf{71.62}} & 11.49                                 \\ \hline
\href{http://twitter-app.mpi-sws.org/who-makes-trends/app.php?query=thebachelor\&date=2017-01-16}{\textit{\#thebachelor}}               & {\color[HTML]{FE0000} \textit{15.61}} & {\color[HTML]{3531FF} \textbf{84.39}} & {\color[HTML]{3531FF} \textbf{84.69}} & {\color[HTML]{FE0000} \textit{4.94}} & 10.37                                & 29.82                                 & 64.94                                 & {\color[HTML]{FE0000} \textit{5.24}}  \\ \hline
\href{http://twitter-app.mpi-sws.org/who-makes-trends/app.php?query=thankyoutrump\&date=2017-01-24}{\textit{\#thankyoutrump}}             & 49.55                                 & 50.45                                 & {\color[HTML]{3531FF} \textbf{81.98}} & 8.11                                 & 9.91                                 & 21.62                                 & {\color[HTML]{FE0000} \textit{54.96}} & {\color[HTML]{3531FF} \textbf{22.52}} \\ \hline
\href{http://twitter-app.mpi-sws.org/who-makes-trends/app.php?query=obamacare\&date=2017-01-04}{\textit{\#obamacare}}                 & {\color[HTML]{3531FF} \textbf{58.11}} & {\color[HTML]{FE0000} \textit{41.89}} & {\color[HTML]{3531FF} \textbf{83.78}} & {\color[HTML]{FE0000} \textit{6.76}} & {\color[HTML]{FE0000} \textit{9.46}} & {\color[HTML]{FE0000} \textit{13.51}} & {\color[HTML]{FE0000} \textit{51.26}} & {\color[HTML]{3531FF} \textbf{32.43}} \\ \hline
\href{http://twitter-app.mpi-sws.org/who-makes-trends/app.php?query=neweditionbet\&date=2017-01-25}{\textit{\#neweditionbet}}             & {\color[HTML]{FE0000} \textit{40.66}} & {\color[HTML]{3531FF} \textbf{59.34}} & {\color[HTML]{FE0000} \textit{28.27}} & {\color[HTML]{3531FF} \textbf{58}}   & 13.73                                & {\color[HTML]{3531FF} \textbf{33.51}} & 59.93                                 & 6.49                                  \\ \hline
\href{http://twitter-app.mpi-sws.org/who-makes-trends/app.php?query=dangerouswomantour\&date=2017-02-03}{\textit{\#dangerouswomantour}}        & {\color[HTML]{FE0000} \textit{36.67}} & {\color[HTML]{3531FF} \textbf{63.33}} & 71.67                                 & 8.33                                 & {\color[HTML]{3531FF} \textbf{20}}   & {\color[HTML]{3531FF} \textbf{43.33}} & {\color[HTML]{FE0000} \textit{50}}    & 6.67                                  \\ \hline
\href{http://twitter-app.mpi-sws.org/who-makes-trends/app.php?query=presidentialtvshows\&date=2017-02-16}{\textit{\#presidentialtvshows}}       & {\color[HTML]{3531FF} \textbf{68.31}} & {\color[HTML]{FE0000} \textit{31.69}} & {\color[HTML]{3531FF} \textbf{80.33}} & 10.93                                & {\color[HTML]{FE0000} \textit{8.74}} & {\color[HTML]{FE0000} \textit{8.20}}  & {\color[HTML]{3531FF} \textbf{72.67}} & {\color[HTML]{3531FF} \textbf{18.58}} \\ \hline
\href{http://twitter-app.mpi-sws.org/who-makes-trends/app.php?query=nationalloveyourpetday\&date=2017-02-20}{\textit{\#nationalloveyourpetday}}    & {\color[HTML]{FE0000} \textit{28.49}} & {\color[HTML]{3531FF} \textbf{71.51}} & {\color[HTML]{3531FF} \textbf{80.27}} & 8.04                                 & 11.69                                & 26.94                                 & 63.29                                 & 9.77                                  \\ \hline
\end{tabular}
\end{table}

\section{Concluding Remarks}

In this chapter, we developed and deployed the \textit{Who Makes Trends?} Web-based service in order to make the demographic biases in Twitter trends more transparent. To accomplish this design, we presented our methodology to build the system. Therefore, we described the data collection and trending topic analysis steps. We also observed that different trends are promoted by user-groups having highly disparate demographics. We classify these disparate demographics into three categories: (i) high gender bias; (ii) high racial bias; and (iii) high age bias. The \textit{Who Makes Trends?} Web-based service is available at \url{http://twitter-app.mpi-sws.org/who-makes-trends/}. In the next chapter, we conclude this dissertation, discuss some limitations, and propose future research directions.

 \chapter{Concluding Discussion and Future Work} \label{chap:conclusion}

Exploring demographic aspects is important both from the perspective of systems and from the sociological point of view. Hence, this dissertation characterizes demographic aspects of users in Twitter, identifying inequalities and asymmetries in gender and race. To do that, we gathered gender and race of Twitter users located in the United States using advanced image processing algorithms from Face++. We collected a large sample of $1,670,863$ users located in the United States with identified demographic information. We have also crawled a large sample of these users' tweets and friends to study connections and interactions among demographic groups of users.

Our analysis corroborates existing evidence about gender inequality in terms of visibility and introduces race as a significant demographic factor, which reveals hiding prejudices between demographic groups. We also show that the Twitter glass ceiling effect typically applied to females. Also, it occurs in Twitter for blacks and asian males. 

We also attempt to characterize the differences in the way of writing for each group pointing the most important linguistic aspects for a specific gender and race. We were able to identify for each demographic group which affective attributes were more prominent in their writing. In the same way, features based on cognitive attributes, temporal references, social and personal concerns, and interpersonal focus showed to have different weights throughout different demographic domains. By analyzing topic interests, we found that each demographic group tends to have its own preferences for the information they share. For instance, we found that males are more into sports, organizations, and more interest in technology while females have more interest in topics related to life, actors, and movie. In the same way, users from different races are also likely to have different interests and preferences. white users are more interested in politics, writers, and organizations when compared to asians, and technology, movie, and politics when compared to black users. On the other hand, black users are more into artists, life, and music topics. When we look into asians, they are more interested in artists, actors, and music than whites and tend to have more interest for movie, companies, and technology when compared to blacks. Another interesting conclusion is based on the most common phrases encountered on each group and their ranking position when compared to different demographic groups. The analysis of these most common phrases led us to conclude that phrases expressing negation figure as one of the most frequent for all domains. Also, the usage of slang, which is common in an environment like Twitter, appears in these frequent phrases too. One interesting point is that, when we compare the difference between the groups, we find interesting trends, like the more prominent interest in zodiac signs among females than among males.

Our effort is the first to quantify to what extent one demographic group follows and interact with each other and the extent to which these connections and interactions reflect in inequalities in Twitter. The connections and interactions among these groups explain part of causes of such inequalities and offer hints for the promotion of inequalities in the online space.

As potential limitations, we have the accuracy of Face++ and our approach to identify users located in the United States. The accuracy of Face++ inferences is an obvious concern in our effort because we rely on this strategy to infer the demographic aspects of Twitter users. Also, our approach to identify users located in the U.S. may bring together some users located in the same U.S. time zone, but from different countries. We believe that these users might represent a small fraction of the users.

Regarding the system design, demographics aspects have a valuable power to provide transparency of algorithms and systems that use user-generated data. Therefore, in this dissertation, we developed and deployed the \textit{Who Makes Trends?} Web-based service in order to make the demographic biases in Twitter trends more transparent and show how powerful are demographic aspects in providing transparency. We believe this strategy can also be applied in different systems such as Google Suggestions and it would help us study stereotype, reappropriation, and many other social concepts.

Finally, there are some future directions we would like to pursue next. First, we plan to explore age as a demographic factor in social media. Second, we plan to examine how diffusion of news happens when it is propagated through demographic group interconnections and communications. Second, we believe that the increasing availability of information about demographics would help the development of systems that promote more diversity and less inequality to users. Then, we want to study the correlation of linguistic differences with other demographic factors e.g age. We plan to use our extracted linguistic characteristics as a feature vector for prediction of gender and race. Moreover, we plan to examine the speed of tweets that are propagated through a specific demographic group.

\ppgccbibliography{references}


\begin{thebibliography}{}

\bibitem[Agrawal et~al.\@, 2009]{agrawal2009diversifying}
\ppgccbibauthordoformat {Agrawal, R.\ppgccbibauthorsep {} Gollapudi,
  S.\ppgccbibauthorsep {} Halverson, A.\ppgccbibauthorlastsep {}
  \ppgccbibauthorand {} Ieong, S.} (\ppgccyeardoformat {2009}).
\newblock \ppgccbibtitledoformat {Diversifying search results}.
\newblock \ppgccbibinstring {} \ppgccbibbooktitledoformat {ACM WSDM}.

\bibitem[An \ppgccciteauthorand {} Weber, 2016]{an2016greysanatomy}
\ppgccbibauthordoformat {An, J. \ppgccbibauthorand {} Weber, I.}
  (\ppgccyeardoformat {2016}).
\newblock \ppgccbibtitledoformat {\#greysanatomy vs. \#yankees: Demographics
  and hashtag use on twitter}.
\newblock \ppgccbibinstring {} \ppgccbibbooktitledoformat {Proceedings of the
  10th International AAAI Conference on Web and Social Media}.

\bibitem[Angwin et~al.\@, 2016]{angwin2016machine}
\ppgccbibauthordoformat {Angwin, J.\ppgccbibauthorsep {} Larson,
  J.\ppgccbibauthorsep {} Mattu, S.\ppgccbibauthorlastsep {} \ppgccbibauthorand
  {} Kirchner, L.} (\ppgccyeardoformat {2016}).
\newblock \ppgccbibtitledoformat {Machine bias: There’s software used across
  the country to predict future criminals. and it’s biased against blacks}.
\newblock \ppgccbibjournaldoformat {ProPublica, May}, \ppgccvolumedoformat
  {23}.

\bibitem[Ara{\'u}jo et~al.\@, 2016]{araujo2016identifying}
\ppgccbibauthordoformat {Ara{\'u}jo, C.~S.\ppgccbibauthorsep {} Meira~Jr,
  W.\ppgccbibauthorlastsep {} \ppgccbibauthorand {} Almeida, V.}
  (\ppgccyeardoformat {2016}).
\newblock \ppgccbibtitledoformat {Identifying stereotypes in the online
  perception of physical attractiveness}.
\newblock \ppgccbibinstring {} \ppgccbibbooktitledoformat {International
  Conference on Social Informatics}, \ppgccbibpagesstring {} \ppgccpagedoformat
  {419}-\ppgccpagedoformat {}-\ppgccpagedoformat {437}. Springer.

\bibitem[Ayres et~al.\@, 2015]{ayres2015race}
\ppgccbibauthordoformat {Ayres, I.\ppgccbibauthorsep {} Banaji,
  M.\ppgccbibauthorlastsep {} \ppgccbibauthorand {} Jolls, C.}
  (\ppgccyeardoformat {2015}).
\newblock \ppgccbibtitledoformat {Race effects on ebay}.
\newblock \ppgccbibjournaldoformat {The RAND Journal of Economics},
  \ppgccvolumedoformat {46}({\ppgccnumberdoformat {4}}):\ppgccpagedoformat
  {891}-\ppgccpagedoformat {}-\ppgccpagedoformat {917}.

\bibitem[Barocas \ppgccciteauthorand {} Selbst, 2016]{barocas2016big}
\ppgccbibauthordoformat {Barocas, S. \ppgccbibauthorand {} Selbst, A.~D.}
  (\ppgccyeardoformat {2016}).
\newblock \ppgccbibtitledoformat {Big data’s disparate impact}.
\newblock \ppgccbibjournaldoformat {California Law Review},
  \ppgccvolumedoformat {104}:\ppgccpagedoformat {671}.

\bibitem[Barr, 2015]{barr2015google}
\ppgccbibauthordoformat {Barr, A.} (\ppgccyeardoformat {2015}).
\newblock \ppgccbibtitledoformat {Google mistakenly tags black people as
  ‘gorillas,’showing limits of algorithms}.
\newblock \ppgccbibjournaldoformat {The Wall Street Journal}.

\bibitem[Bertrand \ppgccciteauthorand {} Mullainathan, 2004]{bertrand2004emily}
\ppgccbibauthordoformat {Bertrand, M. \ppgccbibauthorand {} Mullainathan, S.}
  (\ppgccyeardoformat {2004}).
\newblock \ppgccbibtitledoformat {Are emily and greg more employable than
  lakisha and jamal? a field experiment on labor market discrimination}.
\newblock \ppgccbibjournaldoformat {The American Economic Review},
  \ppgccvolumedoformat {94}({\ppgccnumberdoformat {4}}):\ppgccpagedoformat
  {991}-\ppgccpagedoformat {}-\ppgccpagedoformat {1013}.

\bibitem[Bhattacharya et~al.\@, 2014]{Bhattacharya2014}
\ppgccbibauthordoformat {Bhattacharya, P.\ppgccbibauthorsep {} Zafar,
  M.~B.\ppgccbibauthorsep {} Ganguly, N.\ppgccbibauthorsep {} Ghosh,
  S.\ppgccbibauthorlastsep {} \ppgccbibauthorand {} Gummadi, K.~P.}
  (\ppgccyeardoformat {2014}).
\newblock \ppgccbibtitledoformat {Inferring user interests in the twitter
  social network}.
\newblock \ppgccbibinstring {} \ppgccbibbooktitledoformat {Proceedings of the
  8th ACM Conference on Recommender Systems}, RecSys '14, \ppgccbibpagesstring
  {} \ppgccpagedoformat {357}-\ppgccpagedoformat {}-\ppgccpagedoformat {360},
  New York, NY, USA. ACM.

\bibitem[Blevins \ppgccciteauthorand {} Mullen, 2015]{blevins2015jane}
\ppgccbibauthordoformat {Blevins, C. \ppgccbibauthorand {} Mullen, L.}
  (\ppgccyeardoformat {2015}).
\newblock \ppgccbibtitledoformat {Jane, john... leslie? a historical method for
  algorithmic gender prediction}.
\newblock \ppgccbibjournaldoformat {Digital Humanities Quarterly},
  \ppgccvolumedoformat {9}({\ppgccnumberdoformat {3}}).

\bibitem[Bollen et~al.\@, 2011]{bollen2011twitter}
\ppgccbibauthordoformat {Bollen, J.\ppgccbibauthorsep {} Mao,
  H.\ppgccbibauthorlastsep {} \ppgccbibauthorand {} Zeng, X.}
  (\ppgccyeardoformat {2011}).
\newblock \ppgccbibtitledoformat {Twitter mood predicts the stock market}.
\newblock \ppgccbibjournaldoformat {Journal of Computational Science},
  \ppgccvolumedoformat {2}({\ppgccnumberdoformat {1}}):\ppgccpagedoformat
  {1}-\ppgccpagedoformat {}-\ppgccpagedoformat {8}.

\bibitem[Bonilla-Silva, 2006]{bonilla2006racism}
\ppgccbibauthordoformat {Bonilla-Silva, E.} (\ppgccyeardoformat {2006}).
\newblock \ppgccbibbtitledoformat {Racism without racists: Color-blind racism
  and the persistence of racial inequality in the United States}.
\newblock Rowman \& Littlefield Publishers.

\bibitem[Burger et~al.\@, 2011]{burger2011discriminating}
\ppgccbibauthordoformat {Burger, J.~D.\ppgccbibauthorsep {} Henderson,
  J.\ppgccbibauthorsep {} Kim, G.\ppgccbibauthorlastsep {} \ppgccbibauthorand
  {} Zarrella, G.} (\ppgccyeardoformat {2011}).
\newblock \ppgccbibtitledoformat {Discriminating gender on twitter}.
\newblock \ppgccbibinstring {} \ppgccbibbooktitledoformat {Proceedings of the
  Conference on Empirical Methods in Natural Language Processing},
  \ppgccbibpagesstring {} \ppgccpagedoformat {1301}-\ppgccpagedoformat
  {}-\ppgccpagedoformat {1309}. Association for Computational Linguistics.

\bibitem[Carbonell \ppgccciteauthorand {} Goldstein, 1998]{carbonell199use}
\ppgccbibauthordoformat {Carbonell, J. \ppgccbibauthorand {} Goldstein, J.}
  (\ppgccyeardoformat {1998}).
\newblock \ppgccbibtitledoformat {The use of mmr, diversity-based reranking for
  reordering documents and producing summaries}.
\newblock \ppgccbibinstring {} \ppgccbibbooktitledoformat {Proceedings of the
  21st annual international ACM SIGIR conference on Research and development in
  information retrieval}, \ppgccbibpagesstring {} \ppgccpagedoformat
  {335}-\ppgccpagedoformat {}-\ppgccpagedoformat {336}. ACM.

\bibitem[Carlsson \ppgccciteauthorand {} Rooth, 2007]{carlsson2007evidence}
\ppgccbibauthordoformat {Carlsson, M. \ppgccbibauthorand {} Rooth, D.-O.}
  (\ppgccyeardoformat {2007}).
\newblock \ppgccbibtitledoformat {Evidence of ethnic discrimination in the
  swedish labor market using experimental data}.
\newblock \ppgccbibjournaldoformat {Labour Economics}, \ppgccvolumedoformat
  {14}({\ppgccnumberdoformat {4}}):\ppgccpagedoformat {716}-\ppgccpagedoformat
  {}-\ppgccpagedoformat {729}.

\bibitem[Cha et~al.\@, 2010]{icwsm10cha}
\ppgccbibauthordoformat {Cha, M.\ppgccbibauthorsep {} Haddadi,
  H.\ppgccbibauthorsep {} Benevenuto, F.\ppgccbibauthorlastsep {}
  \ppgccbibauthorand {} Gummadi, K.~P.} (\ppgccyeardoformat {2010}).
\newblock \ppgccbibtitledoformat {{Measuring User Influence in Twitter: The
  Million Follower Fallacy}}.
\newblock \ppgccbibinstring {} \ppgccbibbooktitledoformat {In Proceedings of
  the 4th International AAAI Conference on Weblogs and Social Media}.

\bibitem[Chakraborty et~al.\@, 2017]{chakraborty2017@icwsm}
\ppgccbibauthordoformat {Chakraborty, A.\ppgccbibauthorsep {} Messias,
  J.\ppgccbibauthorsep {} Benevenuto, F.\ppgccbibauthorsep {} Ghosh,
  S.\ppgccbibauthorsep {} Ganguly, N.\ppgccbibauthorlastsep {}
  \ppgccbibauthorand {} Gummadi, K.~P.} (\ppgccyeardoformat {2017}).
\newblock \ppgccbibtitledoformat {Who makes trends? understanding demographic
  biases in crowdsourced recommendations}.
\newblock \ppgccbibinstring {} \ppgccbibbooktitledoformat {Proceedings of the
  11th International AAAI Conference on Web and Social Media}.

\bibitem[Chen \ppgccciteauthorand {} Karger, 2006]{chen2006less}
\ppgccbibauthordoformat {Chen, H. \ppgccbibauthorand {} Karger, D.~R.}
  (\ppgccyeardoformat {2006}).
\newblock \ppgccbibtitledoformat {Less is more: probabilistic models for
  retrieving fewer relevant documents}.
\newblock \ppgccbibinstring {} \ppgccbibbooktitledoformat {Proceedings of the
  29th annual international ACM SIGIR conference on Research and development in
  information retrieval}, \ppgccbibpagesstring {} \ppgccpagedoformat
  {429}-\ppgccpagedoformat {}-\ppgccpagedoformat {436}. ACM.

\bibitem[Chen et~al.\@, 2016]{Chen2016SIP}
\ppgccbibauthordoformat {Chen, L.\ppgccbibauthorsep {} Wu,
  W.\ppgccbibauthorlastsep {} \ppgccbibauthorand {} He, L.} (\ppgccyeardoformat
  {2016}).
\newblock \ppgccbibbtitledoformat {Personality and Recommendation Diversity},
  \ppgccbibpagesstring {} \ppgccpagedoformat {201}-\ppgccpagedoformat
  {}-\ppgccpagedoformat {225}.
\newblock Springer International Publishing, Cham.

\bibitem[Chen et~al.\@, 2015]{chen2015comparative}
\ppgccbibauthordoformat {Chen, X.\ppgccbibauthorsep {} Wang,
  Y.\ppgccbibauthorsep {} Agichtein, E.\ppgccbibauthorlastsep {}
  \ppgccbibauthorand {} Wang, F.} (\ppgccyeardoformat {2015}).
\newblock \ppgccbibtitledoformat {A comparative study of demographic attribute
  inference in twitter}.
\newblock \ppgccbibinstring {} \ppgccbibbooktitledoformat {In Proceedings of
  the 9th International AAAI Conference on Weblogs and Social Media}.

\bibitem[Cheng et~al.\@, 2009]{sysomos}
\ppgccbibauthordoformat {Cheng, A.\ppgccbibauthorsep {} Evans,
  M.\ppgccbibauthorlastsep {} \ppgccbibauthorand {} Singh, H.}
  (\ppgccyeardoformat {2009}).
\newblock \ppgccbibtitledoformat {Inside twitter: An in-depth look inside the
  twitter world}.
\newblock \ppgccbibjournaldoformat {Report of Sysomos, June, Toronto, Canada}.

\bibitem[Comarela et~al.\@, 2012]{Comarela:2012:UFA:2309996.2310017}
\ppgccbibauthordoformat {Comarela, G.\ppgccbibauthorsep {} Crovella,
  M.\ppgccbibauthorsep {} Almeida, V.\ppgccbibauthorlastsep {}
  \ppgccbibauthorand {} Benevenuto, F.} (\ppgccyeardoformat {2012}).
\newblock \ppgccbibtitledoformat {Understanding factors that affect response
  rates in twitter}.
\newblock \ppgccbibinstring {} \ppgccbibbooktitledoformat {Proceedings of the
  23rd ACM Conference on Hypertext and Social Media}.

\bibitem[Cotter et~al.\@, 2001]{cotter2001glass}
\ppgccbibauthordoformat {Cotter, D.~A.\ppgccbibauthorsep {} Hermsen,
  J.~M.\ppgccbibauthorsep {} Ovadia, S.\ppgccbibauthorlastsep {}
  \ppgccbibauthorand {} Vanneman, R.} (\ppgccyeardoformat {2001}).
\newblock \ppgccbibtitledoformat {The glass ceiling effect}.
\newblock \ppgccbibjournaldoformat {Social forces}, \ppgccvolumedoformat
  {80}({\ppgccnumberdoformat {2}}):\ppgccpagedoformat {655}-\ppgccpagedoformat
  {}-\ppgccpagedoformat {681}.

\bibitem[Culotta et~al.\@, 2015]{culotta2015predicting}
\ppgccbibauthordoformat {Culotta, A.\ppgccbibauthorsep {} Kumar,
  N.~R.\ppgccbibauthorlastsep {} \ppgccbibauthorand {} Cutler, J.}
  (\ppgccyeardoformat {2015}).
\newblock \ppgccbibtitledoformat {Predicting the demographics of twitter users
  from website traffic data.}
\newblock \ppgccbibinstring {} \ppgccbibbooktitledoformat {Proceedings of the
  29th AAAI Conference on Artificial Intelligence}.

\bibitem[Cunha et~al.\@, 2012]{Cunha:2012:GBS:2309996.2310055}
\ppgccbibauthordoformat {Cunha, E.\ppgccbibauthorsep {} Magno,
  G.\ppgccbibauthorsep {} Almeida, V.\ppgccbibauthorsep {} Gon\c{c}alves,
  M.~A.\ppgccbibauthorlastsep {} \ppgccbibauthorand {} Benevenuto, F.}
  (\ppgccyeardoformat {2012}).
\newblock \ppgccbibtitledoformat {A gender based study of tagging behavior in
  twitter}.
\newblock \ppgccbibinstring {} \ppgccbibbooktitledoformat {Proceedings of the
  23rd ACM Conference on Hypertext and Social Media}, HT '12,
  \ppgccbibpagesstring {} \ppgccpagedoformat {323}-\ppgccpagedoformat
  {}-\ppgccpagedoformat {324}, New York, NY, USA. ACM.

\bibitem[Datta et~al.\@, 2015]{datta2015automated}
\ppgccbibauthordoformat {Datta, A.\ppgccbibauthorsep {} Tschantz,
  M.~C.\ppgccbibauthorlastsep {} \ppgccbibauthorand {} Datta, A.}
  (\ppgccyeardoformat {2015}).
\newblock \ppgccbibtitledoformat {Automated experiments on ad privacy
  settings}.
\newblock \ppgccbibjournaldoformat {Proceedings on Privacy Enhancing
  Technologies}, \ppgccvolumedoformat {2015}({\ppgccnumberdoformat
  {1}}):\ppgccpagedoformat {92}-\ppgccpagedoformat {}-\ppgccpagedoformat {112}.

\bibitem[De~Choudhury et~al.\@, 2017]{DeChoudhury:2017:GCD:2998181.2998220}
\ppgccbibauthordoformat {De~Choudhury, M.\ppgccbibauthorsep {} Sharma,
  S.~S.\ppgccbibauthorsep {} Logar, T.\ppgccbibauthorsep {} Eekhout,
  W.\ppgccbibauthorlastsep {} \ppgccbibauthorand {} Nielsen, R.~C.}
  (\ppgccyeardoformat {2017}).
\newblock \ppgccbibtitledoformat {Gender and cross-cultural differences in
  social media disclosures of mental illness}.
\newblock \ppgccbibinstring {} \ppgccbibbooktitledoformat {Proceedings of the
  2017 ACM Conference on Computer Supported Cooperative Work and Social
  Computing}, CSCW '17, \ppgccbibpagesstring {} \ppgccpagedoformat
  {353}-\ppgccpagedoformat {}-\ppgccpagedoformat {369}, New York, NY, USA. ACM.

\bibitem[Diakopoulos \ppgccciteauthorand {} Koliska,
  2016]{diakopoulos2016algorithmic}
\ppgccbibauthordoformat {Diakopoulos, N. \ppgccbibauthorand {} Koliska, M.}
  (\ppgccyeardoformat {2016}).
\newblock \ppgccbibtitledoformat {Algorithmic transparency in the news media}.
\newblock \ppgccbibjournaldoformat {Digital Journalism}.

\bibitem[Dovidio \ppgccciteauthorand {} Gaertner, 2000]{dovidio2000aversive}
\ppgccbibauthordoformat {Dovidio, J.~F. \ppgccbibauthorand {} Gaertner, S.~L.}
  (\ppgccyeardoformat {2000}).
\newblock \ppgccbibtitledoformat {Aversive racism and selection decisions: 1989
  and 1999}.
\newblock \ppgccbibjournaldoformat {Psychological science},
  \ppgccvolumedoformat {11}({\ppgccnumberdoformat {4}}):\ppgccpagedoformat
  {315}-\ppgccpagedoformat {}-\ppgccpagedoformat {319}.

\bibitem[Dutta, 2010]{dutta2010s}
\ppgccbibauthordoformat {Dutta, S.} (\ppgccyeardoformat {2010}).
\newblock \ppgccbibtitledoformat {What’s your personal social media
  strategy}.
\newblock \ppgccbibjournaldoformat {Harvard business review},
  \ppgccvolumedoformat {88}({\ppgccnumberdoformat {11}}):\ppgccpagedoformat
  {127}-\ppgccpagedoformat {}-\ppgccpagedoformat {130}.

\bibitem[Dwork et~al.\@, 2012]{dwork2012fairness}
\ppgccbibauthordoformat {Dwork, C.\ppgccbibauthorsep {} Hardt,
  M.\ppgccbibauthorsep {} Pitassi, T.\ppgccbibauthorsep {} Reingold,
  O.\ppgccbibauthorlastsep {} \ppgccbibauthorand {} Zemel, R.}
  (\ppgccyeardoformat {2012}).
\newblock \ppgccbibtitledoformat {Fairness through awareness}.
\newblock \ppgccbibinstring {} \ppgccbibbooktitledoformat {Proceedings of the
  3rd Innovations in Theoretical Computer Science Conference},
  \ppgccbibpagesstring {} \ppgccpagedoformat {214}-\ppgccpagedoformat
  {}-\ppgccpagedoformat {226}. ACM.

\bibitem[Edelman et~al.\@, 2017]{edelman2017racial}
\ppgccbibauthordoformat {Edelman, B.\ppgccbibauthorsep {} Luca,
  M.\ppgccbibauthorlastsep {} \ppgccbibauthorand {} Svirsky, D.}
  (\ppgccyeardoformat {2017}).
\newblock \ppgccbibtitledoformat {Racial discrimination in the sharing economy:
  Evidence from a field experiment}.
\newblock \ppgccbibjournaldoformat {American Economic Journal: Applied
  Economics}, \ppgccvolumedoformat {9}({\ppgccnumberdoformat
  {2}}):\ppgccpagedoformat {1}-\ppgccpagedoformat {}-\ppgccpagedoformat {22}.

\bibitem[Fan et~al.\@, 2014]{fan2014learning}
\ppgccbibauthordoformat {Fan, H.\ppgccbibauthorsep {} Cao, Z.\ppgccbibauthorsep
  {} Jiang, Y.\ppgccbibauthorsep {} Yin, Q.\ppgccbibauthorlastsep {}
  \ppgccbibauthorand {} Doudou, C.} (\ppgccyeardoformat {2014}).
\newblock \ppgccbibtitledoformat {Learning deep face representation}.
\newblock \ppgccbibjournaldoformat {arXiv preprint arXiv:1403.2802}.

\bibitem[Feldman et~al.\@, 2015]{Feldman2015KDD}
\ppgccbibauthordoformat {Feldman, M.\ppgccbibauthorsep {} Friedler,
  S.~A.\ppgccbibauthorsep {} Moeller, J.\ppgccbibauthorsep {} Scheidegger,
  C.\ppgccbibauthorlastsep {} \ppgccbibauthorand {} Venkatasubramanian, S.}
  (\ppgccyeardoformat {2015}).
\newblock \ppgccbibtitledoformat {Certifying and removing disparate impact}.
\newblock \ppgccbibinstring {} \ppgccbibbooktitledoformat {Proceedings of the
  21th ACM SIGKDD International Conference on Knowledge Discovery and Data
  Mining}, KDD '15, \ppgccbibpagesstring {} \ppgccpagedoformat
  {259}-\ppgccpagedoformat {}-\ppgccpagedoformat {268}, New York, NY, USA. ACM.

\bibitem[Felton \ppgccciteauthorand {} Lyons, 2016]{white_house_transparency}
\ppgccbibauthordoformat {Felton, E. \ppgccbibauthorand {} Lyons, T.}
  (\ppgccyeardoformat {2016}).
\newblock \ppgccbibtitledoformat {The administration's report on the future of
  artificial intelligence}.
\newblock
  http://whitehouse.gov/blog/2016/10/12/administrations-report-future-artificial-intelligence.

\bibitem[Freitas et~al.\@, 2015]{Freitas2015@asonam}
\ppgccbibauthordoformat {Freitas, C.\ppgccbibauthorsep {} Benevenuto,
  F.\ppgccbibauthorsep {} Ghosh, S.\ppgccbibauthorlastsep {} \ppgccbibauthorand
  {} Veloso, A.} (\ppgccyeardoformat {2015}).
\newblock \ppgccbibtitledoformat {Reverse engineering socialbot infiltration
  strategies in twitter}.
\newblock \ppgccbibinstring {} \ppgccbibbooktitledoformat {Proceedings of the
  2015 IEEEACM International Conference on Advances in Social Networks Analysis
  and Mining}.

\bibitem[Ge et~al.\@, 2016]{ge2016racial}
\ppgccbibauthordoformat {Ge, Y.\ppgccbibauthorsep {} Knittel,
  C.~R.\ppgccbibauthorsep {} MacKenzie, D.\ppgccbibauthorlastsep {}
  \ppgccbibauthorand {} Zoepf, S.} (\ppgccyeardoformat {2016}).
\newblock \ppgccbibtitledoformat {Racial and gender discrimination in
  transportation network companies}.
\newblock \ppgccbibtechreportstring {}, National Bureau of Economic Research.

\bibitem[Gilbert et~al.\@, 2013]{Gilbert:2013:INT:2470654.2481336}
\ppgccbibauthordoformat {Gilbert, E.\ppgccbibauthorsep {} Bakhshi,
  S.\ppgccbibauthorsep {} Chang, S.\ppgccbibauthorlastsep {} \ppgccbibauthorand
  {} Terveen, L.} (\ppgccyeardoformat {2013}).
\newblock \ppgccbibtitledoformat {"i need to try this"?: A statistical overview
  of pinterest}.
\newblock \ppgccbibinstring {} \ppgccbibbooktitledoformat {Proceedings of the
  SIGCHI Conference on Human Factors in Computing Systems}. ACM.

\bibitem[Goel et~al.\@, 2016]{goel2016precinct}
\ppgccbibauthordoformat {Goel, S.\ppgccbibauthorsep {} Rao,
  J.~M.\ppgccbibauthorsep {} Shroff, R.\ppgccbibauthorlastsep {} et~al.\@}
  (\ppgccyeardoformat {2016}).
\newblock \ppgccbibtitledoformat {Precinct or prejudice? understanding racial
  disparities in new york city’s stop-and-frisk policy}.
\newblock \ppgccbibjournaldoformat {The Annals of Applied Statistics},
  \ppgccvolumedoformat {10}({\ppgccnumberdoformat {1}}):\ppgccpagedoformat
  {365}-\ppgccpagedoformat {}-\ppgccpagedoformat {394}.

\bibitem[Goh et~al.\@, 2016]{goh2016satisfying}
\ppgccbibauthordoformat {Goh, G.\ppgccbibauthorsep {} Cotter,
  A.\ppgccbibauthorsep {} Gupta, M.\ppgccbibauthorlastsep {} \ppgccbibauthorand
  {} Friedlander, M.~P.} (\ppgccyeardoformat {2016}).
\newblock \ppgccbibtitledoformat {Satisfying real-world goals with dataset
  constraints}.
\newblock \ppgccbibinstring {} \ppgccbibbooktitledoformat {Advances in Neural
  Information Processing Systems}, \ppgccbibpagesstring {} \ppgccpagedoformat
  {2415}-\ppgccpagedoformat {}-\ppgccpagedoformat {2423}.

\bibitem[Gollapudi \ppgccciteauthorand {} Sharma, 2009]{gollapudi2009axiomatic}
\ppgccbibauthordoformat {Gollapudi, S. \ppgccbibauthorand {} Sharma, A.}
  (\ppgccyeardoformat {2009}).
\newblock \ppgccbibtitledoformat {An axiomatic approach for result
  diversification}.
\newblock \ppgccbibinstring {} \ppgccbibbooktitledoformat {ACM WWW}.

\bibitem[Graells-Garrido et~al.\@, 2015]{graells2015first}
\ppgccbibauthordoformat {Graells-Garrido, E.\ppgccbibauthorsep {} Lalmas,
  M.\ppgccbibauthorlastsep {} \ppgccbibauthorand {} Menczer, F.}
  (\ppgccyeardoformat {2015}).
\newblock \ppgccbibtitledoformat {First women, second sex: gender bias in
  wikipedia}.
\newblock \ppgccbibinstring {} \ppgccbibbooktitledoformat {Proceedings of the
  26th ACM Conference on Hypertext \& Social Media}.

\bibitem[Guynn, 2015]{guynn2015google}
\ppgccbibauthordoformat {Guynn, J.} (\ppgccyeardoformat {2015}).
\newblock \ppgccbibtitledoformat {Google photos labeled black people as
  ‘gorillas’}.
\newblock \ppgccbibjournaldoformat {USA Today, July}.

\bibitem[Hann\'{a}k et~al.\@, 2017]{Hannak2017CSCW}
\ppgccbibauthordoformat {Hann\'{a}k, A.\ppgccbibauthorsep {} Wagner,
  C.\ppgccbibauthorsep {} Garcia, D.\ppgccbibauthorsep {} Mislove,
  A.\ppgccbibauthorsep {} Strohmaier, M.\ppgccbibauthorlastsep {}
  \ppgccbibauthorand {} Wilson, C.} (\ppgccyeardoformat {2017}).
\newblock \ppgccbibtitledoformat {Bias in online freelance marketplaces:
  Evidence from taskrabbit and fiverr}.
\newblock \ppgccbibinstring {} \ppgccbibbooktitledoformat {Proceedings of the
  2017 ACM Conference on Computer Supported Cooperative Work and Social
  Computing}, CSCW '17, \ppgccbibpagesstring {} \ppgccpagedoformat
  {1914}-\ppgccpagedoformat {}-\ppgccpagedoformat {1933}, New York, NY, USA.
  ACM.

\bibitem[Kamiran \ppgccciteauthorand {} Calders, 2012]{kamiran2012data}
\ppgccbibauthordoformat {Kamiran, F. \ppgccbibauthorand {} Calders, T.}
  (\ppgccyeardoformat {2012}).
\newblock \ppgccbibtitledoformat {Data preprocessing techniques for
  classification without discrimination}.
\newblock \ppgccbibjournaldoformat {Knowledge and Information Systems},
  \ppgccvolumedoformat {33}({\ppgccnumberdoformat {1}}):\ppgccpagedoformat
  {1}-\ppgccpagedoformat {}-\ppgccpagedoformat {33}.

\bibitem[Kamishima et~al.\@, 2012]{kamishima2012fairness}
\ppgccbibauthordoformat {Kamishima, T.\ppgccbibauthorsep {} Akaho,
  S.\ppgccbibauthorsep {} Asoh, H.\ppgccbibauthorlastsep {} \ppgccbibauthorand
  {} Sakuma, J.} (\ppgccyeardoformat {2012}).
\newblock \ppgccbibtitledoformat {Fairness-aware classifier with prejudice
  remover regularizer}.
\newblock \ppgccbibjournaldoformat {Machine Learning and Knowledge Discovery in
  Databases}, \ppgccbibpagesstring {} \ppgccpagedoformat
  {35}-\ppgccpagedoformat {}-\ppgccpagedoformat {50}.

\bibitem[Karimi et~al.\@, 2016]{karimi2016}
\ppgccbibauthordoformat {Karimi, F.\ppgccbibauthorsep {} Wagner,
  C.\ppgccbibauthorsep {} Lemmerich, F.\ppgccbibauthorsep {} Jadidi,
  M.\ppgccbibauthorlastsep {} \ppgccbibauthorand {} Strohmaier, M.}
  (\ppgccyeardoformat {2016}).
\newblock \ppgccbibtitledoformat {Inferring gender from names on the web: A
  comparative evaluation of gender detection methods}.
\newblock \ppgccbibinstring {} \ppgccbibbooktitledoformat {Proceedings of the
  25th International Conference on World Wide Web}.

\bibitem[Koumchatzky \ppgccciteauthorand {} Andryeyev,
  2017]{twitter_timeline_deep}
\ppgccbibauthordoformat {Koumchatzky, N. \ppgccbibauthorand {} Andryeyev, A.}
  (\ppgccyeardoformat {2017}).
\newblock \ppgccbibtitledoformat {Using deep learning at scale in twitter’s
  timelines}.
\newblock
  http://blog.twitter.com/engineering/en\_us/topics/insights/2017/using-deep-learning-at-scale-in-witters-timelines.html.

\bibitem[Kricheli-Katz \ppgccciteauthorand {} Regev, 2016]{kricheli2016many}
\ppgccbibauthordoformat {Kricheli-Katz, T. \ppgccbibauthorand {} Regev, T.}
  (\ppgccyeardoformat {2016}).
\newblock \ppgccbibtitledoformat {How many cents on the dollar? women and men
  in product markets}.
\newblock \ppgccbibjournaldoformat {Science advances}, \ppgccvolumedoformat
  {2}({\ppgccnumberdoformat {2}}):\ppgccpagedoformat {e1500599}.

\bibitem[Kulshrestha et~al.\@, 2017]{juhi2017}
\ppgccbibauthordoformat {Kulshrestha, J.\ppgccbibauthorsep {} Eslami,
  M.\ppgccbibauthorsep {} Messias, J.\ppgccbibauthorsep {} Zafar,
  M.~B.\ppgccbibauthorsep {} Ghosh, S.\ppgccbibauthorsep {} Gummadi,
  K.~P.\ppgccbibauthorlastsep {} \ppgccbibauthorand {} Karahalios, K.}
  (\ppgccyeardoformat {2017}).
\newblock \ppgccbibtitledoformat {Quantifying search bias: Investigating
  sources of bias for political searches in social media.}
\newblock \ppgccbibinstring {} \ppgccbibbooktitledoformat {Proceedings of the
  20th ACM Conference on Computer-Supported Cooperative Work and Social
  Computing}.

\bibitem[Kulshrestha et~al.\@, 2012]{kulshrestha2012geographic}
\ppgccbibauthordoformat {Kulshrestha, J.\ppgccbibauthorsep {} Kooti,
  F.\ppgccbibauthorsep {} Nikravesh, A.\ppgccbibauthorlastsep {}
  \ppgccbibauthorand {} Gummadi, P.~K.} (\ppgccyeardoformat {2012}).
\newblock \ppgccbibtitledoformat {Geographic dissection of the twitter
  network.}
\newblock \ppgccbibinstring {} \ppgccbibbooktitledoformat {In Proceedings of
  the 6th International AAAI Conference on Weblogs and Social Media}.

\bibitem[Liu \ppgccciteauthorand {} Ruths, 2013]{liu2013s}
\ppgccbibauthordoformat {Liu, W. \ppgccbibauthorand {} Ruths, D.}
  (\ppgccyeardoformat {2013}).
\newblock \ppgccbibtitledoformat {What's in a name? using first names as
  features for gender inference in twitter.}
\newblock \ppgccbibinstring {} \ppgccbibbooktitledoformat {AAAI Spring
  Symposium Series}, volume~\ppgccvolumedoformat {13}, \ppgccbibpagestring
  {}~01.

\bibitem[Luong et~al.\@, 2011]{Luong2011KDD}
\ppgccbibauthordoformat {Luong, B.~T.\ppgccbibauthorsep {} Ruggieri,
  S.\ppgccbibauthorlastsep {} \ppgccbibauthorand {} Turini, F.}
  (\ppgccyeardoformat {2011}).
\newblock \ppgccbibtitledoformat {k-nn as an implementation of situation
  testing for discrimination discovery and prevention}.
\newblock \ppgccbibinstring {} \ppgccbibbooktitledoformat {Proceedings of the
  17th ACM SIGKDD International Conference on Knowledge Discovery and Data
  Mining}, KDD '11, \ppgccbibpagesstring {} \ppgccpagedoformat
  {502}-\ppgccpagedoformat {}-\ppgccpagedoformat {510}, New York, NY, USA. ACM.

\bibitem[McPherson et~al.\@, 2001]{mcpherson2001birds}
\ppgccbibauthordoformat {McPherson, M.\ppgccbibauthorsep {} Smith-Lovin,
  L.\ppgccbibauthorlastsep {} \ppgccbibauthorand {} Cook, J.~M.}
  (\ppgccyeardoformat {2001}).
\newblock \ppgccbibtitledoformat {Birds of a feather: Homophily in social
  networks}.
\newblock \ppgccbibjournaldoformat {Annual review of sociology},
  \ppgccbibpagesstring {} \ppgccpagedoformat {415}-\ppgccpagedoformat
  {}-\ppgccpagedoformat {444}.

\bibitem[Merluzzi \ppgccciteauthorand {} Dobrev, 2015]{merluzzi2015unequal}
\ppgccbibauthordoformat {Merluzzi, J. \ppgccbibauthorand {} Dobrev, S.~D.}
  (\ppgccyeardoformat {2015}).
\newblock \ppgccbibtitledoformat {Unequal on top: Gender profiling and the
  income gap among high earner male and female professionals}.
\newblock \ppgccbibjournaldoformat {Social science research},
  \ppgccvolumedoformat {53}:\ppgccpagedoformat {45}-\ppgccpagedoformat
  {}-\ppgccpagedoformat {58}.

\bibitem[Messias et~al.\@, 2016]{Messias2016@asonam}
\ppgccbibauthordoformat {Messias, J.\ppgccbibauthorsep {} Benevenuto,
  F.\ppgccbibauthorsep {} Weber, I.\ppgccbibauthorlastsep {} \ppgccbibauthorand
  {} Zagheni, E.} (\ppgccyeardoformat {2016}).
\newblock \ppgccbibtitledoformat {From migration corridors to clusters: The
  value of google+ data for migration studies}.
\newblock \ppgccbibinstring {} \ppgccbibbooktitledoformat {Proceedings of the
  2016 IEEE/ACM International Conference on Advances in Social Networks
  Analysis and Mining}.

\bibitem[Messias et~al.\@, 2013]{messias13@firstmonday}
\ppgccbibauthordoformat {Messias, J.\ppgccbibauthorsep {} Schmidt,
  L.\ppgccbibauthorsep {} Rabelo, R.\ppgccbibauthorlastsep {}
  \ppgccbibauthorand {} Benevenuto, F.} (\ppgccyeardoformat {2013}).
\newblock \ppgccbibtitledoformat {You followed my bot! transforming robots into
  influential users in twitter}.
\newblock \ppgccbibjournaldoformat {First Monday}, \ppgccvolumedoformat
  {18}({\ppgccnumberdoformat {7}}).

\bibitem[Messias et~al.\@, 2017]{messiasWI2017}
\ppgccbibauthordoformat {Messias, J.\ppgccbibauthorsep {} Vikatos,
  P.\ppgccbibauthorlastsep {} \ppgccbibauthorand {} Benevenuto, F.}
  (\ppgccyeardoformat {2017}).
\newblock \ppgccbibtitledoformat {White, man, and highly followed: Gender and
  race inequalities in twitter}.
\newblock \ppgccbibinstring {} \ppgccbibbooktitledoformat {Proceedings of the
  IEEE/WIC/ACM International Conference on Web Intelligence}, WI '17, New York,
  NY, USA. ACM.

\bibitem[Mislove et~al.\@, 2011]{mislove2011understanding}
\ppgccbibauthordoformat {Mislove, A.\ppgccbibauthorsep {} Lehmann,
  S.\ppgccbibauthorsep {} Ahn, Y.-Y.\ppgccbibauthorsep {} Onnela,
  J.-P.\ppgccbibauthorlastsep {} \ppgccbibauthorand {} Rosenquist, J.~N.}
  (\ppgccyeardoformat {2011}).
\newblock \ppgccbibtitledoformat {Understanding the demographics of twitter
  users.}
\newblock \ppgccbibinstring {} \ppgccbibbooktitledoformat {In Proceedings of
  the 5th International AAAI Conference on Weblogs and Social Media}.

\bibitem[Morales et~al.\@, 2014]{morales2014efficiency}
\ppgccbibauthordoformat {Morales, A.\ppgccbibauthorsep {} Borondo,
  J.\ppgccbibauthorsep {} Losada, J.~C.\ppgccbibauthorlastsep {}
  \ppgccbibauthorand {} Benito, R.~M.} (\ppgccyeardoformat {2014}).
\newblock \ppgccbibtitledoformat {Efficiency of human activity on information
  spreading on twitter}.
\newblock \ppgccbibjournaldoformat {Social Networks}, \ppgccvolumedoformat
  {39}:\ppgccpagedoformat {1}-\ppgccpagedoformat {}-\ppgccpagedoformat {11}.

\bibitem[Morstatter et~al.\@, 2014]{morstatter2014biased}
\ppgccbibauthordoformat {Morstatter, F.\ppgccbibauthorsep {} Pfeffer,
  J.\ppgccbibauthorlastsep {} \ppgccbibauthorand {} Liu, H.}
  (\ppgccyeardoformat {2014}).
\newblock \ppgccbibtitledoformat {When is it biased?: assessing the
  representativeness of twitter's streaming api}.
\newblock \ppgccbibinstring {} \ppgccbibbooktitledoformat {Proceedings of the
  23rd International Conference on World Wide Web}.

\bibitem[Nilizadeh et~al.\@, 2016]{nilizadeh2016twitter}
\ppgccbibauthordoformat {Nilizadeh, S.\ppgccbibauthorsep {} Groggel,
  A.\ppgccbibauthorsep {} Lista, P.\ppgccbibauthorsep {} Das,
  S.\ppgccbibauthorsep {} Ahn, Y.-Y.\ppgccbibauthorsep {} Kapadia,
  A.\ppgccbibauthorlastsep {} \ppgccbibauthorand {} Rojas, F.}
  (\ppgccyeardoformat {2016}).
\newblock \ppgccbibtitledoformat {Twitter's glass ceiling: The effect of
  perceived gender on online visibility}.
\newblock \ppgccbibinstring {} \ppgccbibbooktitledoformat {In Proceedings of
  the 10th International AAAI Conference on Weblogs and Social Media}.

\bibitem[Nunez, 2016]{facebook_bias}
\ppgccbibauthordoformat {Nunez, M.} (\ppgccyeardoformat {2016}).
\newblock \ppgccbibtitledoformat {{Former Facebook Workers: We Routinely
  Suppressed Conservative News}}.
\newblock
  gizmodo.com/former-facebook-workers-we-routinely-suppressed-conser-1775461006.

\bibitem[O'Connor et~al.\@, 2010]{o2010tweets}
\ppgccbibauthordoformat {O'Connor, B.\ppgccbibauthorsep {} Balasubramanyan,
  R.\ppgccbibauthorsep {} Routledge, B.~R.\ppgccbibauthorlastsep {}
  \ppgccbibauthorand {} Smith, N.~A.} (\ppgccyeardoformat {2010}).
\newblock \ppgccbibtitledoformat {From tweets to polls: Linking text sentiment
  to public opinion time series.}
\newblock \ppgccbibinstring {} \ppgccbibbooktitledoformat {In Proceedings of
  the 4th International AAAI Conference on Weblogs and Social Media}.

\bibitem[Ohlheiser, 2016]{Ohlheiser2016}
\ppgccbibauthordoformat {Ohlheiser, A.} (\ppgccyeardoformat {2016}).
\newblock \ppgccbibtitledoformat {Three days after removing human editors,
  facebook is already trending fake news}.
\newblock washingtonpost.com/news/
  the-intersect/wp/2016/08/29/a-fake-headline-about-megyn-kelly-was-trending-on-facebook/.

\bibitem[Oliver \ppgccciteauthorand {} Shapiro, 2006]{oliver2006black}
\ppgccbibauthordoformat {Oliver, M.~L. \ppgccbibauthorand {} Shapiro, T.~M.}
  (\ppgccyeardoformat {2006}).
\newblock \ppgccbibbtitledoformat {Black wealth, white wealth: A new
  perspective on racial inequality}.
\newblock Taylor \& Francis.

\bibitem[Pedreshi et~al.\@, 2008]{pedreshi2008discrimination}
\ppgccbibauthordoformat {Pedreshi, D.\ppgccbibauthorsep {} Ruggieri,
  S.\ppgccbibauthorlastsep {} \ppgccbibauthorand {} Turini, F.}
  (\ppgccyeardoformat {2008}).
\newblock \ppgccbibtitledoformat {Discrimination-aware data mining}.
\newblock \ppgccbibinstring {} \ppgccbibbooktitledoformat {Proceedings of the
  14th ACM SIGKDD international conference on Knowledge discovery and data
  mining}, \ppgccbibpagesstring {} \ppgccpagedoformat {560}-\ppgccpagedoformat
  {}-\ppgccpagedoformat {568}. ACM.

\bibitem[Pennacchiotti \ppgccciteauthorand {} Popescu,
  2011]{pennacchiotti2011machine}
\ppgccbibauthordoformat {Pennacchiotti, M. \ppgccbibauthorand {} Popescu,
  A.-M.} (\ppgccyeardoformat {2011}).
\newblock \ppgccbibtitledoformat {A machine learning approach to twitter user
  classification}.
\newblock \ppgccbibinstring {} \ppgccbibbooktitledoformat {Proceedings of the
  5th International AAAI Conference on Web and Social Media}.

\bibitem[Reis et~al.\@, 2017]{reis2017}
\ppgccbibauthordoformat {Reis, J. C.~S.\ppgccbibauthorsep {} Kwak,
  H.\ppgccbibauthorsep {} An, J.\ppgccbibauthorsep {} Messias,
  J.\ppgccbibauthorlastsep {} \ppgccbibauthorand {} Benevenuto, F.}
  (\ppgccyeardoformat {2017}).
\newblock \ppgccbibtitledoformat {Demographics of news sharing in the u.s.
  twittersphere}.
\newblock \ppgccbibinstring {} \ppgccbibbooktitledoformat {Proceedings of the
  28th ACM Conference on Hypertext and Social Media}. ACM.

\bibitem[Ridgeway \ppgccciteauthorand {} MacDonald, 2009]{ridgeway2009doubly}
\ppgccbibauthordoformat {Ridgeway, G. \ppgccbibauthorand {} MacDonald, J.~M.}
  (\ppgccyeardoformat {2009}).
\newblock \ppgccbibtitledoformat {Doubly robust internal benchmarking and false
  discovery rates for detecting racial bias in police stops}.
\newblock \ppgccbibjournaldoformat {Journal of the American Statistical
  Association}, \ppgccvolumedoformat {104}({\ppgccnumberdoformat
  {486}}):\ppgccpagedoformat {661}-\ppgccpagedoformat {}-\ppgccpagedoformat
  {668}.

\bibitem[Romei \ppgccciteauthorand {} Ruggieri,
  2014]{romei2014multidisciplinary}
\ppgccbibauthordoformat {Romei, A. \ppgccbibauthorand {} Ruggieri, S.}
  (\ppgccyeardoformat {2014}).
\newblock \ppgccbibtitledoformat {A multidisciplinary survey on discrimination
  analysis}.
\newblock \ppgccbibjournaldoformat {The Knowledge Engineering Review},
  \ppgccvolumedoformat {29}({\ppgccnumberdoformat {05}}):\ppgccpagedoformat
  {582}-\ppgccpagedoformat {}-\ppgccpagedoformat {638}.

\bibitem[Romero et~al.\@, 2011]{romero2011influence}
\ppgccbibauthordoformat {Romero, D.~M.\ppgccbibauthorsep {} Galuba,
  W.\ppgccbibauthorsep {} Asur, S.\ppgccbibauthorlastsep {} \ppgccbibauthorand
  {} Huberman, B.~A.} (\ppgccyeardoformat {2011}).
\newblock \ppgccbibtitledoformat {Influence and passivity in social media}.
\newblock \ppgccbibinstring {} \ppgccbibbooktitledoformat {Joint European
  Conference on Machine Learning and Knowledge Discovery in Databases},
  \ppgccbibpagesstring {} \ppgccpagedoformat {18}-\ppgccpagedoformat
  {}-\ppgccpagedoformat {33}. Springer.

\bibitem[Ronson, 2016]{ronson2016so}
\ppgccbibauthordoformat {Ronson, J.} (\ppgccyeardoformat {2016}).
\newblock \ppgccbibbtitledoformat {So you've been publicly shamed}.
\newblock Riverhead Books (Hardcover).

\bibitem[Sharma et~al.\@, 2012]{sharma2012inferring}
\ppgccbibauthordoformat {Sharma, N.~K.\ppgccbibauthorsep {} Ghosh,
  S.\ppgccbibauthorsep {} Benevenuto, F.\ppgccbibauthorsep {} Ganguly,
  N.\ppgccbibauthorlastsep {} \ppgccbibauthorand {} Gummadi, K.}
  (\ppgccyeardoformat {2012}).
\newblock \ppgccbibtitledoformat {Inferring who-is-who in the twitter social
  network}.
\newblock \ppgccbibjournaldoformat {ACM SIGCOMM Computer Communication Review},
  \ppgccvolumedoformat {42}({\ppgccnumberdoformat {4}}):\ppgccpagedoformat
  {533}-\ppgccpagedoformat {}-\ppgccpagedoformat {538}.

\bibitem[Sheth et~al.\@, 2011]{sheth2011towards}
\ppgccbibauthordoformat {Sheth, S.~K.\ppgccbibauthorsep {} Bell,
  J.~S.\ppgccbibauthorsep {} Arora, N.\ppgccbibauthorlastsep {}
  \ppgccbibauthorand {} Kaiser, G.~E.} (\ppgccyeardoformat {2011}).
\newblock \ppgccbibtitledoformat {Towards diversity in recommendations using
  social networks}.

\bibitem[Sloan et~al.\@, 2015]{sloan2015tweets}
\ppgccbibauthordoformat {Sloan, L.\ppgccbibauthorsep {} Morgan,
  J.\ppgccbibauthorsep {} Burnap, P.\ppgccbibauthorlastsep {}
  \ppgccbibauthorand {} Williams, M.} (\ppgccyeardoformat {2015}).
\newblock \ppgccbibtitledoformat {Who tweets? deriving the demographic
  characteristics of age, occupation and social class from twitter user
  meta-data}.
\newblock \ppgccbibjournaldoformat {PloS one}, \ppgccvolumedoformat
  {10}({\ppgccnumberdoformat {3}}):\ppgccpagedoformat {e0115545}.

\bibitem[Sugimoto et~al.\@, 2013]{sugimoto2013global}
\ppgccbibauthordoformat {Sugimoto, C.~R.\ppgccbibauthorsep {} Lariviere,
  V.\ppgccbibauthorsep {} Ni, C.\ppgccbibauthorsep {} Gingras,
  Y.\ppgccbibauthorsep {} Cronin, B.\ppgccbibauthorlastsep {} et~al.\@}
  (\ppgccyeardoformat {2013}).
\newblock \ppgccbibtitledoformat {Global gender disparities in science}.
\newblock \ppgccbibjournaldoformat {Nature}, \ppgccvolumedoformat
  {504}({\ppgccnumberdoformat {7479}}):\ppgccpagedoformat
  {211}-\ppgccpagedoformat {}-\ppgccpagedoformat {213}.

\bibitem[Sugimoto et~al.\@, 2015]{sugimoto2015relationship}
\ppgccbibauthordoformat {Sugimoto, C.~R.\ppgccbibauthorsep {} Ni,
  C.\ppgccbibauthorlastsep {} \ppgccbibauthorand {} Larivi{\`e}re, V.}
  (\ppgccyeardoformat {2015}).
\newblock \ppgccbibtitledoformat {On the relationship between gender
  disparities in scholarly communication and country-level development
  indicators}.
\newblock \ppgccbibjournaldoformat {Science and Public Policy},
  \ppgccvolumedoformat {42}({\ppgccnumberdoformat {6}}):\ppgccpagedoformat
  {789}-\ppgccpagedoformat {}-\ppgccpagedoformat {810}.

\bibitem[Sweeney, 2013]{sweeney2013discrimination}
\ppgccbibauthordoformat {Sweeney, L.} (\ppgccyeardoformat {2013}).
\newblock \ppgccbibtitledoformat {Discrimination in online ad delivery}.
\newblock \ppgccbibjournaldoformat {Queue}, \ppgccvolumedoformat
  {11}({\ppgccnumberdoformat {3}}):\ppgccpagedoformat
  {10:10}-\ppgccpagedoformat {}-\ppgccpagedoformat {10:29}.
\newblock ISSN \ppgccissndoformat {1542-7730}.

\bibitem[Tausczik \ppgccciteauthorand {} Pennebaker,
  2010]{tausczik2010psychological}
\ppgccbibauthordoformat {Tausczik, Y.~R. \ppgccbibauthorand {} Pennebaker,
  J.~W.} (\ppgccyeardoformat {2010}).
\newblock \ppgccbibtitledoformat {The psychological meaning of words: Liwc and
  computerized text analysis methods}.
\newblock \ppgccbibjournaldoformat {Journal of language and social psychology},
  \ppgccvolumedoformat {29}({\ppgccnumberdoformat {1}}):\ppgccpagedoformat
  {24}-\ppgccpagedoformat {}-\ppgccpagedoformat {54}.

\bibitem[Todisco, 2014]{todisco2014share}
\ppgccbibauthordoformat {Todisco, M.} (\ppgccyeardoformat {2014}).
\newblock \ppgccbibtitledoformat {Share and share alike: Considering racial
  discrimination in the nascent room-sharing economy}.
\newblock \ppgccbibjournaldoformat {Stan. L. Rev. Online}, \ppgccvolumedoformat
  {67}:\ppgccpagedoformat {121}-\ppgccpagedoformat {}-\ppgccpagedoformat {129}.

\bibitem[Tumasjan et~al.\@, 2010]{tumasjan2010predicting}
\ppgccbibauthordoformat {Tumasjan, A.\ppgccbibauthorsep {} Sprenger,
  T.~O.\ppgccbibauthorsep {} Sandner, P.~G.\ppgccbibauthorlastsep {}
  \ppgccbibauthorand {} Welpe, I.~M.} (\ppgccyeardoformat {2010}).
\newblock \ppgccbibtitledoformat {Predicting elections with twitter: What 140
  characters reveal about political sentiment}.
\newblock \ppgccbibinstring {} \ppgccbibbooktitledoformat {Proceedings of the
  4th International AAAI Conference on Web and Social Media}.

\bibitem[Victor, 2016]{victor2017nyt}
\ppgccbibauthordoformat {Victor, D.} (\ppgccyeardoformat {2016}).
\newblock \ppgccbibtitledoformat {Microsoft created a twitter bot to learn from
  users. it quickly became a racist jerk.}
\newblock \ppgccbibjournaldoformat {The New York Times}.

\bibitem[Vikatos et~al.\@, 2017]{vikatos2017}
\ppgccbibauthordoformat {Vikatos, P.\ppgccbibauthorsep {} Messias,
  J.\ppgccbibauthorsep {} Miranda, M.\ppgccbibauthorlastsep {}
  \ppgccbibauthorand {} Benevenuto, F.} (\ppgccyeardoformat {2017}).
\newblock \ppgccbibtitledoformat {Linguistic diversities of demographic groups
  in twitter}.
\newblock \ppgccbibinstring {} \ppgccbibbooktitledoformat {Proceedings of the
  28th ACM Conference on Hypertext and Social Media}. ACM.

\bibitem[Wachs et~al.\@, 2017]{hannak2017@icwsm}
\ppgccbibauthordoformat {Wachs, J.\ppgccbibauthorsep {} Hannak,
  A.\ppgccbibauthorsep {} Voros, A.\ppgccbibauthorlastsep {} \ppgccbibauthorand
  {} Daroczy, B.} (\ppgccyeardoformat {2017}).
\newblock \ppgccbibtitledoformat {Why do men get more attention? exploring
  factors behind success in an online design community}.
\newblock \ppgccbibinstring {} \ppgccbibbooktitledoformat {Proceedings of the
  11th International AAAI Conference on Web and Social Media}, ICWSM'17,
  Montreal, Canada.

\bibitem[Wagner et~al.\@, 2016]{wagner2016women}
\ppgccbibauthordoformat {Wagner, C.\ppgccbibauthorsep {} Graells-Garrido,
  E.\ppgccbibauthorsep {} Garcia, D.\ppgccbibauthorlastsep {}
  \ppgccbibauthorand {} Menczer, F.} (\ppgccyeardoformat {2016}).
\newblock \ppgccbibtitledoformat {Women through the glass ceiling: gender
  asymmetries in wikipedia}.
\newblock \ppgccbibjournaldoformat {EPJ Data Science}, \ppgccvolumedoformat
  {5}({\ppgccnumberdoformat {1}}):\ppgccpagedoformat {1}.

\bibitem[Willis \ppgccciteauthorand {} Todorov, 2006]{willis2006first}
\ppgccbibauthordoformat {Willis, J. \ppgccbibauthorand {} Todorov, A.}
  (\ppgccyeardoformat {2006}).
\newblock \ppgccbibtitledoformat {First impressions making up your mind after a
  100-ms exposure to a face}.
\newblock \ppgccbibjournaldoformat {Psychological science},
  \ppgccvolumedoformat {17}({\ppgccnumberdoformat {7}}):\ppgccpagedoformat
  {592}-\ppgccpagedoformat {}-\ppgccpagedoformat {598}.

\bibitem[Wu et~al.\@, 2011]{wu2011says}
\ppgccbibauthordoformat {Wu, S.\ppgccbibauthorsep {} Hofman,
  J.~M.\ppgccbibauthorsep {} Mason, W.~A.\ppgccbibauthorlastsep {}
  \ppgccbibauthorand {} Watts, D.~J.} (\ppgccyeardoformat {2011}).
\newblock \ppgccbibtitledoformat {Who says what to whom on twitter}.
\newblock \ppgccbibinstring {} \ppgccbibbooktitledoformat {Proceedings of the
  20th International Conference on World Wide Web}.

\bibitem[Yin et~al.\@, 2015]{yin2015learning}
\ppgccbibauthordoformat {Yin, Q.\ppgccbibauthorsep {} Cao, Z.\ppgccbibauthorsep
  {} Jiang, Y.\ppgccbibauthorlastsep {} \ppgccbibauthorand {} Fan, H.}
  (\ppgccyeardoformat {2015}).
\newblock \ppgccbibtitledoformat {Learning deep face representation}.
\newblock US Patent 20,150,347,820.

\bibitem[Zafar et~al.\@, 2017a]{Zafar2017WWW}
\ppgccbibauthordoformat {Zafar, M.~B.\ppgccbibauthorsep {} Valera,
  I.\ppgccbibauthorsep {} Gomez~Rodriguez, M.\ppgccbibauthorlastsep {}
  \ppgccbibauthorand {} Gummadi, K.~P.} (\ppgccyeardoformat {2017a}).
\newblock \ppgccbibtitledoformat {Fairness beyond disparate treatment \&\
  disparate impact: Learning classification without disparate mistreatment}.
\newblock \ppgccbibinstring {} \ppgccbibbooktitledoformat {Proceedings of the
  26th International Conference on World Wide Web}, WWW '17,
  \ppgccbibpagesstring {} \ppgccpagedoformat {1171}-\ppgccpagedoformat
  {}-\ppgccpagedoformat {1180}, Republic and Canton of Geneva, Switzerland.
  International World Wide Web Conferences Steering Committee.

\bibitem[Zafar et~al.\@, 2017b]{zafar2017AISTATS}
\ppgccbibauthordoformat {Zafar, M.~B.\ppgccbibauthorsep {} Valera,
  I.\ppgccbibauthorsep {} Rodriguez, M.~G.\ppgccbibauthorlastsep {}
  \ppgccbibauthorand {} Gummadi, K.~P.} (\ppgccyeardoformat {2017b}).
\newblock \ppgccbibtitledoformat {Fairness constraints: Mechanisms for fair
  classification}.
\newblock \ppgccbibinstring {} \ppgccbibbooktitledoformat {Proceedings of the
  20th International Conference on Artificial Intelligence and Statistics},
  AISTATS '17.

\bibitem[Zagheni et~al.\@, 2014]{Zagheni2014}
\ppgccbibauthordoformat {Zagheni, E.\ppgccbibauthorsep {} Garimella, V.
  R.~K.\ppgccbibauthorsep {} Weber, I.\ppgccbibauthorlastsep {}
  \ppgccbibauthorand {} State, B.} (\ppgccyeardoformat {2014}).
\newblock \ppgccbibtitledoformat {Inferring international and internal
  migration patterns from twitter data}.
\newblock \ppgccbibinstring {} \ppgccbibbooktitledoformat {Proceedings of the
  23rd International Conference on World Wide Web}.

\bibitem[Zemel et~al.\@, 2013]{zemel2013learning}
\ppgccbibauthordoformat {Zemel, R.\ppgccbibauthorsep {} Wu,
  Y.\ppgccbibauthorsep {} Swersky, K.\ppgccbibauthorsep {} Pitassi,
  T.\ppgccbibauthorlastsep {} \ppgccbibauthorand {} Dwork, C.}
  (\ppgccyeardoformat {2013}).
\newblock \ppgccbibtitledoformat {Learning fair representations}.
\newblock \ppgccbibinstring {} \ppgccbibbooktitledoformat {Proceedings of the
  30th International Conference on Machine Learning}, ICML '13,
  \ppgccbibpagesstring {} \ppgccpagedoformat {325}-\ppgccpagedoformat
  {}-\ppgccpagedoformat {333}.

\bibitem[Ziegler et~al.\@, 2005]{ziegler2005improving}
\ppgccbibauthordoformat {Ziegler, C.-N.\ppgccbibauthorsep {} McNee,
  S.~M.\ppgccbibauthorsep {} Konstan, J.~A.\ppgccbibauthorlastsep {}
  \ppgccbibauthorand {} Lausen, G.} (\ppgccyeardoformat {2005}).
\newblock \ppgccbibtitledoformat {Improving recommendation lists through topic
  diversification}.
\newblock \ppgccbibinstring {} \ppgccbibbooktitledoformat {ACM WWW}.

\end{thebibliography}
\end{document}